\documentclass[floatfix,twocolumn,showpacs,preprintnumbers,amsmath,amssymb,pra,superscriptaddress,longbibliography]{revtex4-1}
\usepackage{color}
\usepackage[usenames,dvipsnames,svgnames,table]{xcolor}
\usepackage[colorlinks=true,linkcolor=blue,urlcolor=blue,citecolor=blue]{hyperref}
\usepackage{mathtools}
\usepackage{graphicx}
\usepackage{dcolumn}
\usepackage{array}
\usepackage{lipsum}
\usepackage{bm}
\usepackage{subfigure}
\usepackage{amssymb}
\usepackage{multirow}
\usepackage{tabularx}
\usepackage{amsmath}
\usepackage{braket}
\graphicspath{{plots/}}
 \usepackage{lipsum}
\usepackage{mathrsfs}

\newcommand{\beq}{\begin{equation}}
\newcommand{\eeq}{\end{equation}}
\newcommand{\bea}{\begin{eqnarray}}
\newcommand{\eea}{\end{eqnarray}}

\begin{document}
\title{
\textit{Ab initio} path integral Monte Carlo approach to the momentum distribution of the uniform electron gas at finite temperature without fixed nodes
}

\author{Tobias Dornheim}
\email{t.dornheim@hzdr.de}

\affiliation{Center for Advanced Systems Understanding (CASUS), D-02826 G\"orlitz, Germany}
\affiliation{Helmholtz-Zentrum Dresden-Rossendorf (HZDR), D-01328 Dresden, Germany}

\author{Maximilian B\"ohme}
\affiliation{Center for Advanced Systems Understanding (CASUS), D-02826 G\"orlitz, Germany}
\affiliation{Helmholtz-Zentrum Dresden-Rossendorf (HZDR), D-01328 Dresden, Germany}
\affiliation{Technische  Universit\"at  Dresden,  D-01062  Dresden,  Germany}

\author{Burkhard Militzer}
\affiliation{Department of Earth and Planetary Science, University of California, Berkeley, California 94720, USA}
\affiliation{Department of Astronomy, University of California, Berkeley, California 94720, USA}

\author{Jan Vorberger}
\affiliation{Helmholtz-Zentrum Dresden-Rossendorf (HZDR), D-01328 Dresden, Germany}

\begin{abstract}
We present extensive new \textit{ab intio} path integral Monte Carlo results for the momentum distribution function $n(\mathbf{k})$ of the uniform electron gas (UEG) in the warm dense matter (WDM) regime over a broad range of densities and temperatures. This allows us to study the nontrivial exchange--correlation induced increase of low-momentum states around the Fermi temperature, and to investigate its connection to the related lowering of the kinetic energy compared to the ideal Fermi gas. In addition, we investigate the impact of quantum statistics on both $n(\mathbf{k})$ and the off-diagonal density matrix in coordinate space, and find that it cannot be neglected even in the strongly coupled electron liquid regime.
Our results were derived without any nodal constraints, and thus constitute a benchmark for other methods and approximations.
\end{abstract}

\maketitle

\section{Introduction\label{sec:introduction}}

Warm dense matter (WDM) -- an exotic state with extreme densities and temperatures -- has emerged as an active frontier in plasma physics and material science~\cite{fortov_review,wdm_book,new_POP}. In nature, such conditions occur in astrophysical objects such as giant planet interiors~\cite{Nettelmann2008,Militzer_2008,militzer1,Benuzzi_Mounaix_2014}, brown dwarfs~\cite{saumon1,becker}, and neutron star crusts~\cite{Chamel2008}.
In addition, WDM has been predicted to occur on the pathway of a fuel capsule towards inertial confinement fusion~\cite{hu_ICF}. It may also be important as a catalyst for hot-electron chemistry~\cite{Brongersma2015}. Finally, these extreme conditions harbor potential for the discovery of novel materials like lonsdaleite~\cite{Kraus2016}, nanodiamonds~\cite{Kraus2017}, or superhard BC8 allotropes of carbon or silicon~\cite{Lazicki2021}.

 WDM states are now routinely generated in large laboratory research facilities using different compression techniques~\cite{falk_wdm}. At the same time, a rigorous theoretical description is challenging to obtain due to the nontrivial interplay of Coulomb correlations, thermal excitations, and fermionic quantum degeneracy effects of the electrons~\cite{wdm_book,new_POP,review}. This is often characterized by two parameters that are both of the order of unity at WDM conditions: a) the density parameter $r_s=\overline{r}/a_\textnormal{B}$ (with $\overline{r}$ being the average inter-particle distance and $a_\textnormal{B}$ being the Bohr radius) and b) the degeneracy temperature $\theta=k_\textnormal{B}T/E_\textnormal{F}$, where $E_\textnormal{F}$ denotes the noninteracting Fermi energy~\cite{quantum_theory,Ott2018}.

In this situation, \textit{ab initio} path integral Monte Carlo (PIMC) techniques~\cite{cep} constitute a promising method as they in principle allow for an exact solution of the fully correlated quantum many-body problem without any empirical input (like the exchange--correlation functional in density functional theory). Unfortunately, PIMC simulations of electrons are severely hampered by the well-known fermion sign problem~\cite{Ce91,binder,dornheim_sign_problem}, which leads to an exponential increase in computation time with increasing system size and decreasing temperature. In fact, the sign problem has been revealed to be $NP$-hard by Troyer and Wiese for some Hamiltonians under specific assumptions~\cite{troyer}, which makes it less likely that an exact and general solution can be found.

In practice, there are two different ways to proceed: Ceperley and colleagues~\cite{PC94} have introduced a nodal constraint on the thermal density matrix (commonly known as \emph{fixed-node} approximation), which completely removes the sign problem for the diagonal elements of the density matrix. The resulting restricted PIMC (RPIMC) simulations have enabled one to study systems with nuclei and hundreds of electrons. Starting with hydrogen~\cite{Ma96,MC01} and helium~\cite{Mi06}, the RPIMC method has been extended heavier elements. First, free-article nodes were employed to simulate elements up to neon~\cite{Driver2012,Driver2015} and later Hartree-Fock nodes were introduced to perform simulations of hot, dense aluminum and silicon~\cite{Driver2018,MilitzerDriver2015}. The predictions were subsequently combined into an EOS database~\cite{FPEOS}. 
Recently, shock experiments on CH plastic reached gigabar pressures~\cite{Kritcher2020}. The findings were in good agreement with earlier PIMC predictions~\cite{ZhangCH2017,ZhangCH2018}. However, these practical advantages come at a cost. For interacting systems, the nodes are not exactly known and therefore introduce an uncontrolled approximation. For high electronic densities and low temperatures, the RPIMC sampling becomes inefficient and some RPIMC predictions have been shown to be inaccurate~\cite{lee2020phaseless,dornheim_POP,Schoof_PRL_2015,review,Malone_PRL_2016}.

A second possible route towards describing WDM with simulations is to employ unbiased PIMC methods without any nodal constraints to exactly solve the more basic uniform electron gas (UEG), where the nuclei are replaced by a positive homogeneous background~\cite{Fraser_Foulkes_PRB_1996,review,loos}. In fact, the PIMC simulation of the UEG at WDM conditions has been a highly active field of research over the last years~\cite{Brown_PRL_2013,lee2020phaseless,dornheim_POP,Schoof_PRL_2015,review,Malone_PRL_2016,Malone_JCP_2015,Dornheim_PRL_2020,dornheim_prb_2016,dornheim_prl,dynamic_folgepaper,dornheim_dynamic,groth_jcp,dornheim_pre}, which finally culminated in the first parametrization of the exchange-correlation free energy $f_\textnormal{xc}$~\cite{groth_prl,ksdt} over a wide range of conditions. Such a consistently temperature-dependent XC-functional can then be used in thermal DFT~\cite{Mermin_DFT_1965} calculations of WDM. The importance of thermal XC-effects has been independently verified with theoretical methods by different groups~\cite{Sjostrom_PRB_2014,kushal,karasiev_importance}.

A number of properties of the warm dense UEG have been studied based on PIMC methods. This includes the static density response and the related local field correction~\cite{dornheim_ML,dornheim_electron_liquid,dornheim_HEDP,Dornheim_PRL_2020_ESA}, dynamic properties like the dynamic structure factor $S(\mathbf{k},\omega)$~\cite{dornheim_dynamic,dynamic_folgepaper,Hamann_CPP_2020,Hamann_PRB_2020,Dornheim_PRE_2020}, and even the nonlinear density response to a strong perturbation~\cite{Dornheim_PRL_2020}. At the same time, one fundamental property of the UEG -- the momentum distribution function $n(\mathbf{k})$ -- has been substantially less understood. RPIMC results for $1/16 \leq \theta\leq1$ for $r_s = 4$ and 40 were presented in Ref.~\cite{Militzer_PRL_2002,Militzer_momentum_HEDP_2019}. Hunger \textit{et al.}~\cite{Hunger_PRE_2021} recently presented results from configuration PIMC (CPIMC) simulations  covering the high-density regime of $r_s\lesssim1$.

In this work, we employ the direct PIMC simulation method that samples contributions from positive and negative permutations without imposing any nodal constraints. We present an extensive set of new PIMC results for $n(\mathbf{k})$ covering the entire range of relevant densities for low and moderate degeneracy, $\theta \geq 0.75$. This allows us to study the nontrivial increase in the occupation of low-momentum states due to XC-effects that were first reported in Ref.~\cite{Militzer_PRL_2002}. This occupation change is directly connected, though not equal, to an interaction-induced lowering of the kinetic energy. Furthermore, we study the impact of quantum statistics on $n(\mathbf{k})$. We find that the impact of Fermi statistics cannot be neglected even in the strongly coupled regime at $r_s=50$. We again stress that we do not impose any nodal restrictions, which makes our new data set an ideal benchmark for other approaches, like the fixed-node approximation.

The paper is organized as follows: In Sec.~\ref{sec:theory}, we introduce the relevant theoretical background, including the PIMC method (\ref{sec:PIMC}) and our approach to the estimation of $n(\mathbf{k})$ (\ref{sec:nk_theory}). Sec.~\ref{sec:results} contains all our new results, starting with a verification of our implementation by benchmarking against independent CPIMC results~(\ref{sec:verification}) and an analysis of the fermion sign problem and the related issue of finite-size effects (\ref{sec:FSC}). This is followed by a detailed physical discussion of the effects of density and temperature on $n(\mathbf{k})$ in Sec.~\ref{sec:rs_and_theta}, which also includes an investigation of the related lowering of the kinetic energy. Finally, we briefly touch upon the importance of quantum statistics (\ref{sec:statistics}) and compare our new simulation results to previous RPIMC data from Ref.~\cite{Militzer_momentum_HEDP_2019} (\ref{sec:RPIMC}). The paper is concluded by a summary and outlook in Sec.~\ref{sec:summary}.

\section{Theory\label{sec:theory}}
We assume Hartree atomic units throughout this work.

\subsection{Path integral Monte Carlo\label{sec:PIMC}}

We consider $N=N_\uparrow+N_\downarrow$ electrons in the canonical ensemble, i.e., the volume $V=L^3$, number density $n=N/V$, and inverse temperature $\beta=1/k_\textnormal{B}T$ are fixed. The expectation value of an arbitrary observable $\hat A$ is then given by
\begin{eqnarray}\label{eq:expectation}
\braket{\hat A} = \frac{1}{Z} \textnormal{Tr} \left( e^{-\beta\hat H} \hat A \right) \ ,
\end{eqnarray}
where the normalization $Z$ is given by the partition function
\begin{eqnarray}\label{eq:trace}
Z = \textnormal{Tr} \left( e^{-\beta\hat H} \right) \ .
\end{eqnarray}
We note that $\hat H$ denotes the usual Hamiltonian of the UEG which is given by an ideal kinetic part $\hat K$ and the Ewald interaction, see e.g. Ref.~\cite{review} for details. The basic idea of the PIMC method~\cite{cep} is to evaluate Eq.~(\ref{eq:trace}) in coordinate space, and taking into account the proper anti-symmetrization due to Fermi statistics leads to
\begin{eqnarray}\nonumber
Z &=& \frac{1}{N_\uparrow !N_\downarrow !}\sum_{\sigma_\uparrow\in S_{N_\uparrow}}\sum_{\sigma_\downarrow\in S_{N_\downarrow}}\textnormal{sgn}(\sigma_\downarrow)\textnormal{sgn}(\sigma_\uparrow)\\
& &\times \int \textnormal{d}\mathbf{R} \bra{\mathbf{R}}  e^{-\beta\hat H} \ket{\hat\pi_{\sigma_\uparrow}\hat\pi_{\sigma_\downarrow}\mathbf{R}}\ ,\label{eq:antisymmetry}
\end{eqnarray}
where $\mathbf{R}=(\mathbf{r}_0,\mathbf{r}_1,\dots ,\mathbf{r}_{N_1})^T$ includes the coordinates of all $N$ particles.
Evidently, Eq.~(\ref{eq:antisymmetry}) incorporates the summation over all permutation elements of the permutation groups of spin-up and -down fermions, with $\hat\pi_{\sigma_\uparrow}$ and $\hat\pi_{\sigma_\downarrow}$ being the corresponding permutation operators.
Yet, a direct evaluation of the matrix elements in Eq.~(\ref{eq:antisymmetry}) is not possible as the kinetic and interaction contribution to the full Hamiltonian do not commute. For the direct PIMC method, this problem is overcome by using an exact semi-group property of the density matrix, which allows to re-write the partition function as the integral over paths of particle coordinates in imaginary time, where the coordinates of all $N$ particles are evaluated on $P$ imaginary-time slices. 
The final expression for the partition function can then simply be written as the integral over all possible paths $\mathbf{X}$
\begin{eqnarray}\label{eq:PIMC_Z}
Z=\int\textnormal{d}\mathbf{X} W(\mathbf{X})\ ,
\end{eqnarray}
and the weight function can be readily evaluated; see Refs.~\cite{review,cep,boninsegni1} for a more detailed introduction to the PIMC method.

The Monte Carlo expectation value for Eq.~(\ref{eq:expectation}) takes the form
\begin{eqnarray}\label{eq:Monte_Carlo}
\braket{\hat A}_\textnormal{MC} = \frac{1}{N_\textnormal{MC}} \sum_{l=1}^{N_\textnormal{MC}} A(\mathbf{X}_l)\ ,
\end{eqnarray}
where the Monte Carlo estimator of the observable $\hat A$ is defined by 
\begin{eqnarray}
\braket{\hat A} = \frac{1}{Z} \int \textnormal{d}\mathbf{X}\ W(\mathbf{X}) A(\mathbf{X}) \ ,
\end{eqnarray}
and the configurations $\mathbf{X}_l$ in Eq.~(\ref{eq:Monte_Carlo}) are randomly generated according to the probability distribution $P(\mathbf{X})=W(\mathbf{X})/Z$ using the celebrated Metropolis algorithm~\cite{metropolis}.

For electrons, however, the weight function $W(\mathbf{X})$ is not strictly positive due to the fermionic antisymmetry of the density matrix under the exchange of particle coordinates. Therefore, $P(\mathbf{X})$ does not qualify as a legitimate probability distribution, and we instead randomly generate the paths according to
\begin{eqnarray}\label{eq:modified}
P'(\mathbf{X}) = \frac{1}{Z'} |W(\mathbf{X})| \ ,
\end{eqnarray}
with the modified normalization
\begin{eqnarray}\label{eq:Z_bose}
Z' = \int \textnormal{d}\mathbf{X}\ |W(\mathbf{X})| \ .
\end{eqnarray}
We note that in the case of direct PIMC, Eq.~(\ref{eq:Z_bose}) constitutes to the exact partition function of a Bose-system at the same conditions.
The exact fermionic expectation value can then be extracted by evaluating the ratio
\begin{eqnarray}\label{eq:ratio}
\braket{\hat A} = \frac{\braket{\hat A \hat S}'}{\braket{\hat S}'}\ ,
\end{eqnarray}
where $\braket{\dots}'$ indicates the expectation value computed from the probability distribution defined in Eq.~(\ref{eq:modified}). Here $S(\mathbf{X})=W(\mathbf{X})/|W(\mathbf{X})|$ is the estimator of the sign operator $\hat S$, and the denominator of Eq.~(\ref{eq:ratio}) is correspondingly simply being referred to as the \emph{average sign}.

In particular, $S \equiv \braket{\hat S}'$ constitutes a measure for the degree of cancellations between positive and negative contributions to the partition function $Z$, and exponentially decreases both toward low temperature and with increasing system size $N$. At the same time, the statistical uncertainty of the expectation value Eq.~(\ref{eq:ratio}) is inversely proportional to $S$~\cite{binder}, 
\begin{eqnarray}
\frac{\Delta A}{A}\sim \frac{1}{S\sqrt{N_\textnormal{MC}}}\ ,
\end{eqnarray}
which leads to an exponentially increasing error bar that can only be reduced by increasing the number of Monte Carlo samples (with scales linear in compute time) as $1/\sqrt{N_\textnormal{MC}}$. This \emph{exponential wall} is the notorious fermion sign problem, which constitutes the main limitation of our approach. A detailed yet accessible analysis of the sign problem has recently been presented in Ref.~\cite{dornheim_sign_problem}.

\subsection{PIMC evaluation of the momentum distribution\label{sec:nk_theory}}

The momentum distribution of $N_\sigma$ (with $\sigma\in[\uparrow,\downarrow]$ denoting the spin) electrons is defined as~\cite{Militzer_momentum_HEDP_2019}
\begin{eqnarray}\label{eq:momentum_distribution}
n_\sigma(\mathbf{k}) = \frac{(2\pi)^d}{V} \left<\sum_{l=1}^{N_\sigma}\delta\left({\mathbf{\hat{k}_l}}-\mathbf{k}\right) \right>\ ,
\end{eqnarray}
with the normalization
\begin{eqnarray}\label{eq:normalization}
\sum_\mathbf{k}n_\sigma(\mathbf{k}) = N_\sigma\ .
\end{eqnarray}
For an ideal (i.e., noninteracting) Fermi system, Eq.~(\ref{eq:momentum_distribution}) is given by the Fermi distribution
\begin{eqnarray}\label{eq:Fermi}
n_0(\mathbf{k}) = \frac{1}{1+\textnormal{exp}\left(\beta(E_\mathbf{k}-\mu)\right)}\ ,
\end{eqnarray}
with $E_\mathbf{k}=k^2/2$ and $\mu$ being the usual chemical potential~\cite{quantum_theory}.
Regarding the PIMC formalism, Eq.~(\ref{eq:momentum_distribution}) is off-diagonal in coordinate space and requires the presence of a single \emph{open trajectory} in the simulation scheme~\cite{cep}.

\begin{figure}\centering
\includegraphics[width=0.475\textwidth]{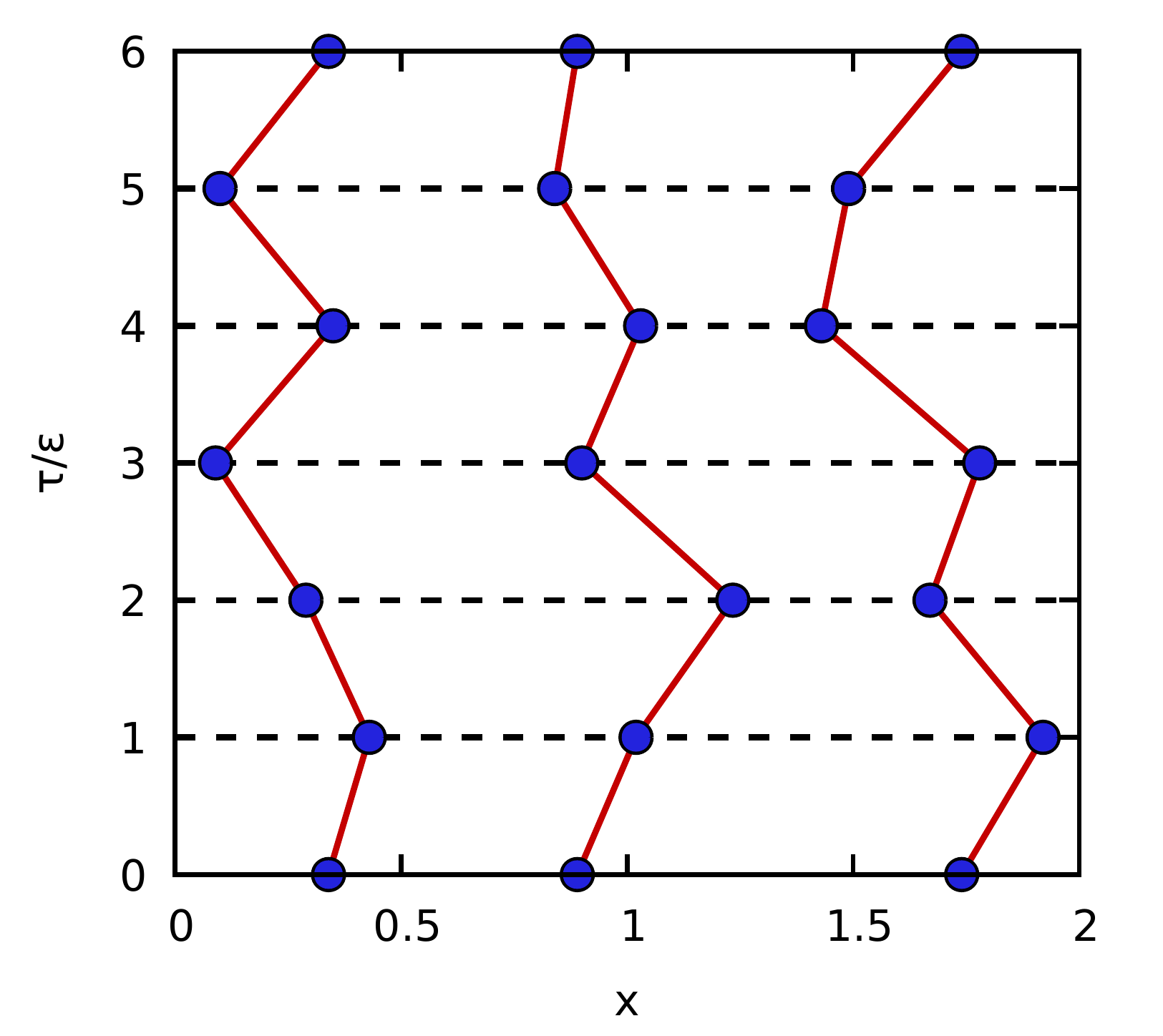}\\\includegraphics[width=0.475\textwidth]{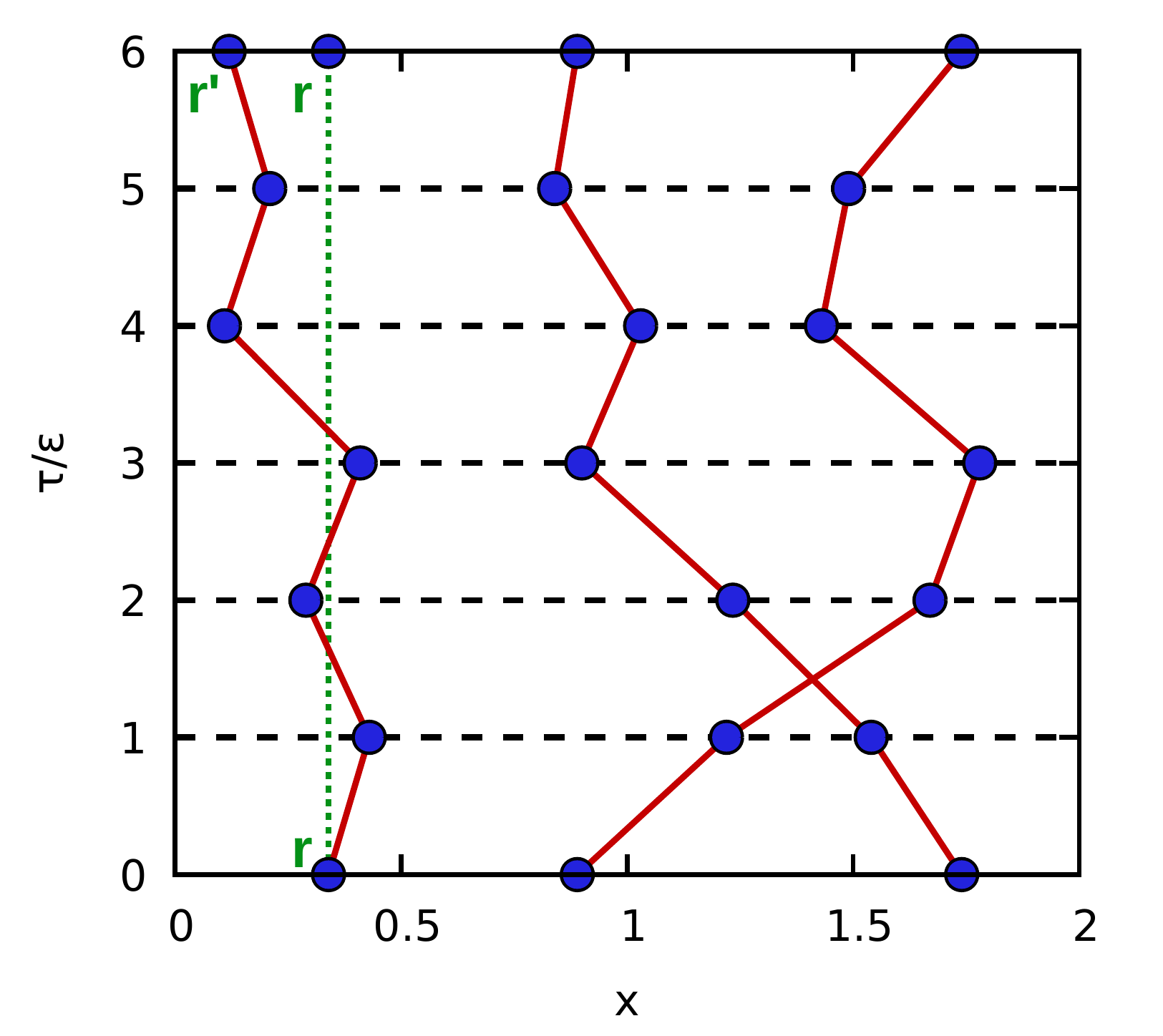}
\caption{\label{fig:illustration}
Schematic illustration of the canonical configuration space (top panel) defined by the canonical partition function $Z$, and the off-diagonal configuration space $Z_{\mathbf{r},\mathbf{r'};\sigma}$ including a single open trajectory (bottom panel).
}
\end{figure} 

This is illustrated in Fig.~\ref{fig:illustration}, where we show PIMC configurations with $N=3$ particles and $P=6$, with $\tau\in[0,\beta]$ being the imaginary time (and $\epsilon=\beta/P$ the imaginary-time step). As mentioned above, each particle is represented by an entire \emph{path} of particle coordinates, and the integration is carried out over the coordinates on all slices. The top panel shows a \emph{closed} configuration (the coordinates on slices 0 and $P$ are equal), which contributes to the usual canonical partition function $Z$.
In contrast, the bottom panel depicts an \emph{open} configuration, where the coordinates of one particle deviate between the last ($\mathbf{r}'$) and first ($\mathbf{r}$) time slice.
Let us define this extended configuration space (with $\sigma$ being the spin-orientation of $\mathbf{r}$ and $\mathbf{r}'$) as  
\begin{eqnarray}\label{eq:Z_open}
Z_{\mathbf{r},\mathbf{r'};\sigma} = \int \textnormal{d}\mathbf{\tilde{X}} W(\mathbf{\tilde{X}})\ ,
\end{eqnarray}
where $\mathbf{\tilde{X}}=(\mathbf{X},\mathbf{r}')^T$ includes the off-diagonal end-coordinate $\mathbf{r}'$ as an additional degree of freedom. In addition, we note that the bottom panel contains a permutation cycle of two particles, leading to a negative sign of the configuration weight.

Following Refs.~\cite{cep,Militzer_momentum_HEDP_2019}, the translation of Eq.~(\ref{eq:momentum_distribution}) into the imaginary-time path integral picture then leads to the expression
\begin{eqnarray}\label{eq:nk_formula}
n_\sigma(\mathbf{k}) = \frac{1}{V} \frac{Z_{\mathbf{r},\mathbf{r'};\sigma}}{Z} \left<
e^{i\mathbf{k}(\mathbf{r}-\mathbf{r}')}
\right>_{\mathbf{r},\mathbf{r'};\sigma}\ ,
\end{eqnarray}
where the expectation value is computed with respect to the configuration space defined by Eq.~(\ref{eq:Z_open}) according to the probability density $P(\mathbf{\tilde{X}})=W(\mathbf{\tilde{X}})/Z_{\mathbf{r},\mathbf{r'};\sigma} $.
While the direct Monte Carlo sampling of paths within this off-diagonal configuration space is straightforward using slightly modified standard techniques, this would not allow the complete evaluation of Eq.~(\ref{eq:nk_formula}) as the ratio of the normalizations is a-priori unknown. In principle, the missing factor can be obtained either from Eq.~(\ref{eq:normalization}) or the normalization of the related off-diagonal single-particle density matrix $n(\mathbf{r},\mathbf{r}')$; see Ref.~\cite{Militzer_momentum_HEDP_2019} for details.
Yet, this introduces an additional source of uncertainty and potentially significantly reduces the attainable degree of accuracy of $n(\mathbf{k})$ itself.

To avoid this issue, we follow the basic idea of the continuous-space \emph{worm algorithm} by Boninsegni \textit{et al.}~\cite{boninsegni1}, and introduce the extended configuration space
\begin{eqnarray}\label{eq:Z_tot}
Z_\textnormal{tot} = Z + cP \sum_\sigma Z_{\mathbf{r},\mathbf{r'};\sigma}\ ,
\end{eqnarray}
which combines the canonical (closed) configurations $\mathbf{X}$ with off-diagonal (open) configurations $\mathbf{\tilde{X}}$. Here the factor $c$ can be freely chosen to optimize the ratio of $\mathbf{X}$- and $\mathbf{\tilde{X}}$ configurations, and the factor $P$ comes from the fact that, within our simulations, $\mathbf{r}$ and $\mathbf{r'}$ can be located on any time slice.

It is straightforward to see that
\begin{eqnarray}\label{eq:count_G}
\braket{\delta(\mathbf{r},\mathbf{r'};\sigma)}_\textnormal{tot} = \frac{cP Z_{\mathbf{r},\mathbf{r'};\sigma}}{Z_\textnormal{tot}}\ ,
\end{eqnarray}
and, similarly,
\begin{eqnarray}\label{eq:count_Z}
\braket{\delta_Z}_\textnormal{tot} = \frac{ Z}{Z_\textnormal{tot}}\ ,
\end{eqnarray}
where the expectation values are computed in the combined ensemble, and the estimators are defined as
\begin{eqnarray}\nonumber 
\delta(\mathbf{r},\mathbf{r'};\sigma) &=& \begin{cases}
      1, & \textnormal{configuration with open path of spin $\sigma$} \\
      0, & \text{otherwise}
    \end{cases}\\
\delta_Z &=& \begin{cases}
      1, & \textnormal{closed configuration} \\
      0, & \text{otherwise}\ .
    \end{cases}
\end{eqnarray}
In combination, Eqs.~(\ref{eq:count_G}) and (\ref{eq:count_Z}) allow to exactly estimate the previously unknown ratio of the normalizations as
\begin{eqnarray}\label{eq:proportionality}
\frac{Z_{\mathbf{r},\mathbf{r'};\sigma}}{Z} = \frac{1}{cP} \frac{\braket{\delta(\mathbf{r},\mathbf{r'};\sigma)}_\textnormal{tot}}{\braket{\delta_Z}_\textnormal{tot}}\ .
\end{eqnarray}
The final expression for the momentum distribution of spin-component $\sigma$ is thus given by
\begin{eqnarray}\label{eq:nk_final}
n(\mathbf{k}) = \frac{1}{cPV} \frac{\braket{\delta(\mathbf{r},\mathbf{r'};\sigma)}_\textnormal{tot}}{\braket{\delta_Z}_\textnormal{tot}} \left<
e^{i\mathbf{k}(\mathbf{r}-\mathbf{r}')}
\right>_{\mathbf{r},\mathbf{r'};\sigma}\ .
\end{eqnarray}
Furthermore, we always sample configurations according to the modulus value of the configuration space in either $Z$, $Z_{\mathbf{r},\mathbf{r'};\sigma}$, or $Z_\textnormal{tot}$, and the evaluation of the ratio Eq.~(\ref{eq:ratio}) then transforms Eq.~(\ref{eq:nk_final}) to
\begin{eqnarray}\label{eq:nk_final_final}
n(\mathbf{k}) = \frac{1}{cPV} \frac{\braket{\delta(\mathbf{r},\mathbf{r'};\sigma)\hat S}_\textnormal{tot}'}{\braket{\delta_Z\hat S}_\textnormal{tot}'} \frac{\left<
e^{i\mathbf{k}(\mathbf{r}-\mathbf{r}')} \hat S
\right>_{\mathbf{r},\mathbf{r'};\sigma}'}{\braket{\hat S}_{\mathbf{r},\mathbf{r'};\sigma}'}\ .
\end{eqnarray}
Let us conclude this section with a few practical remarks. In principle, our simulation scheme can be viewed as a truncated, canonical version of the grand-canonical \emph{worm algorithm} described in Ref.~\cite{boninsegni1}. Correspondingly, we sample particle exchange in the open sector by using the \emph{Swap}-update, which allows for a large acceptance ratio due to its local nature. 
Furthermore, this approach allows us to compute both standard canonical observables (like different energies, the static structure factor $S(\mathbf{k})$, etc.) and off-diagonal observables like $n_\sigma(\mathbf{k})$ and $n(\mathbf{r},\mathbf{r}')$ within the same run, without any computational extra cost. In fact, even if one was only interested in closed configurations, the extended configuration space constitutes an efficient way to switch between different permutation sectors and simulation time within open configurations is, thus, rarely wasted.

\section{Results\label{sec:results}}

\subsection{Verification of the implementation\label{sec:verification}}

\begin{figure*}\centering
\includegraphics[width=0.475\textwidth]{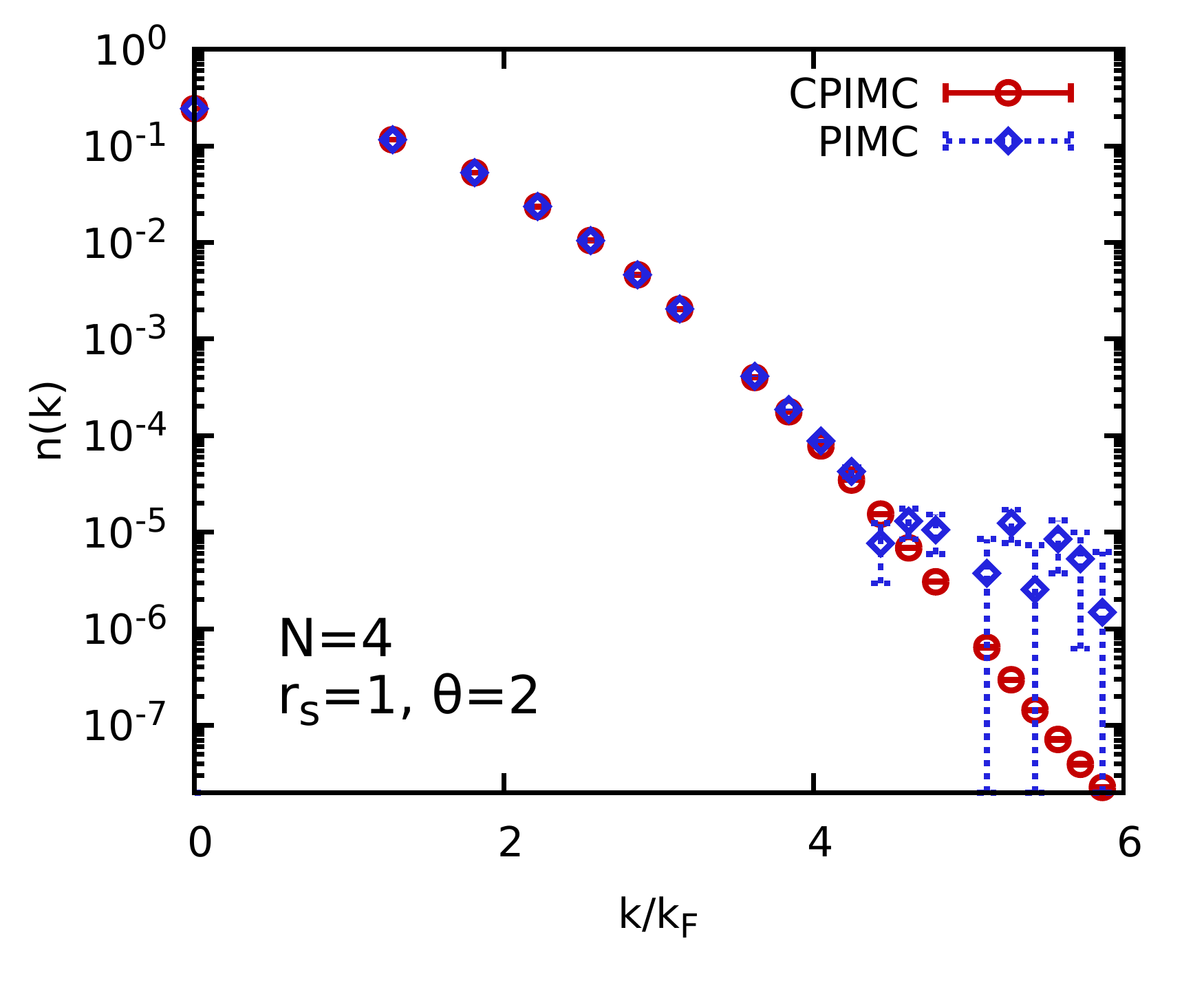}
\includegraphics[width=0.475\textwidth]{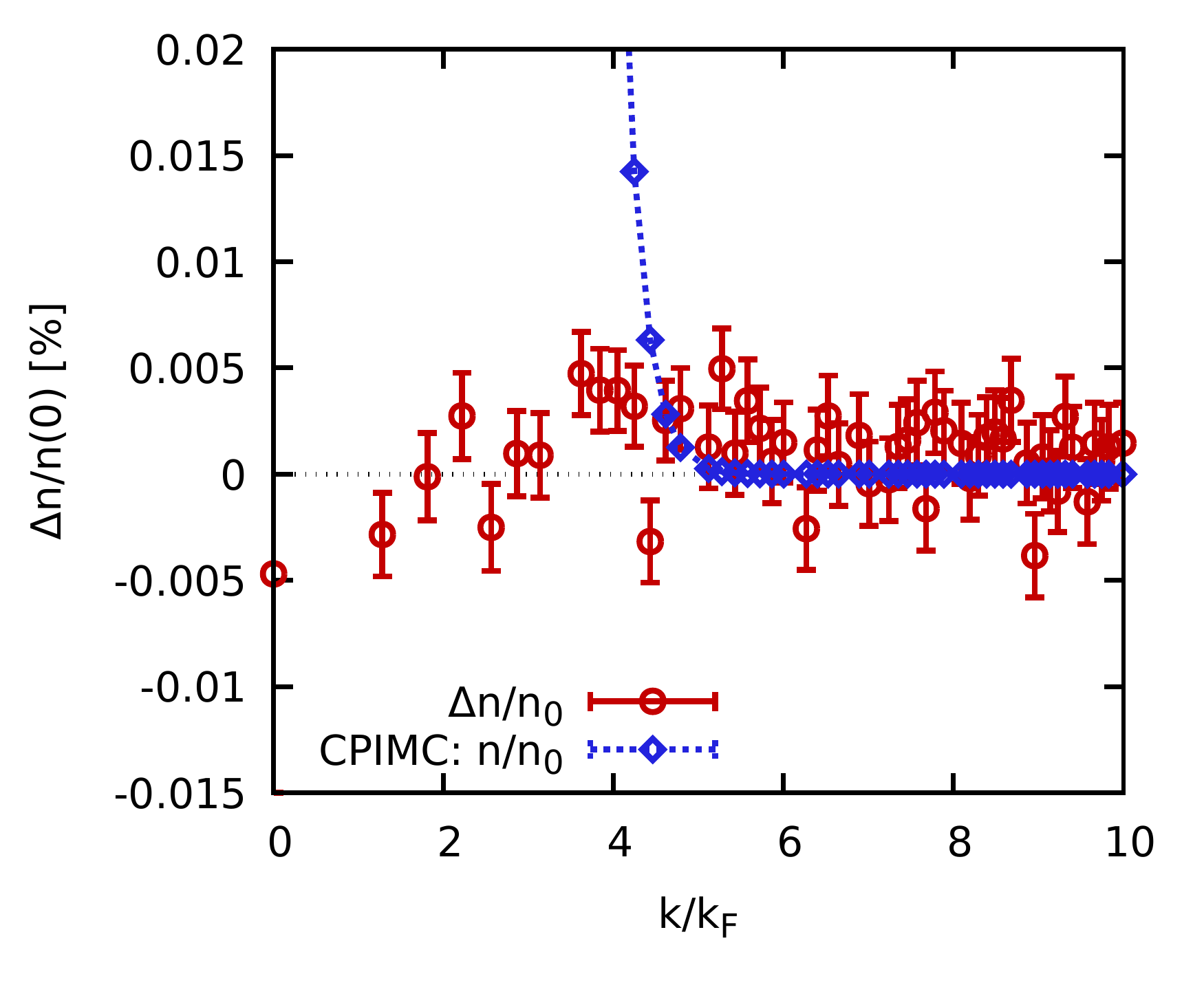}\\\vspace*{-1.45cm}
\includegraphics[width=0.475\textwidth]{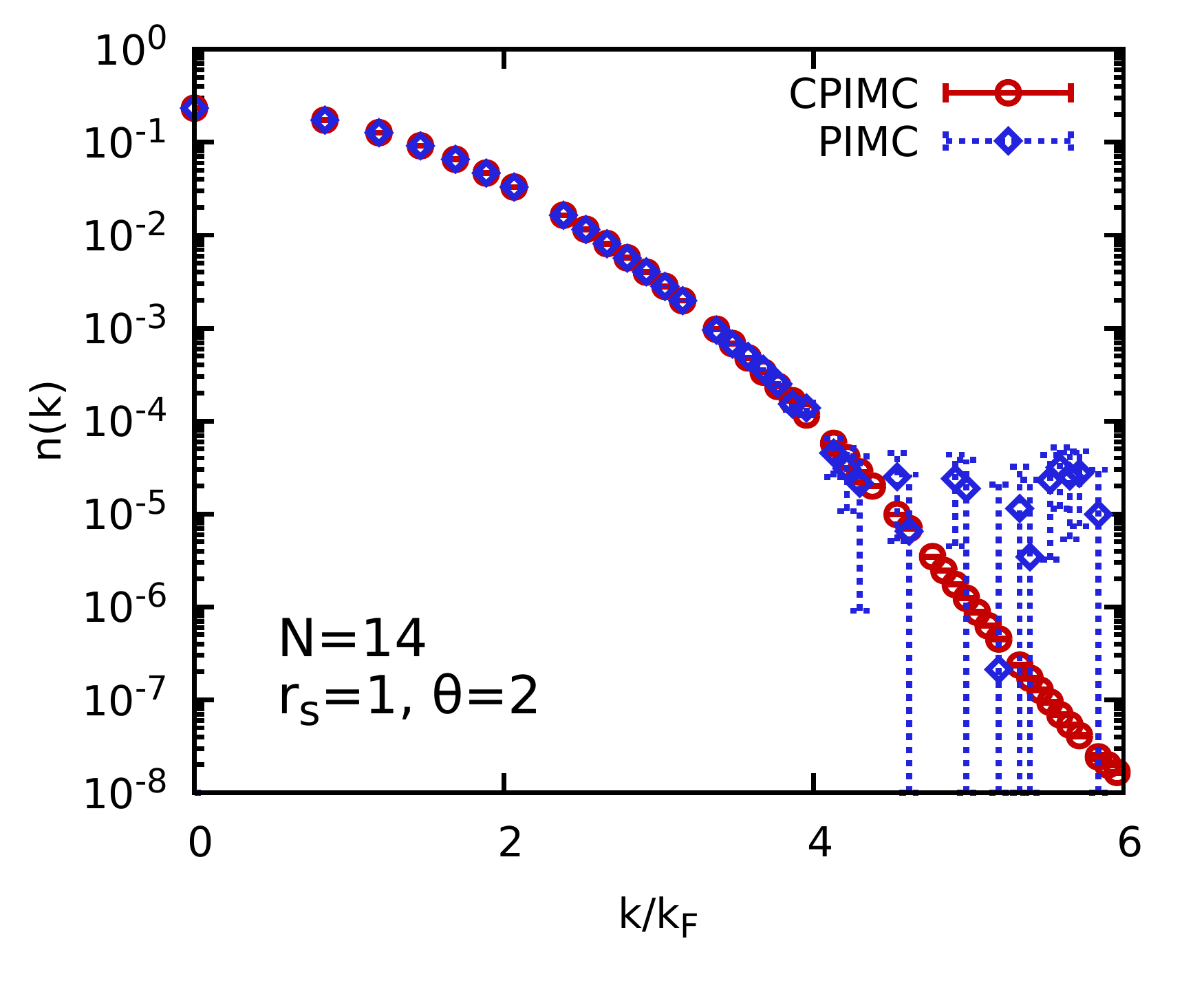}\hspace*{0.012\textwidth}
\includegraphics[width=0.463\textwidth]{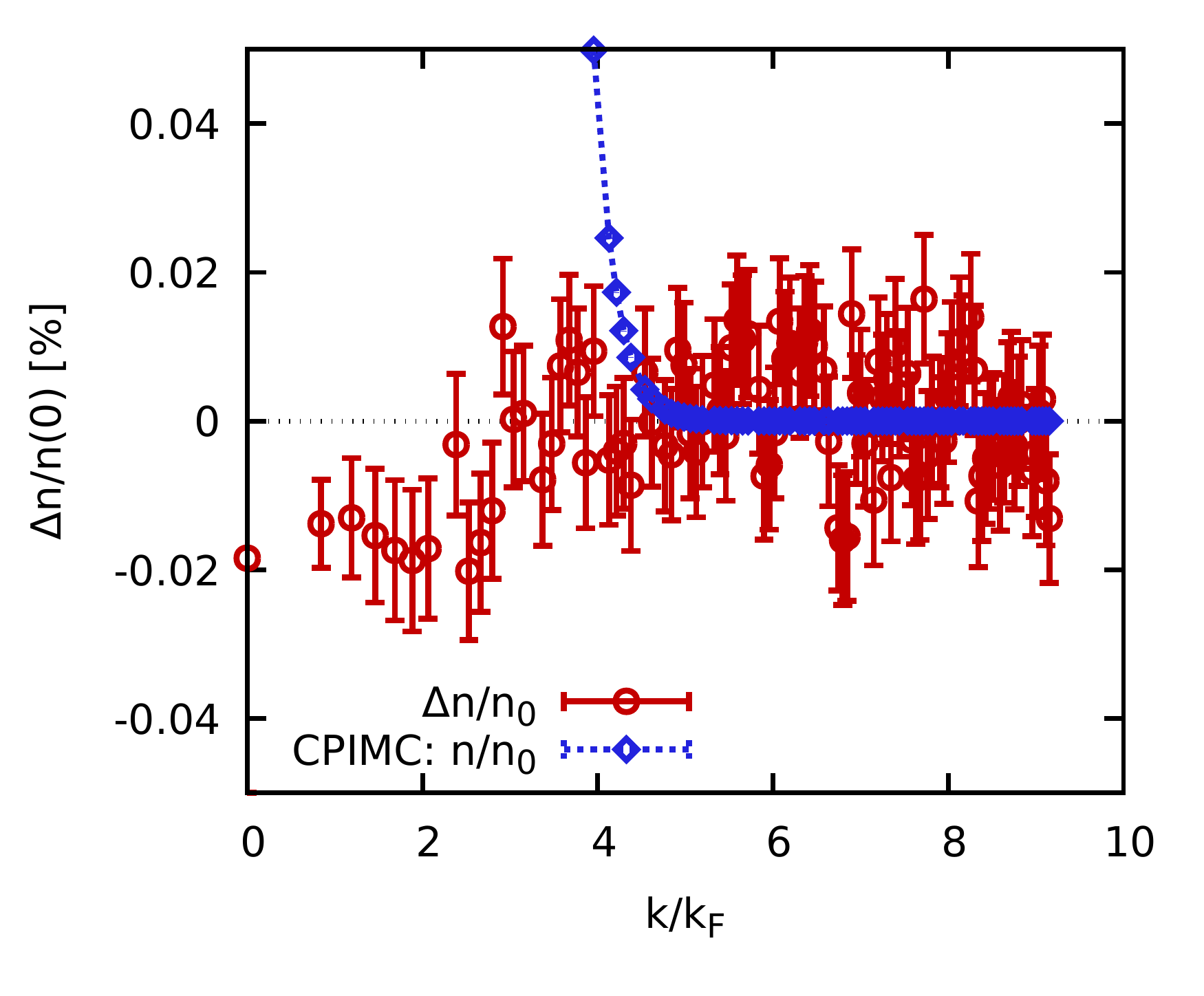}
\caption{\label{fig:spectrum_rs1_theta2}
Left: Momentum distribution function $n(\mathbf{k})$ of the UEG with $N=4$ (top row) and $N=14$ (bottom row) unpolarized electrons at $r_s=1$ and $\theta=2$; red circles: CPIMC~\cite{Hunger_PRE_2021}; blue diamonds: PIMC results, cf.~Eq.~(\ref{eq:nk_final}). Right panel: difference between CPIMC and PIMC (red circles) as a percentage of $n(\mathbf{0})$; blue diamonds: actual value of $n/n(\mathbf{0})$ [also as a percentage of $n(0)$] computed from CPIMC.
}
\end{figure*}

Let us begin the discussion of our new simulation results by benchmarking against exact CPIMC data~\cite{Hunger_PRE_2021}. This is shown in Fig.~\ref{fig:spectrum_rs1_theta2}, where we compare the momentum distribution for $N=4$ (top row) and $N=14$ (bottom row) unpolarized electrons at $r_s=1$ and $\theta=2$. The left column directly shows $n(\mathbf{k})$ on a logarithmic scale, and the red circles (blue diamonds) correspond to the CPIMC (direct PIMC) results. Evidently, we find excellent agreement between the two independent data sets over more than four orders of magnitude in the momentum distribution for both system sizes. Around $k\approx 4k_\textnormal{F}$, the relative error bars of the PIMC data become large, and $n(\mathbf{k})$ can no longer be clearly resolved.

This can be seen particularly well in the right column where we show the relative deviation between PIMC and CPIMC in units of $n(0)$ as a percentage,
\begin{eqnarray}\label{eq:percent}
\frac{\Delta n}{n(0)} [\%] = \frac{n_\textnormal{PIMC}(\mathbf{k})-n_\textnormal{CPIMC}(\mathbf{k})}{n_\textnormal{CPIMC}(\mathbf{0})}\times 100\ ,
\end{eqnarray}
as the red circles. First and foremost, we stress the high accuracy of our PIMC data, with a relative statistical uncertainty of $\lesssim10^{-4}$ for small wave numbers. In addition, we note that the absolute magnitude of the PIMC error bars stays approximately constant over the entire $k$-range. At the same time, $n(\mathbf{k})$ rapidly decreases with increasing $k$, which in turn means that the signal-to-noise ratio in the PIMC data eventually vanishes.

The CPIMC method is directly formulated in momentum-space and thus has a substantially smaller intrinsic variance in the estimator of $n(\mathbf{k})$. This straightforwardly translates into the reduced CPIMC error bars, which even allows for the resolution of \emph{quantum tail} effects (see Eq.~(\ref{eq:k8-g0}) below) at $k\gtrsim7k_\textnormal{F}$ reported by Hunger \textit{et al.}~\cite{Hunger_PRE_2021}.
Still, the CPIMC method breaks down at large $r_s$, which makes the investigation of $n(\mathbf{k})$ using complementary methods that are formulated in coordinate space indispensable.

The blue diamonds in the right column of Fig.~\ref{fig:spectrum_rs1_theta2} show the CPIMC data for $n(\mathbf{k})/n(0)$ [also in percent of $n(0)$], which clearly shows the rapid decay of the momentum distribution mentioned earlier. In particular, the CPIMC results for $n(\mathbf{k})$ become comparable to the PIMC error bars around $k=4k_\textnormal{F}$, which thus constitutes the natural limit up to which PIMC results are reliable.

\begin{figure*}\centering
\includegraphics[width=0.475\textwidth]{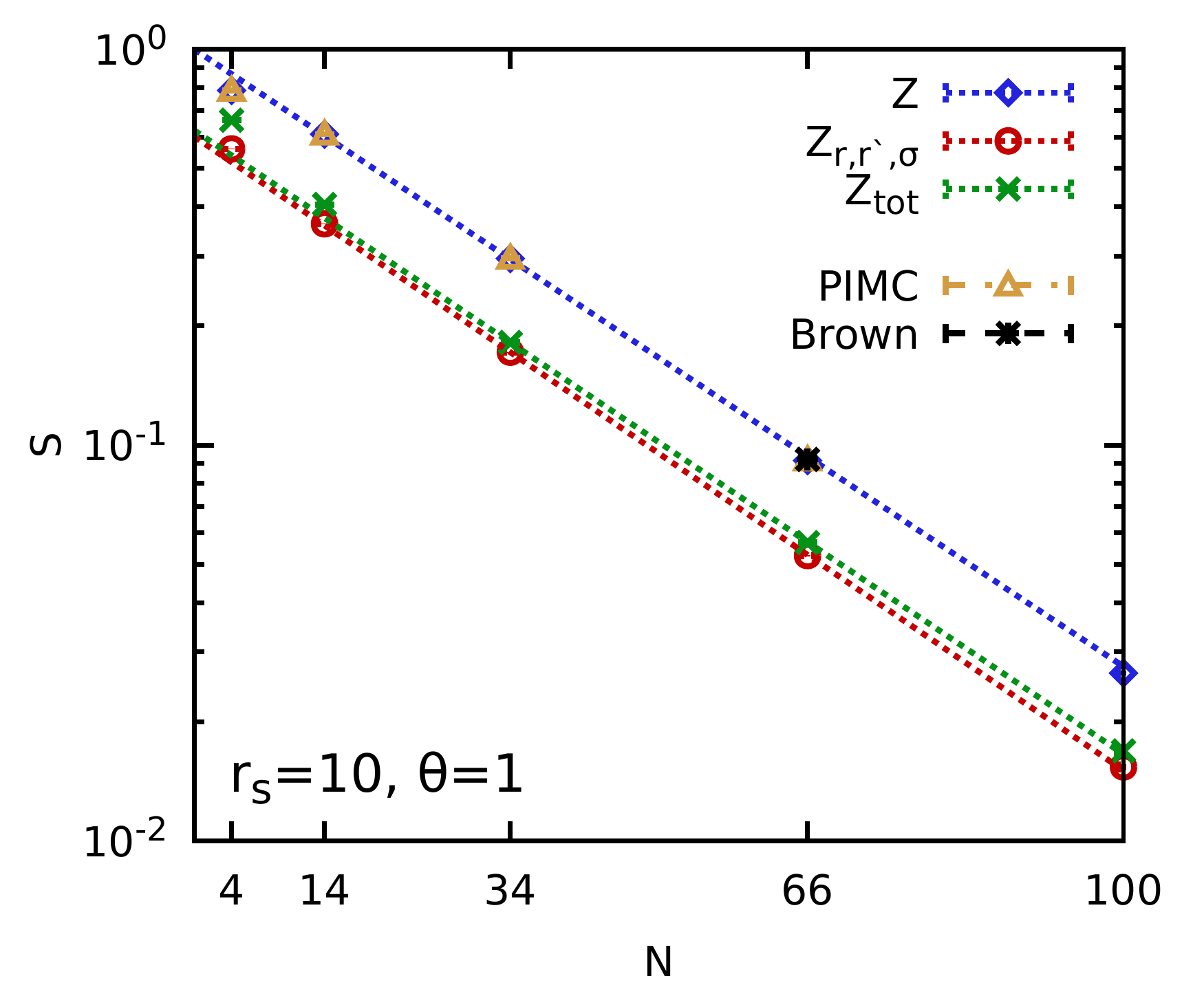}\includegraphics[width=0.475\textwidth]{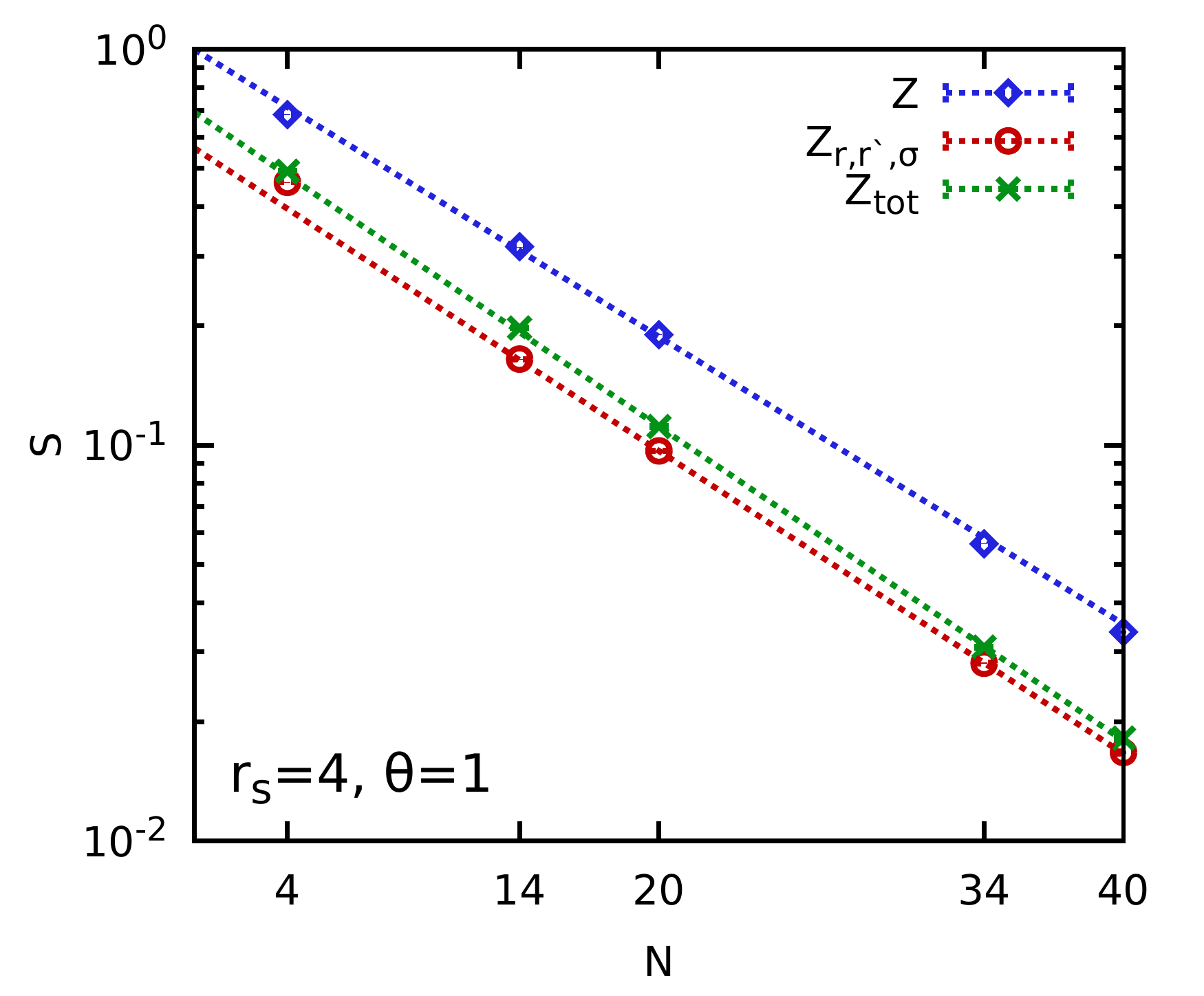}
\caption{\label{fig:sign_rs10_theta1}
System-size dependence of the average sign at $\theta=1$ for $r_s=10$ (left) and $r_s=4$ (right). The blue diamonds, red circles, and green crosses have been obtained within the canonical ensemble ($Z$), the off-diagonal ensemble ($Z_{\mathbf{r},\mathbf{r'},\sigma}$), and the combined ensemble ($Z_\textnormal{tot}$), respectively. The dotted lines show exponential fits~\cite{dornheim_sign_problem} taking into account the three largest values of $N$. Yellow triangles: independent PIMC results from canonical simulations; black star: taken from Brown \textit{et al.}~\cite{Brown_PRL_2013}.
}
\end{figure*}

\subsection{System size dependence\label{sec:FSC}}

Throughout this work, we use the direct PIMC method without any nodal constraints. Therefore, our simulations are afflicted with the fermion sign problem~\cite{dornheim_sign_problem}, which strongly limits the feasible number of electrons. This is demonstrated in Fig.~\ref{fig:sign_rs10_theta1}, where we show the $N$-dependence of the average sign $S$ computed within the three different configuration spaces at $\theta=1$ for $r_s=10$ (left panel) and $r_s=4$ (right panel). 
In particular, $r_s=10$ constitutes a rather low density and can be viewed as the boundary towards the strongly coupled electron liquid regime~\cite{dornheim_dynamic,dornheim_electron_liquid}. While such conditions are exotic in the sense that they are currently beyond the reach of even evaporation experiments~\cite{Zastrau}, they are still interesting from a theoretical perspective and offer exciting physical phenomena like a possible incipient excitonic mode~\cite{dornheim_dynamic,dynamic_folgepaper,Takada_PRB_2016,Takada_PRL_2002}. Moreover, they offer the rare possibility to study strongly correlated quantum systems, which is interesting in its own right.
The density in the right panel ($r_s=4$), on the other hand, can be realized in experiments with metals such as sodium~\cite{Huotari_PRL_2010}, or via hydrogen jets~\cite{Zastrau}.

The blue diamonds have been obtained within the canonical ensemble, i.e., using only closed paths within the simulation. As an additional verification of our implementation, we have also included previous PIMC results for the sign at $r_s=10$ for $4\leq N \leq 66$ as the yellow triangles, which are in excellent agreement to our new data. In addition, the single black star at $N=66$ in the left panel has been taken from Ref.~\cite{Brown_PRL_2013} and, too, agrees to both aforementioned data sets. The dotted blue lines have been obtained from an exponential fit (see Ref.~\cite{dornheim_sign_problem} for details) taking into account the three largest system-sizes in both cases. This indeed confirms the exponential decrease of $S$ with $N$ and thus the exponential increase of computation time mentioned in Sec.~\ref{sec:PIMC} above. In practice, this means that simulations are limited to $N=100$ ($N=40$) electrons for $r_s=10$ ($r_s=4$) at $\theta=1$. A similar scaling has recently been presented~\cite{dornheim_sign_problem} upon decreasing the temperature $\theta$, which makes the thorough analysis of finite-size effects~\cite{dornheim_prl,Holzmann_PRB_FSC_2016,Chiesa_PRL_2006,dornheim2021overcoming} presented below indispensable.

We next discuss the red circles, which have been obtained in the configuration space of open paths corresponding to $Z_{\mathbf{r},\mathbf{r'},\sigma}$. Firstly, we find that these data exhibit a qualitatively similar dependence on $N$ as the blue diamonds, i.e., an exponential decrease towards larger system size. Further, the red circles consistently attain lower values in $S$ compared to the blue diamonds, which is due to the increased probability of particle exchange due to the presence of the open trajectory. From a practical perspective, this means that the simulation in the offdiagonal configuration space are computationally more involved, which makes the computation of $n(\mathbf{k})$ for fermions particularly challenging. In fact, even the nodal constraint in the RPIMC method~\cite{Militzer_momentum_HEDP_2019} does not fully remove the sign problem for off-diagonal configurations (whereas there is no sign problem in RPIMC in the space of closed configurations).

Finally, the green crosses depict the average signs in the combined configuration space of both open and closed configurations defined by $Z_\textnormal{tot}$ in Eq.~(\ref{eq:Z_tot}) above. Naturally, the respective average value of the sign is located between those two cases and exhibits a similar exponential decay with $N$.

\begin{figure*}\centering
\includegraphics[width=0.475\textwidth]{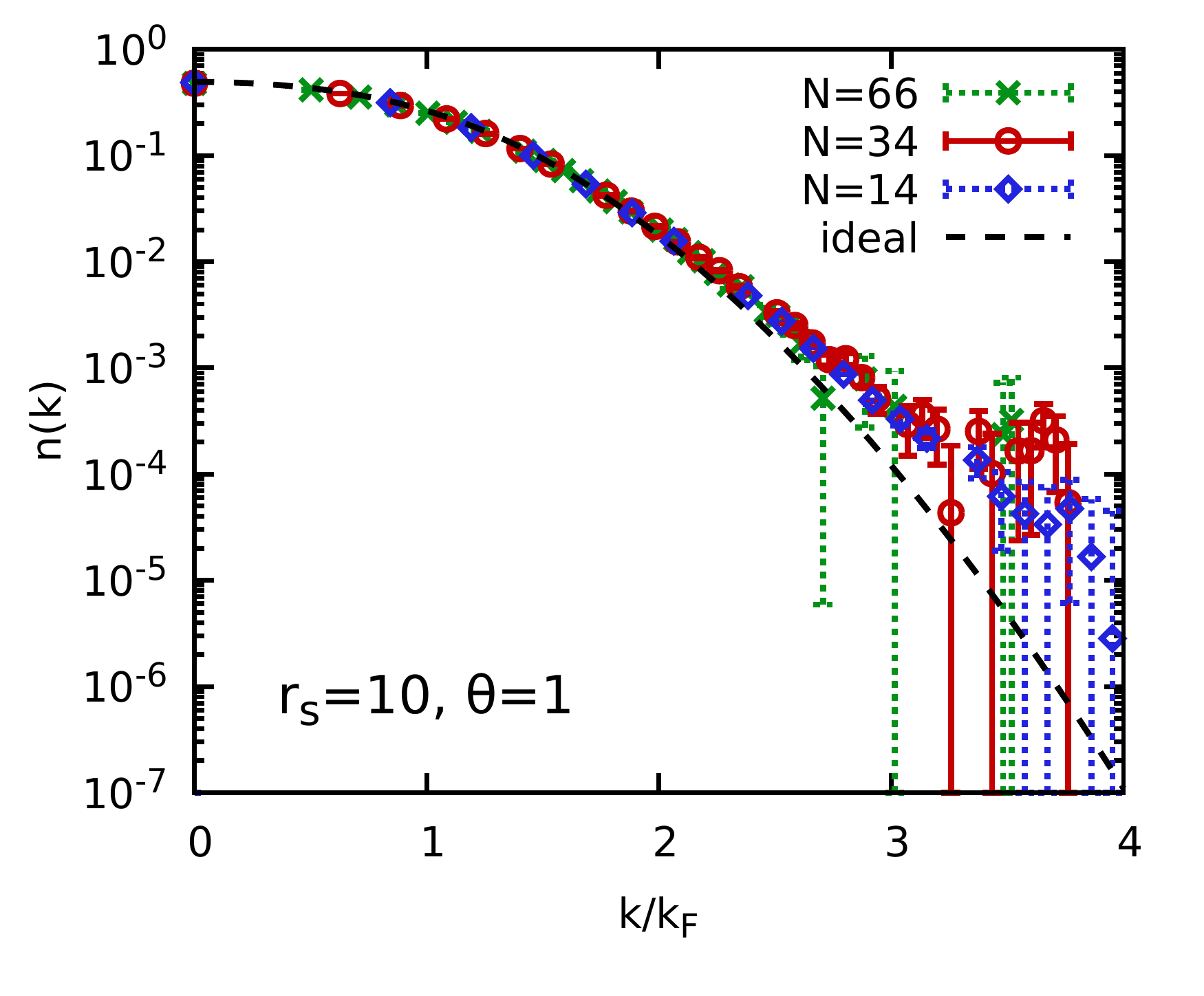}\includegraphics[width=0.475\textwidth]{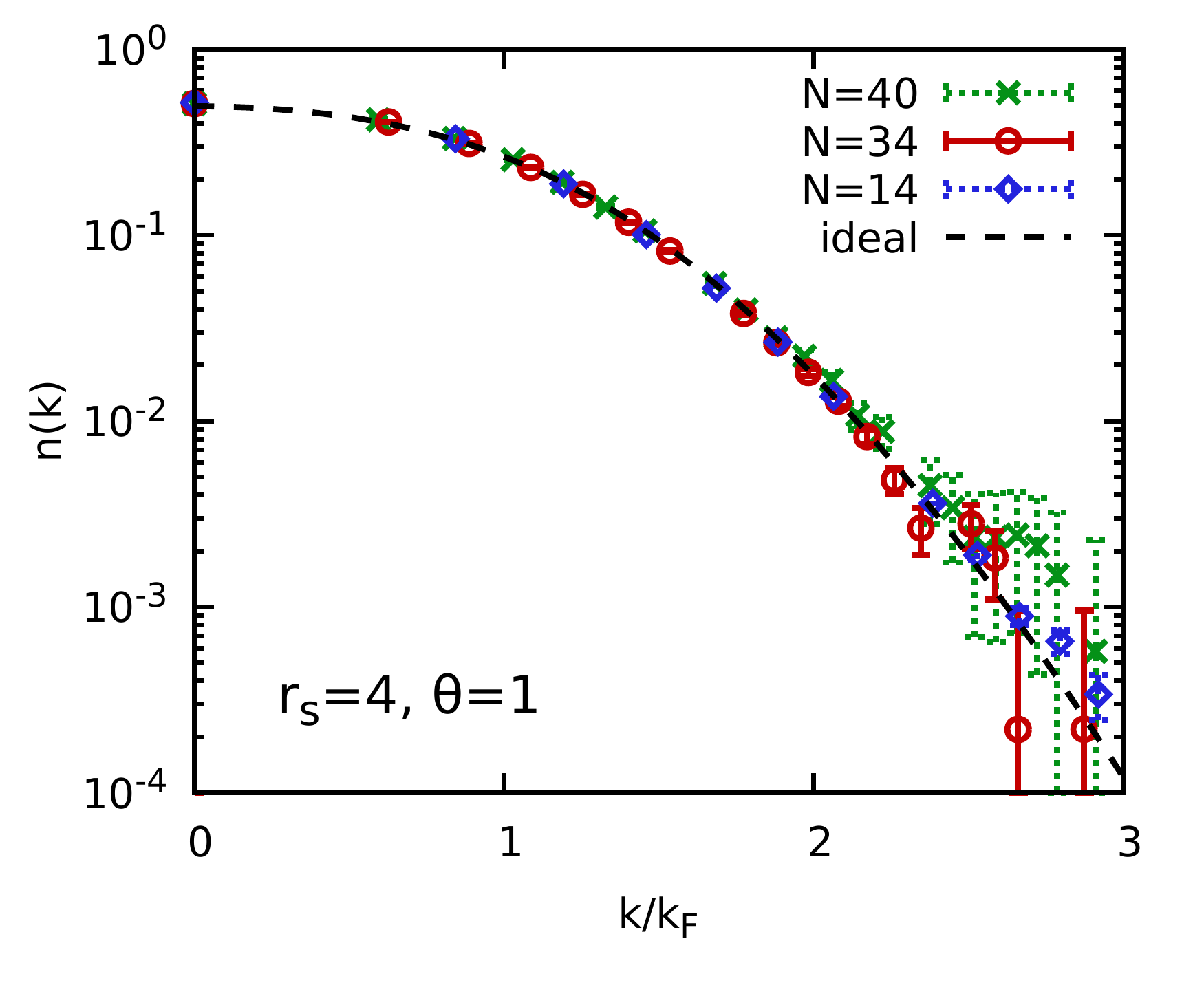}\\\vspace*{-1.3cm}
\hspace*{-0.01\textwidth}\includegraphics[width=0.49\textwidth]{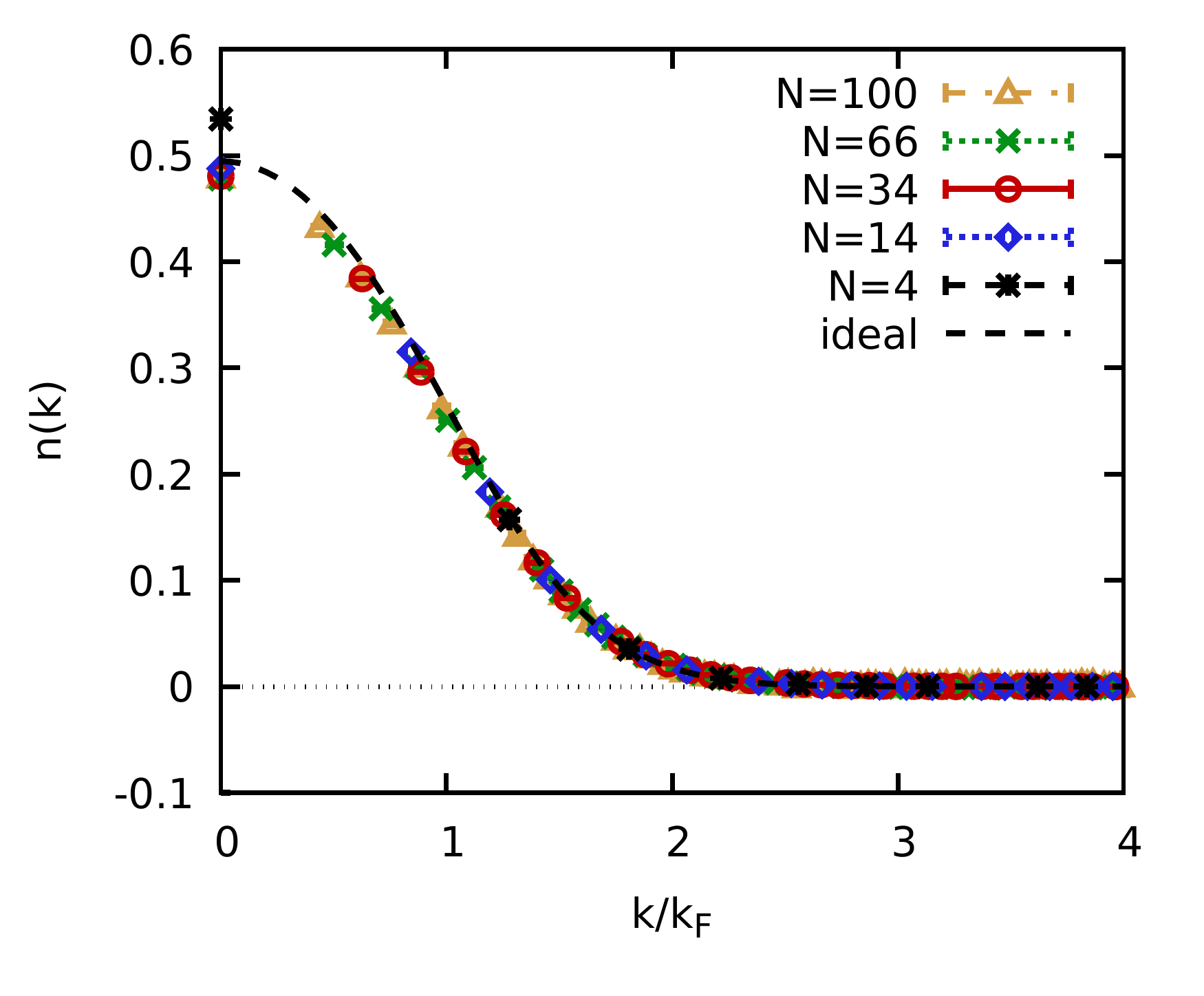}\hspace*{-0.015\textwidth}\includegraphics[width=0.49\textwidth]{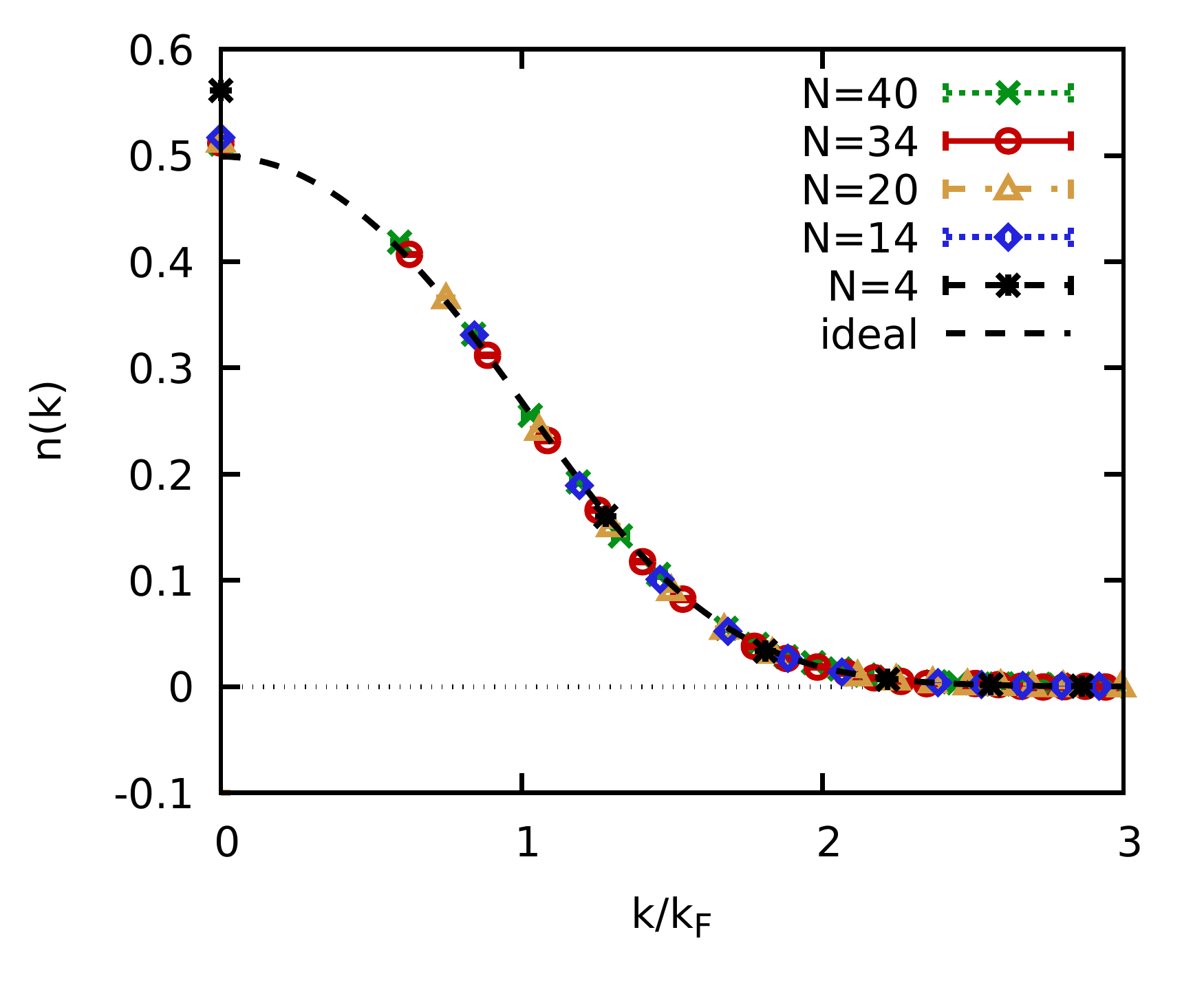}\\\vspace*{-1.3cm}
\hspace*{-0.015\textwidth}\includegraphics[width=0.502\textwidth]{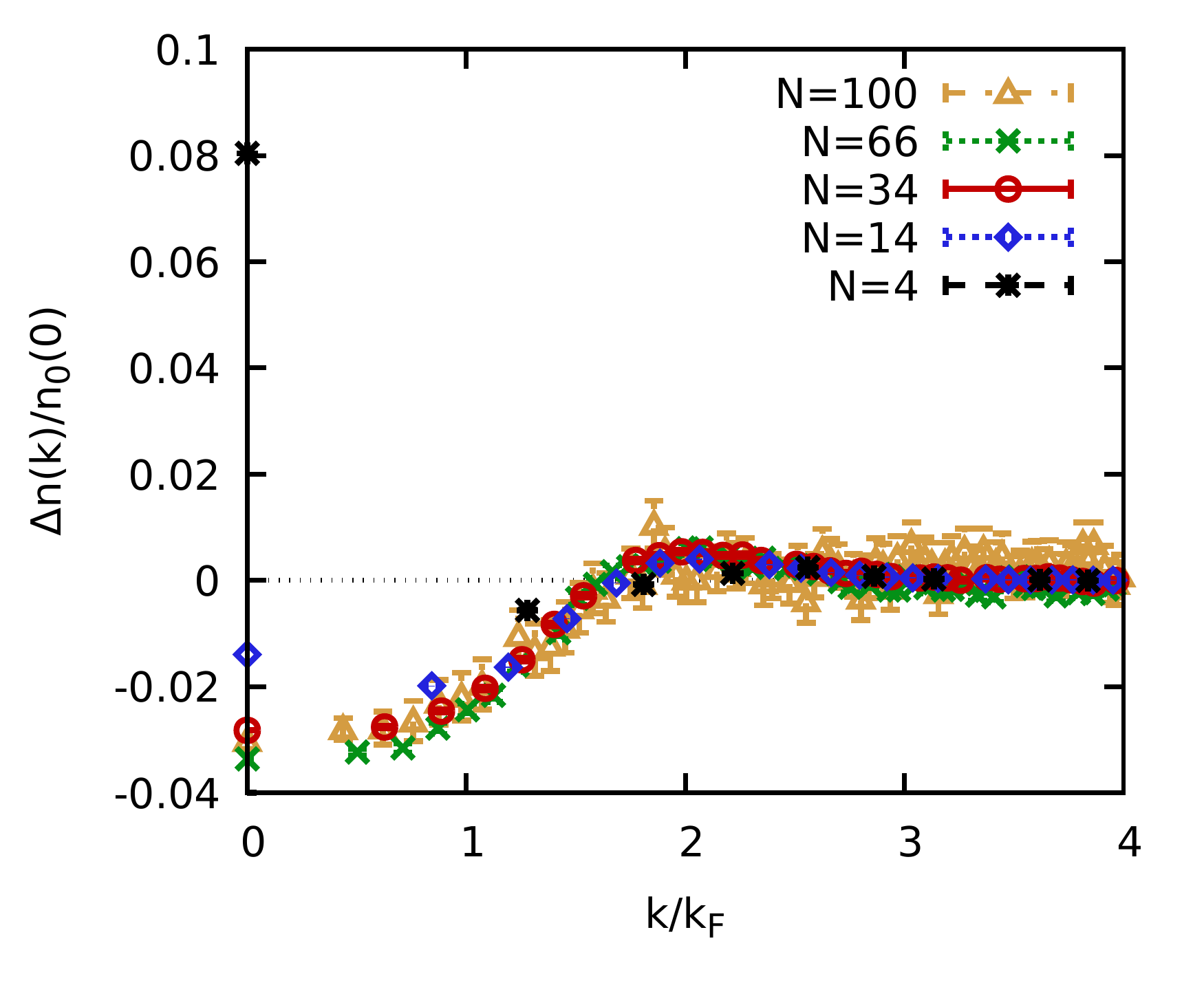}\hspace*{-0.028\textwidth}\includegraphics[width=0.502\textwidth]{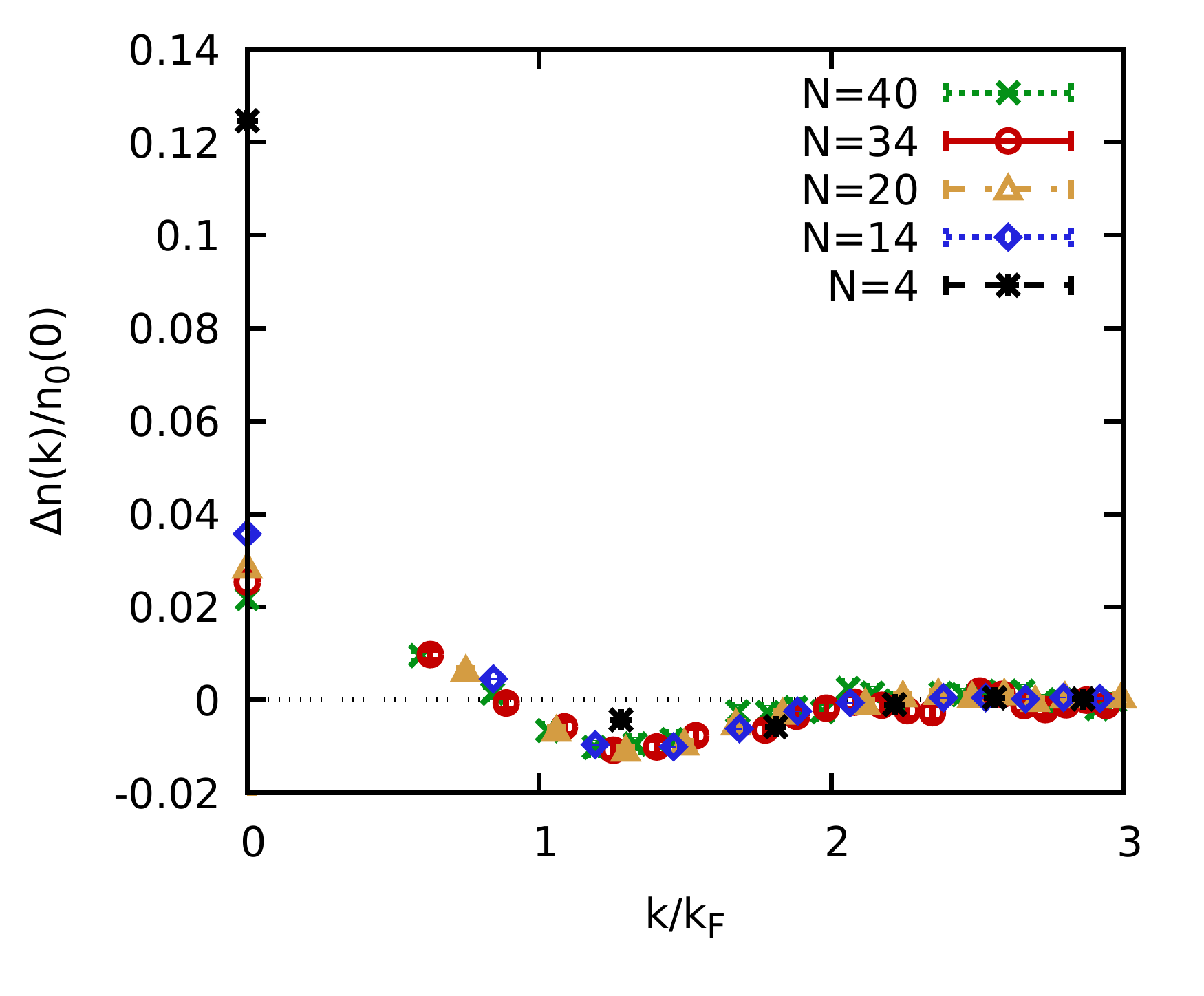}
\caption{\label{fig:FSC}
PIMC results for the momentum distribution function of the unpolarized UEG at $r_s=10$ (left column) and $r_s=4$ (right column) and $\theta=1$ for $N=14$ (blue diamonds) and $N=34$ for different system-sizes $N$ on a logarithmic (top row) and linear (center row) scale. Dashed black: ideal Fermi gas $n_0(k)$, Eq.~(\ref{eq:Fermi}). Bottom panel: relative deviation between PIMC data for $n(\mathbf{k})$ and $n_0(\mathbf{k})$, divided by $n_0(0)$.
}
\end{figure*}

Let us next analyze the system-size dependence of $n(\mathbf{k})$ itself, which is shown in Fig.~\ref{fig:FSC}. The left column
shows our PIMC results for $r_s=10$ and the top and center panels correspond to a logarithmic and linear $y$-axis, respectively.
The former case is particularly well suited to analyze the behaviour at large momenta, where our PIMC data for $N=14$ (blue diamonds), $N=34$ (red circles), and $N=66$ (green crosses) electrons substantially deviate from the ideal Fermi function [dashed black curve, see Eq.~(\ref{eq:Fermi})]. At the same time, we find that the PIMC data sets for different $N$ cannot be distinguished from each other within the given statistical uncertainty at this scale.

In contrast, the linear scale in the center panel is well suited to observe the behaviour of $n(\mathbf{k})$ for small $k$, and finite-size effects are substantial for $\mathbf{k}=\mathbf{0}$. In fact, the PIMC simulation with only $N=4$ electrons suggests an increase of the occupation at the center compared to the Fermi function, whereas it is decreased for all other values of $N$.
To analyze this effect in more detail, we show the relative deviation between our PIMC data and $n_0(\mathbf{k})$ in the bottom panel of Fig.~\ref{fig:FSC}. Firstly, we note that finite-size effects are indeed most pronounced for small momenta and quickly decrease with increasing $k$. In fact, even as few as $N=14$ electrons appear to give a sufficient description of $n(k)$ for $k\gtrsim k_\textnormal{F}$, which is further substantiated by logarithmic scale in the top panel.
At the same time, finite-size effects for $N\geq34$ are of the order of $\sim0.1\%$ even for $k=0$, which is sufficient for practical applications.
A heuristical explanation for the observed manifestation of finite-size effects is given by the length scales, as small $k$ correspond to large wave-lengths and, hence, the long-range behaviour of the electrons. Yet, the box length within a PIMC simulation is finite, which constitutes a well known source of bias. Following the same logic, the large-$k$ behaviour of the momentum distribution function is dominated by short-range single-particle and exchange-correlation effects, which are well described by a QMC simulation with finite $N$~\cite{dornheim2021overcoming}.

A similar analysis for $r_s=4$ is presented in the right column of Fig.~\ref{fig:FSC}.
Since the higher density corresponds to a decreased coupling strength, we are not able to resolve any differences between our PIMC data and the ideal Fermi distribution $n_0(k)$ at large $k$. At the same time, we note that there are also no differences between the PIMC data points for different $N$ within the given level of accuracy visible on the logarithmic scale.
This changes in the center panel, where $n(0)$ for $N=4$ particles is again substantially higher than the results from the other values of $N$. This is confirmed by examining the relative deviation to the ideal Fermi function $n_0(k)$ shown in the bottom panel, where the data point for $N=4$ exhibits a finite-size error of $\sim10\%$ at $k=0$. Yet, this error appears to rapidly decrease upon increasing the system size, and we find an error of approximately $1\%$ for $N=14$, whereas the other data points cannot be distinguished within the given Monte Carlo error bars. Furthermore, we find excellent agreement between all data sets (excluding $N=4$) for all other $k$, which means that our PIMC simulations are capable to give real insight into the momentum distribution of the warm dense UEG The physical interpretation of $n(\mathbf{k})$ itself is extensively covered in the next section.

\subsection{Density and temperature dependence\label{sec:rs_and_theta}}

\begin{figure*}\centering
\includegraphics[width=0.475\textwidth]{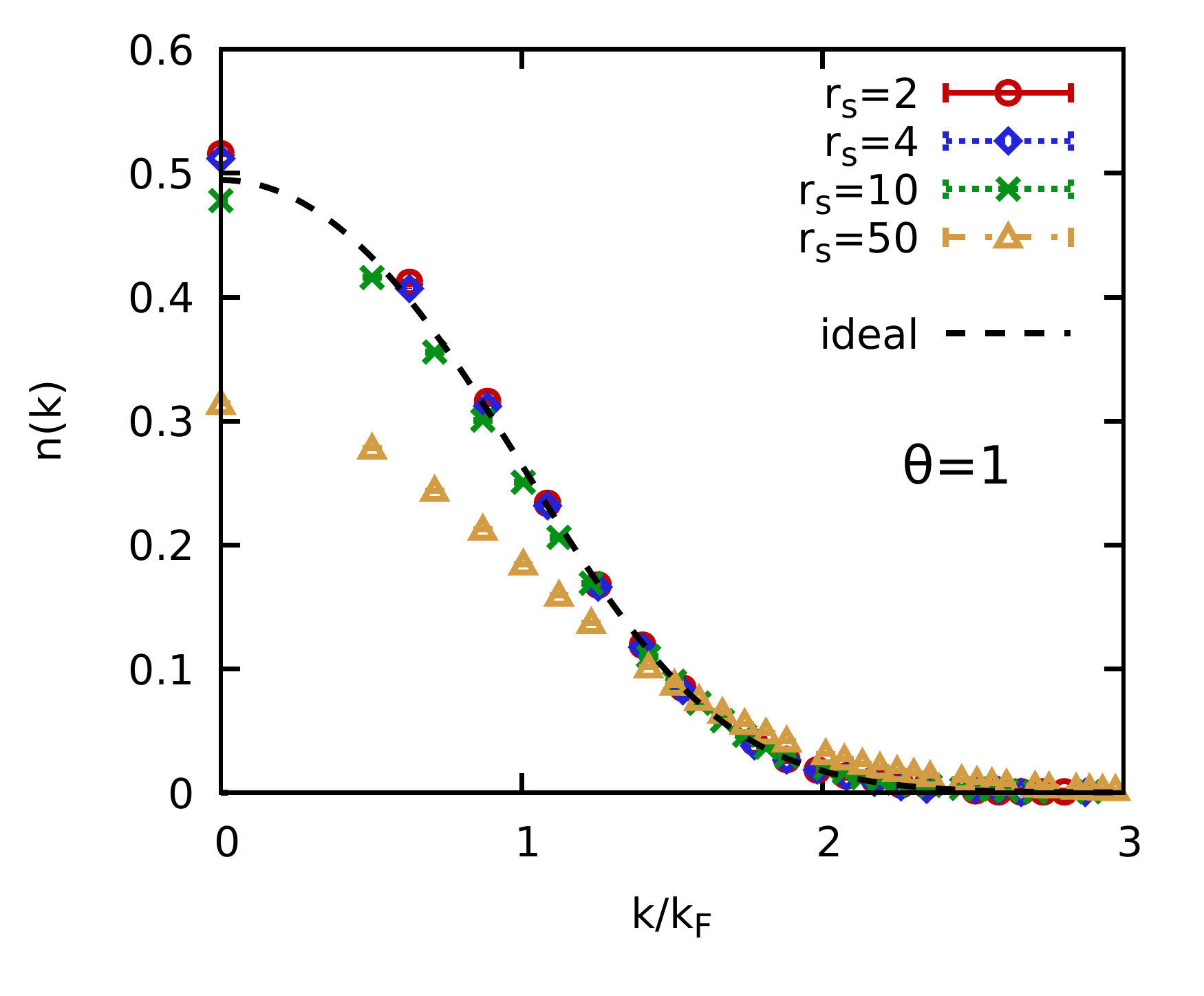}\includegraphics[width=0.475\textwidth]{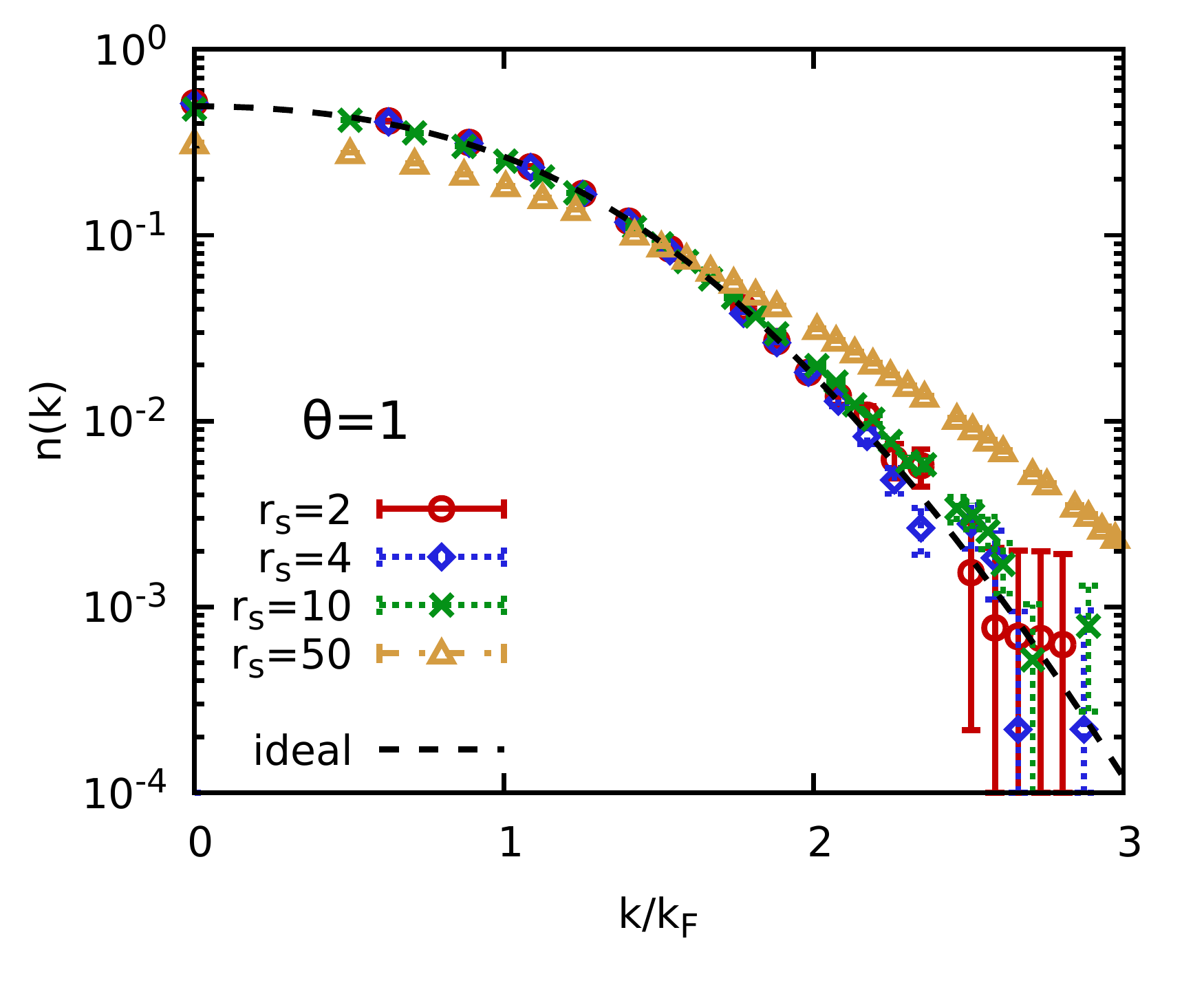}
\caption{\label{fig:rs_dependence_theta1}
Density dependence of the momentum distribution at $\theta=1$ obtained for $N=66$ ($r_s=10,50$) and $N=34$ ($r_s=2,4$) unpolarized electrons. The left and right panels correspond to a linear and logarithmic scale.
}
\end{figure*}

A particularly interesting topic of investigation is the behaviour of the momentum distribution for different values of the density parameter $r_s$. In the limit of $r_s\to0$, the kinetic energy dominates and $n(\mathbf{k})$ converges towards the Fermi distribution function, Eq.~(\ref{eq:Fermi}). For some temperatures, the occupation of $n(0)$ is actually increased compared to $n_0(0)$ upon increasing $r_s$, which leads to a lowering of the kinetic energy due to exchange--correlation effects; see Refs.~\cite{Militzer_PRL_2002,Hunger_PRE_2021} for a more detailed explanation.
In the vicinity of metallic densities (depending on the temperature), this effects vanishes, $n(0)$ decreases, and the kinetic energy is increased compared to the ideal Fermi gas~\cite{Militzer_PRL_2002,Hunger_PRE_2021}. While the behaviour of $K_\textnormal{xc}=K-E_0$ can be straightforwardly investigated using an accurate parametrization of the exchange--correlation free energy $f_\textnormal{xc}$ via the relation~\cite{review}
 \begin{eqnarray}\label{eq:Kxc}
 K_\textnormal{xc} 
 &=& - f_\textnormal{xc}(r_s,\theta)
 -\theta \frac{\partial f_\textnormal{xc}(r_s,\theta)}{\partial\theta}\Bigg|_{r_s}\\ \nonumber
& & - r_s \frac{\partial f_\textnormal{xc}(r_s,\theta)}{\partial r_s}\Bigg|_\theta \quad ,
 \end{eqnarray}
reliable data for $n(k)$ in the transition region are sparse.
To remedy this unsatisfactory situation, we have carried out extensive PIMC calculations of $n(\mathbf{k})$ at the Fermi temperature ($\theta=1$) for $N=34$ and $N=66$ unpolarized electrons for $2\leq r_s \leq 50$, using up to $\mathcal{O}\left(10^4\right)$ CPU-hours for the highest densities. The results are shown in Fig.~\ref{fig:rs_dependence_theta1}, with the different point-styles corresponding to different values of $r_s$, and the dashed black curve showing the ideal Fermi distribution. Let us first consider the left panel, which shows our results on a linear scale. For $r_s=2$ and $r_s=4$, we find that $n(0)$ is indeed increased compared to $n_0(k)$, which further confirms previous findings in Ref.~\cite{Militzer_PRL_2002,Militzer_momentum_HEDP_2019} and also the recent study by Hunger \textit{et al.}~\cite{Hunger_PRE_2021}. For $r_s=10$ (green crosses), the UEG starts to approach a strongly coupled electron liquid and the pronounced exchange--correlation effects push the electrons to occupy larger momenta. Finally, the yellow triangles correspond to $r_s=50$, which constitutes a strongly coupled system where the Coulomb repulsion between the electrons actually dominates~\cite{dornheim_electron_liquid}. This results in a momentum distribution function that is substantially different to $n_0(0)$ for all $k$, and a pronounced increase in the kinetic energy compared to the ideal system.

Let us next examine the density dependence of the momentum distribution function for large momenta shown in the right panel of Fig.~\ref{fig:rs_dependence_theta1}. For $r_s=2$ and $r_s=4$, the PIMC results for $n(\mathbf{k})$ cannot be distinguished from the dashed black curve within the given level of accuracy. For completeness, we mention that, even at small $r_s$, the momentum distribution function of an interacting electron gas will eventually deviate from $n_0(k)$ due to the emergence of a \emph{quantum tail}, which scales as~\cite{hofmann_short-distance_2013,yasuhara_note_1976}
\begin{align}
    \lim_{k\to \infty} n(k)  = \frac{4}{9}\left(\frac{4}{9\pi}\right)^{2/3}\left( \frac{r_s}{\pi}\right)^2 \frac{k^8_F}{k^8}g^{\uparrow\downarrow}(0)\,,
      \label{eq:k8-g0}
\end{align}
where $g^{\uparrow\downarrow}(0)$ is the pair distribution function at zero-distance that has recently been parametrized by Dornheim \textit{et al.}~\cite{Dornheim_PRL_2020_ESA}. Yet, this tail cannot be resolved using PIMC methods formulated in coordinate space, as Eq.~(\ref{eq:k8-g0}) only starts to hold for very large values of $k$ and, thus, very small values of $n(\mathbf{k})$; see Ref.~\cite{Hunger_PRE_2021} for a topical investigation of \emph{quantum tails} for $r_s\lesssim1$.
For $r_s=10$, we observe a small, yet significant increase of $n(\mathbf{k})$ compared to $n_0(\mathbf{k})$, which is directly responsible for the increase in $K$ shown in Fig.~\ref{fig:rs_dependence_kxc} below.
Lastly, the yellow triangles are qualitatively different from all other depicted curves, and we find a large increase in the occupation of large momenta.

\begin{figure}\centering
\includegraphics[width=0.475\textwidth]{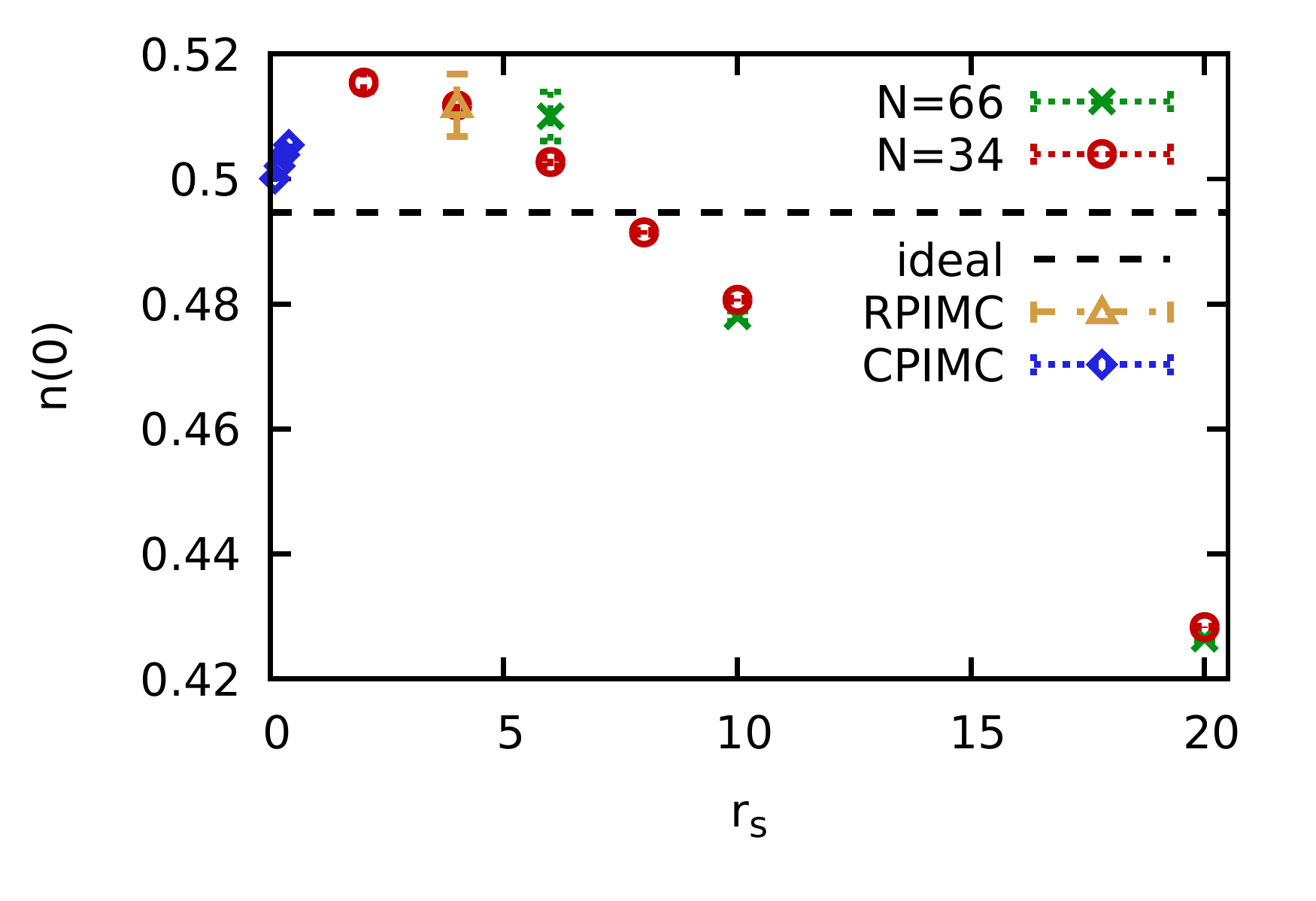}\\\vspace*{-1.25cm}\hspace*{-0.01\textwidth}\includegraphics[width=0.485\textwidth]{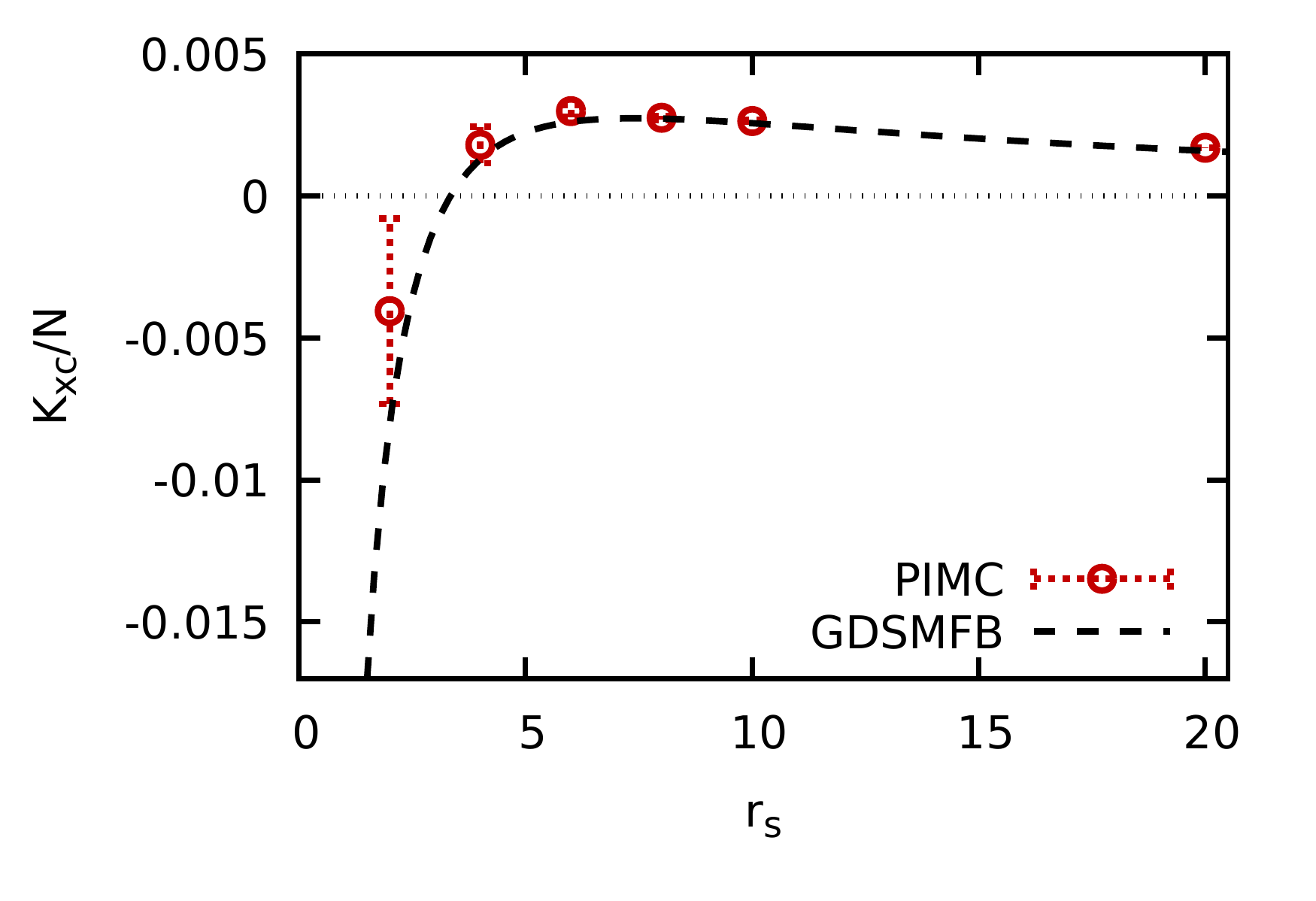}
\caption{\label{fig:rs_dependence_kxc}
Top panel: Density dependence of $n(0)$. Green crosses and red circles: new PIMC results for $N=66$ and $N=34$; yellow triangle: RPIMC data from Ref.~\cite{Militzer_momentum_HEDP_2019}; blue diamonds: CPIMC data from Ref.~\cite{Hunger_PRE_2021}; dashed line: ideal value. Bottom panel: Exchange--correlation contribution to the kinetic energy $K_\textnormal{xc}=K-E_0$. Symbols: PIMC results for $N=34$ (using $E_0$ for $N=34$ obtained with CPIMC); dashed line: $K_\textnormal{xc}$ computed via Eq.~(\ref{eq:Kxc}) from the parametrization of $f_\textnormal{xc}$ by Groth \textit{et al.}~\cite{groth_prl} (GDSMFB).
}
\end{figure}

As the next step, we investigate the occupation of the lowest momentum state $n(0)$ and its connection to exchange--correlation contribution to the kinetic energy $K_\textnormal{xc}$ in Fig.~\ref{fig:rs_dependence_kxc}.
More specifically, the top panel shows the $r_s$-dependence of $n(0)$ with the green crosses and red circles showing our new PIMC results for $N=66$ and $N=34$ unpolarized electrons, respectively. In addition, we have included a data point at $r_s=4$ from Ref.~\cite{Militzer_momentum_HEDP_2019} (yellow triangle), and CPIMC results for $r_s\leq0.5$ by Hunger \textit{et al.}~\cite{Hunger_PRE_2021}.

First and foremost, we note that the results from different, independent methods by different groups give a consistent picture, which further corroborates the high quality of our current picture of the UEG as a fundamental model system~\cite{review}. Secondly, $n(0)$ does indeed exhibit the nontrivial density dependence motivated in the beginning of this section, and $n(\mathbf{k})-n_0(\mathbf{k})$ changes its sign from positive to negative around $r_s\approx7.5$.
The corresponding values of $K_\textnormal{xc}$ are shown in the bottom panel of Fig.~\ref{fig:rs_dependence_kxc}, where the dashed black line has been obtained by evaluating Eq.~(\ref{eq:Kxc}) using as input the accurate parametrization of $f_\textnormal{xc}$ by Groth \textit{et al.}~\cite{groth_prl} (GDSMFB). In addition, the red circles are PIMC data for $N=34$ which have been obtained by consistently subtracting $E_0$ for the same system size (obtained with CPIMC~\cite{dornheim_prb_2016}).
Evidently, finite-size effects in $K$ are small at these conditions, and the PIMC data are in perfect agreement to the theoretical curve within the respective error bars. Further, we note that the comparably large error bar of $K_\textnormal{xc}$ at $r_s=2$ is a direct consequence of the fermion sign problem described earlier, and we find $S\approx0.01$ ($S\approx0.006$) for canonical (off-diagonal) configurations. For completeness, we note that a negative value of $K_\textnormal{xc}$ is directly related to a negative value of the electronic local field correction of the UEG in the limit of large wave numbers~\cite{dornheim_ML}.

Comparing the location of the sign change of $K_\textnormal{xc}$ around $r_s=3$ to the respective sign change of $n(\mathbf{k})-n_0(\mathbf{k})$ shown in the top panel for $r_s\sim7.5$, we find that they clearly do not coincide. In other words, an increase of the momentum distribution function at $\mathbf{k}=\mathbf{0}$ does not automatically lead to a decrease in the kinetic energy $K$.

\begin{figure}\centering
\includegraphics[width=0.475\textwidth]{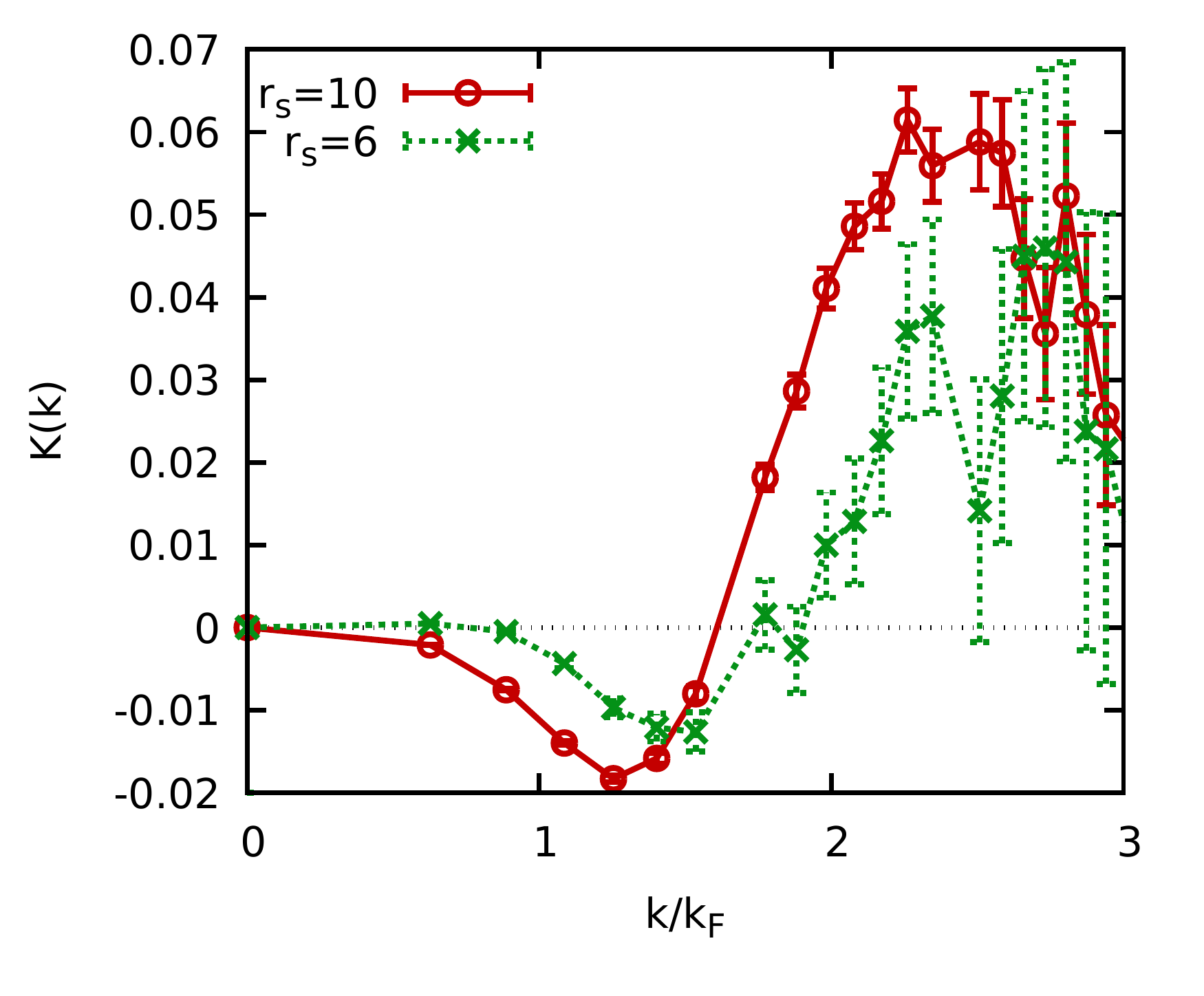}
\caption{\label{fig:contribution}
Wave-number resolved difference between $n(\mathbf{k})$ and $n_0(\mathbf{k})$ regarding the proportional contribution to the kinetic energy [cf.~Eq.~(\ref{eq:contribution})] at $\theta=1$ for $r_s=10$ (red circles) and $r_s=6$ (green crosses).
}
\end{figure}

\begin{figure}\centering
\includegraphics[width=0.475\textwidth]{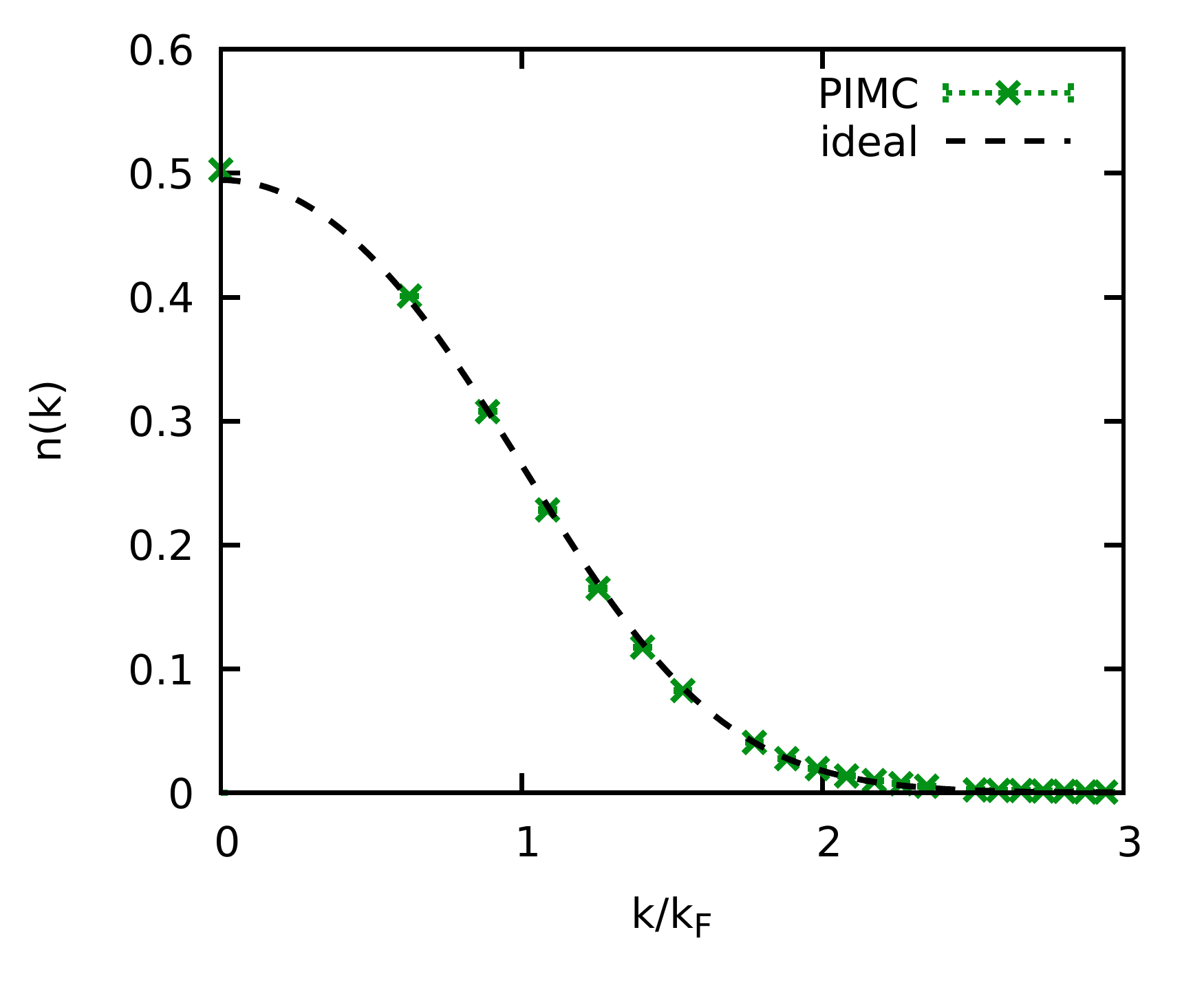}
\caption{\label{fig:rs6}
Momentum distribution function of the UEG for $r_s=6$ and $\theta=1$ for $N=34$ unpolarized electrons.
}
\end{figure}

This certainly calls for an explanation, which we provide in Fig.~\ref{fig:contribution} by showing the respective (proportional) difference between $n(k)$ and $n_0(k)$ regarding the contribution of each $k$ to the total kinetic energy $K$,
\begin{eqnarray}\label{eq:contribution}
K(k) = \left(n(k)-n_0(k)\right)\frac{k^4}{k_\textnormal{F}^4}\ .
\end{eqnarray}
The red circles have been obtained for $r_s=10$, where $n(0)<n_0(0)$ and $K_\textnormal{xc}>0$. Correspondingly, we find that the positive contributions to $K(k)$ for large $k$ clearly dominate over the negative contributions at small $k$. 
The green crosses correspond to $r_s=6$, an interesting case with $K_\textnormal{xc}>0$ but $n(0)>n_0(0)$. 
For the two smallest $k$-vectors, the occupation is increased compared to $n_0$ (see also the depiction of $n(\mathbf{k})$ itself shown in Fig.~\ref{fig:rs6}), followed by more pronounced negative contributions to $K(k)$ for $k_\textnormal{F}\lesssim k \lesssim 2k_\textnormal{F}$. Yet, this effect is overcompensated by the contributions from large momenta $k\gtrsim 2k_\textnormal{F}$, which are the reason for the positive value of $K_\textnormal{xc}$ at $r_s=6$ observed in Fig.~\ref{fig:rs_dependence_kxc}. 
In a nutshell, we conclude that the increase in $n(0)$ compared to $n_0(0)$ is a necessary, but not sufficient criterion for a negative value of $K_\textnormal{xc}$, as negative contributions to $K$ (from intermediate $k$) can be compensated by an increased occupation at large $k$.

\begin{figure*}\centering
\includegraphics[width=0.475\textwidth]{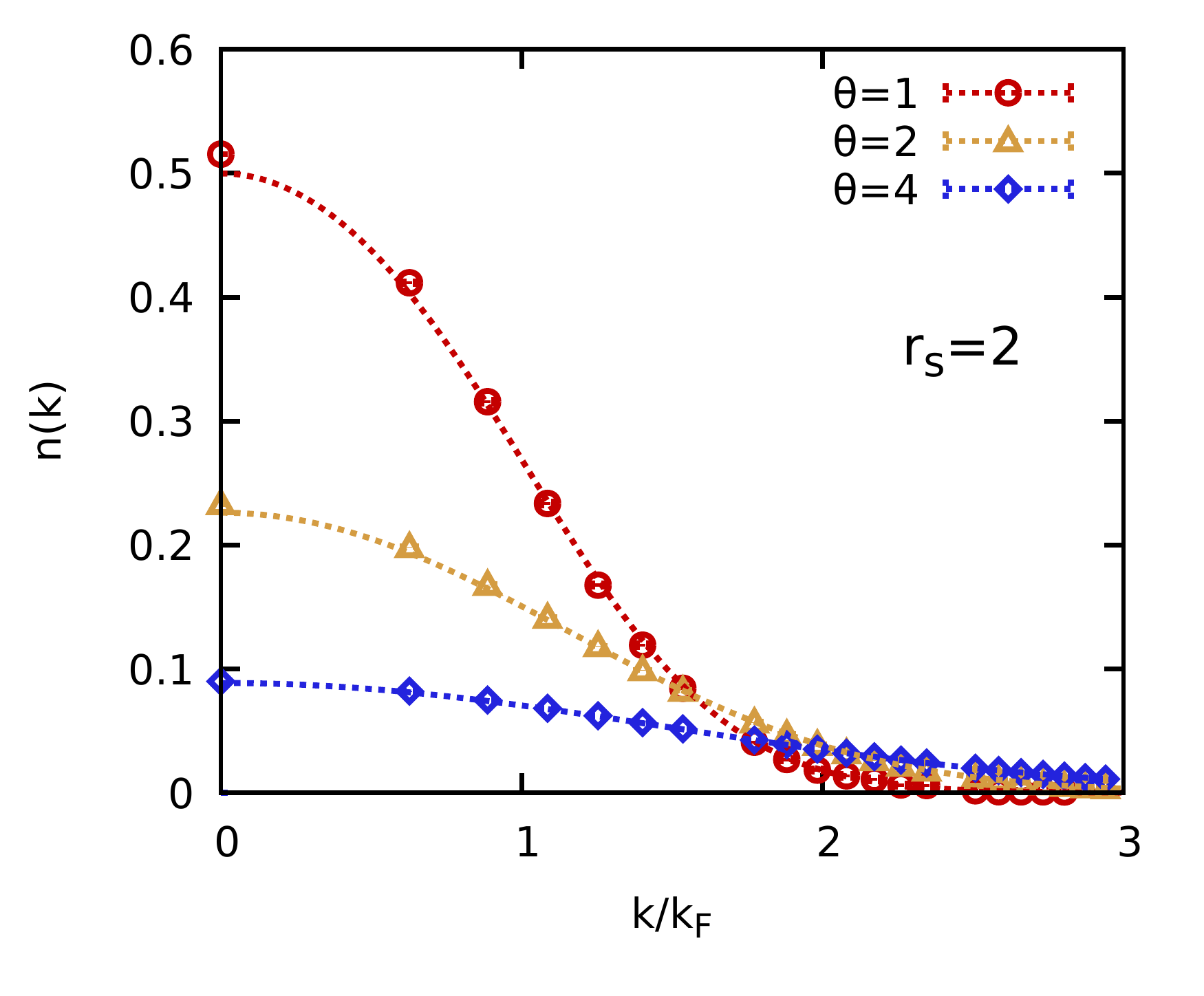}\includegraphics[width=0.475\textwidth]{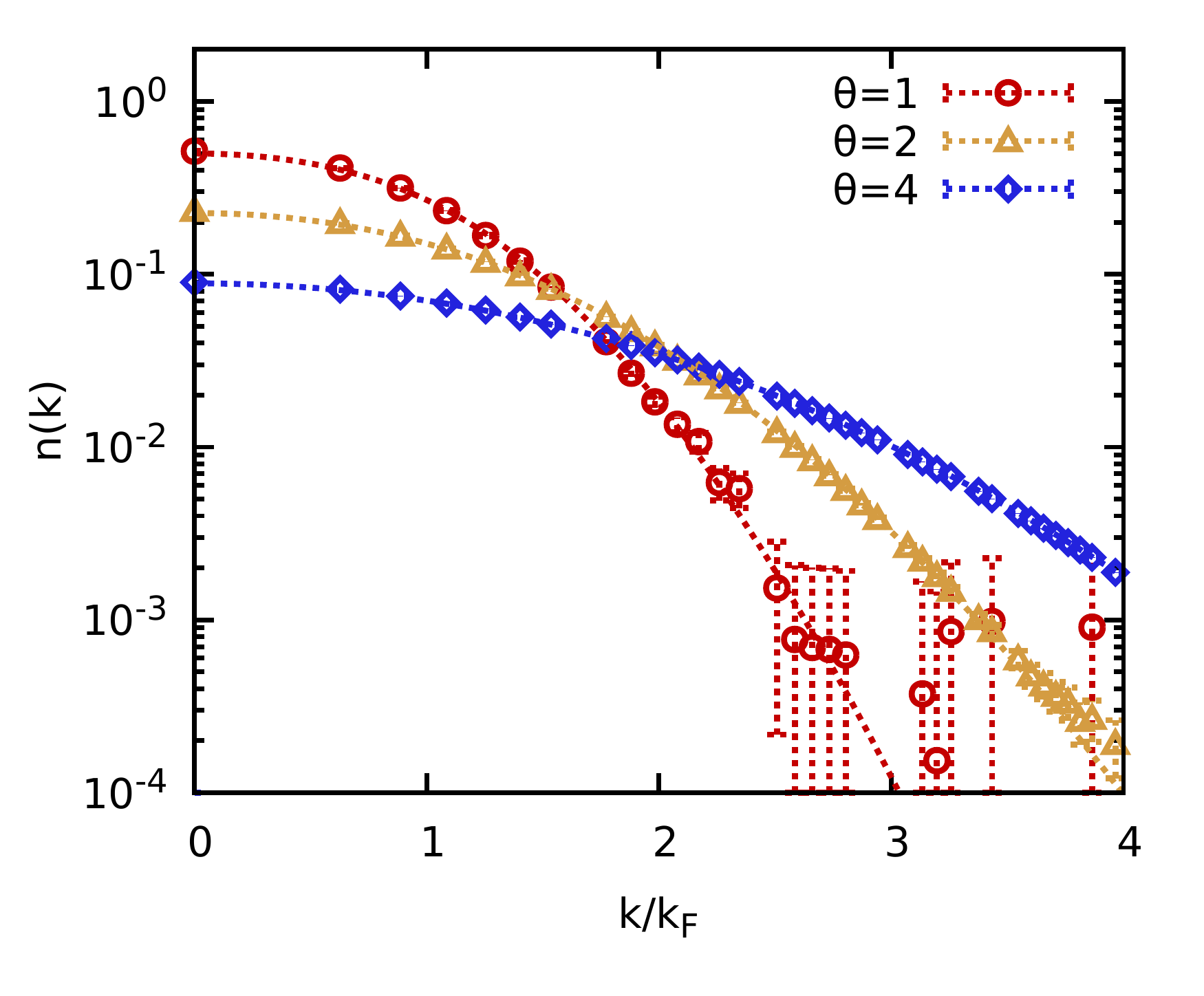}\\
\includegraphics[width=0.475\textwidth]{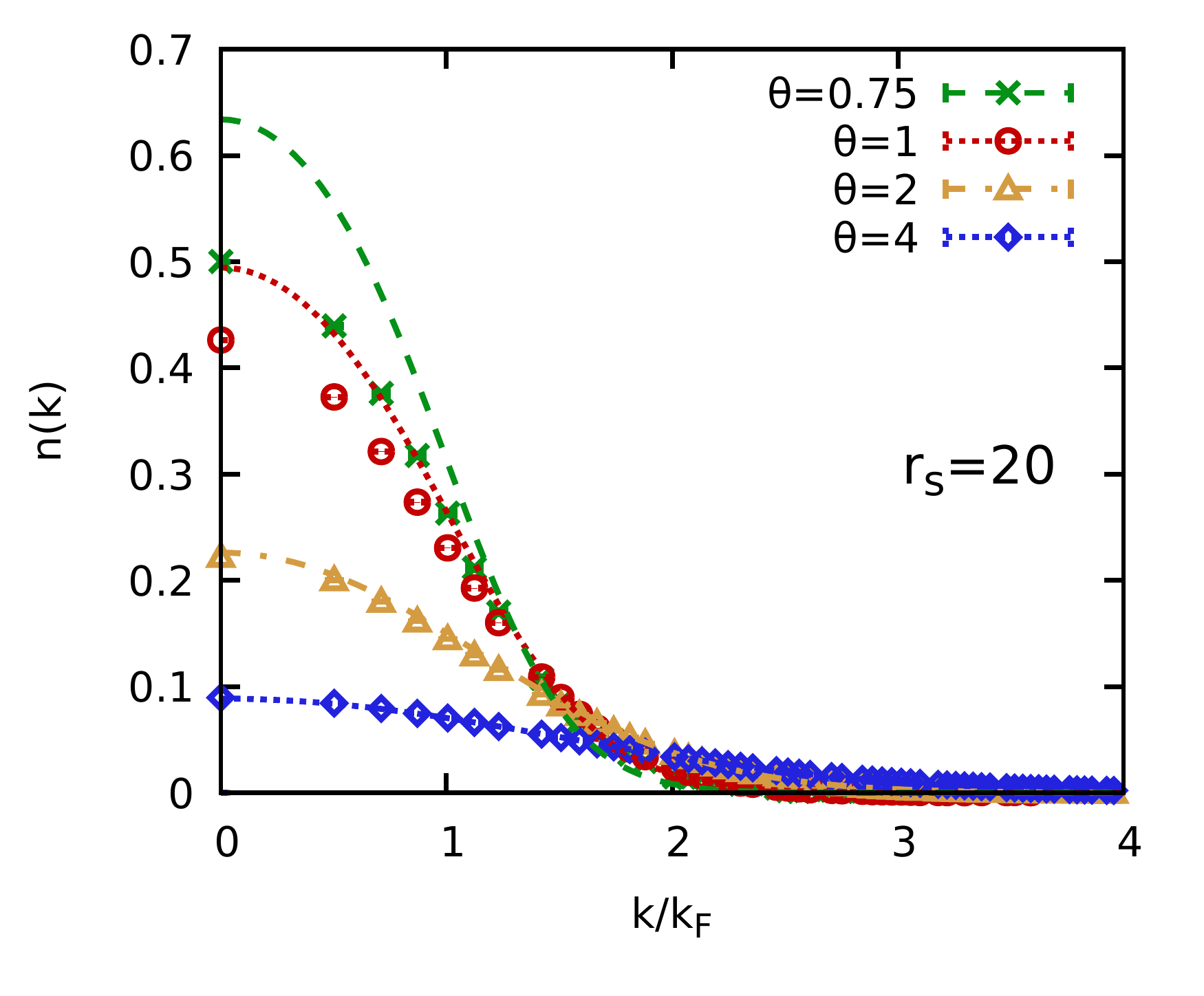}\includegraphics[width=0.475\textwidth]{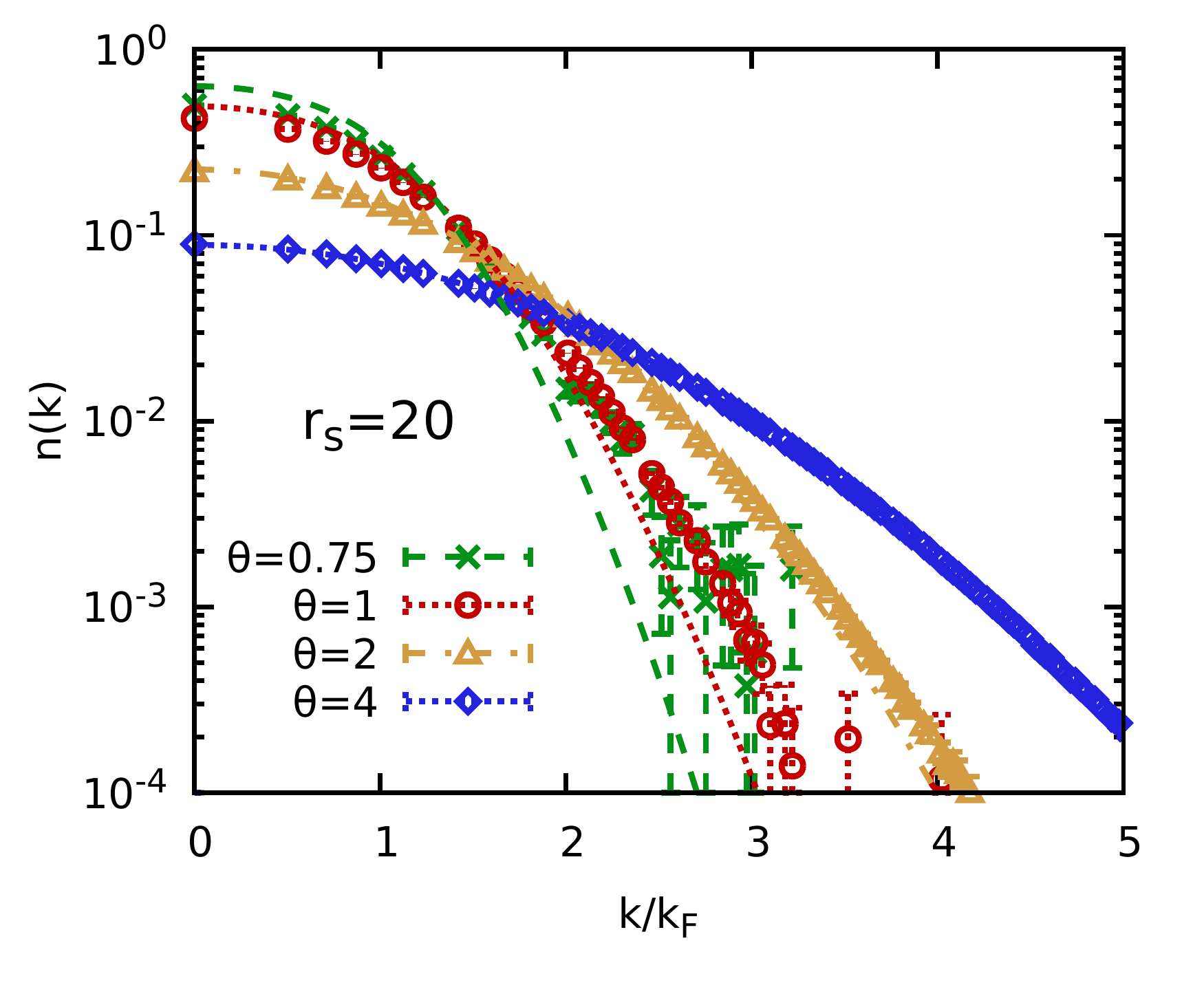}
\caption{\label{fig:theta_dependence}
Temperature dependence of the momentum distribution function for $r_s=2$ (top row) and $r_s=20$ (bottom row). The left and right columns show results on a linear and logarithmic scale, respectively.
}
\end{figure*}

A second interesting dimension to be investigated is the dependence of the momentum distribution on the reduced temperature $\theta$. This is shown in Fig.~\ref{fig:theta_dependence} for $r_s=2$ (top row) and $r_s=20$ (bottom row), with the left and right columns showing our PIMC results on a linear and logarithmic scale. Let us start by considering the higher density, which approximately corresponds to the density of conduction electrons in metals such as aluminum~\cite{Sperling_PRL_2015}. The red circles, yellow triangles, and blue diamonds correspond to PIMC data for $\theta=1$, $\theta=2$, and $\theta=4$, respectively, and the dotted curves to the ideal Fermi distribution, Eq.~(\ref{eq:Fermi}).
While there appear pronounced deviations between $n(\mathbf{k})$ and $n_0(\mathbf{k})$ for small $k$ at $\theta=1$, these difference decrease towards higher temperature, as it is expected. Consequently, no deviations can be resolved for $\theta=4$ with the bare eye.
Proceeding to the right panel, we see that no differences between the PIMC data and the Fermi function can be resolved for large $k$ within the given level of statistical uncertainty. Still, we re-iterate our earlier point that $n(\mathbf{k})$ and $n_0(\mathbf{k})$ will eventually diverge from each other due to the \emph{quantum tail}, cf.~Eq.~(\ref{eq:k8-g0}).

The bottom row shows results for the temperature dependence of the UEG in the electron liquid regime ($r_s=20$), where simulations with $N=66$ unpolarized electrons are feasible even for $\theta=0.75$, and we find an average sign of $S\approx0.1$ ($S\approx0.05$) for the canonical (off-diagonal) configuration space.
Negative signs also enter RPIMC simulations with open paths. However, their impact is less severe. For a system of $N=66$ unpolarized electrons at $r_s$=40, the average sign decreases from 0.83 only to 0.20 as the temperature as lowered from $\theta=1$ to $1/16$ rendering such RPIMC simulation feasible.

At electron liquid conditions, the impact of exchange-correlation effects is substantially more pronounced compared to $r_s=2$, and the difference between $n(\mathbf{k})$ and $n_0(\mathbf{k})$ can easily be seen with the bare eye on the linear scale even for $\theta=2$. Looking at the logarithmic scale, we can clearly resolve an increased occupation of the PIMC data compared to the Fermi function for $\theta\leq2$, whereas no such difference is visible for $\theta=4$.

\subsection{Impact of quantum statistics\label{sec:statistics}}

\begin{figure*}\centering
\includegraphics[width=0.475\textwidth]{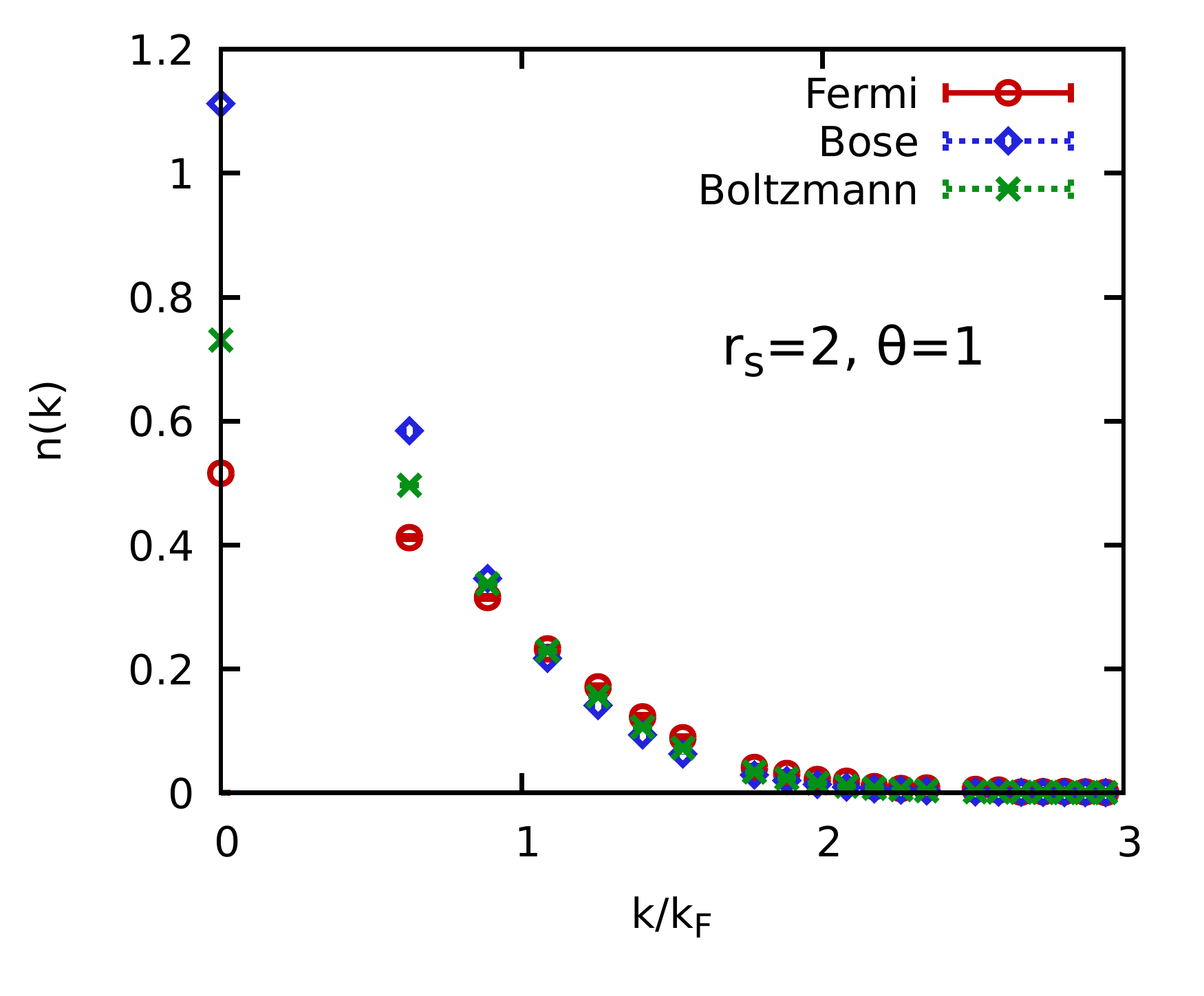}\includegraphics[width=0.475\textwidth]{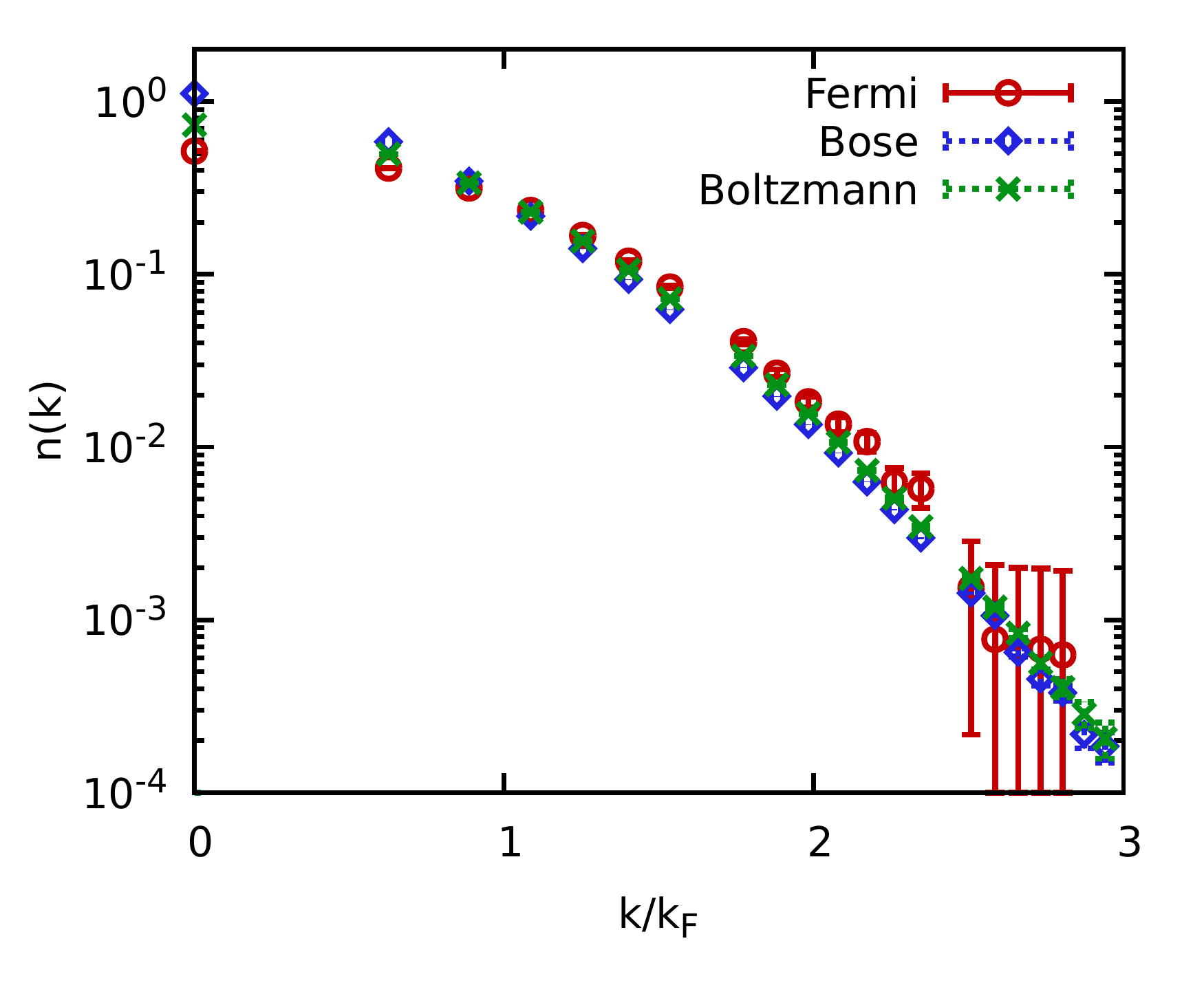}
\\ \vspace*{-1cm}
\hspace*{0.01\textwidth}\includegraphics[width=0.465\textwidth]{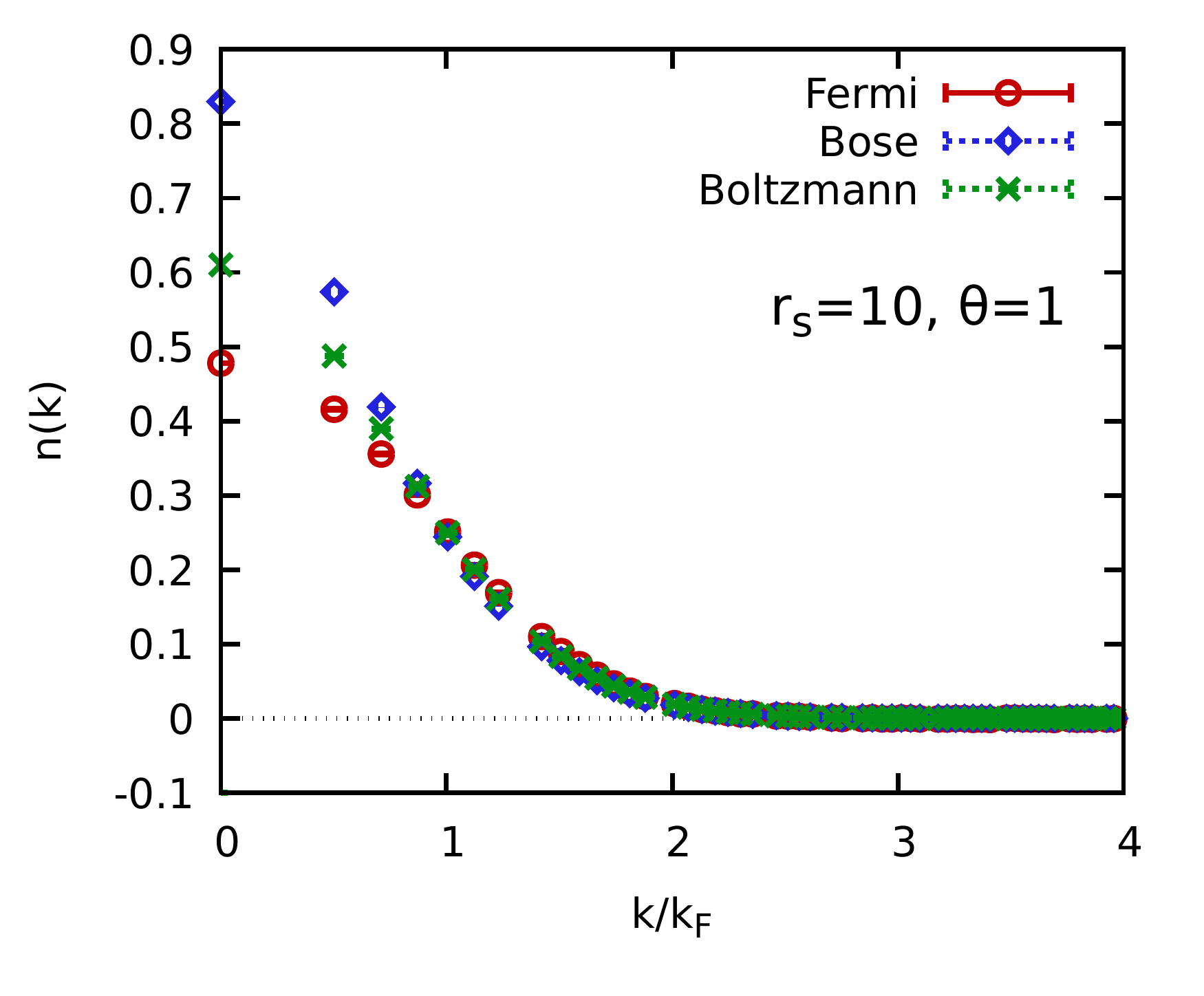}\includegraphics[width=0.475\textwidth]{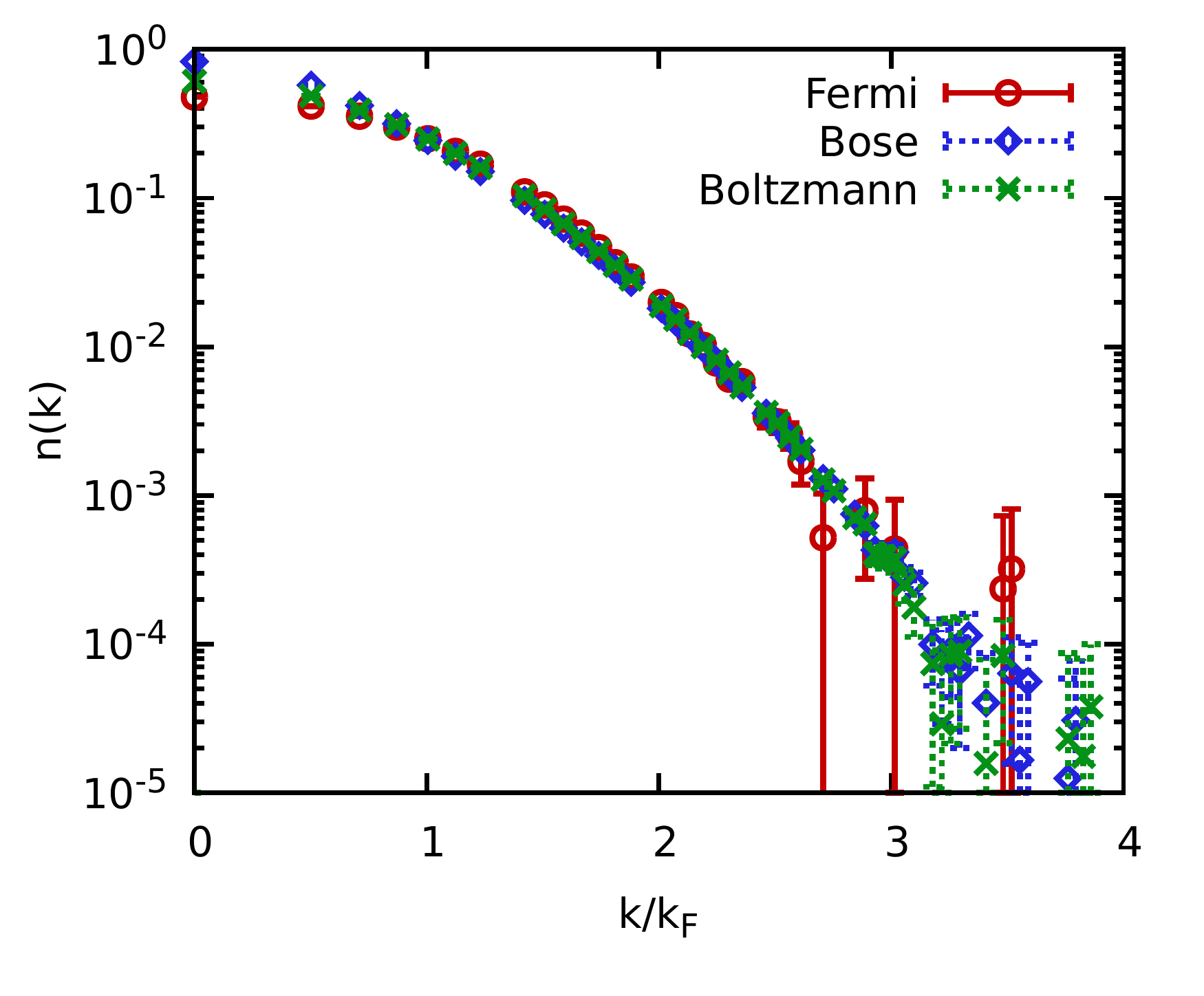}\\ \vspace*{-1cm}
\includegraphics[width=0.475\textwidth]{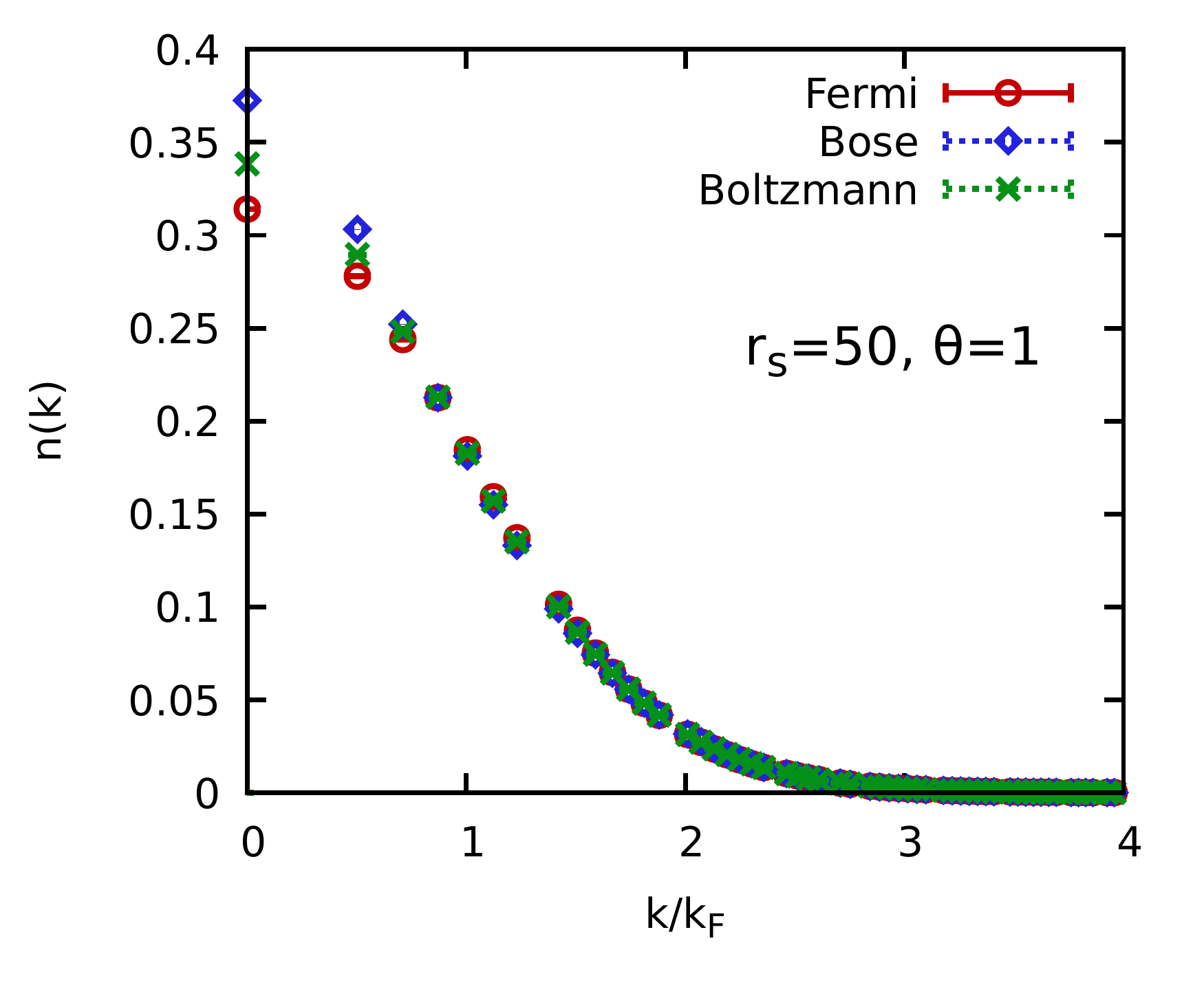}\includegraphics[width=0.475\textwidth]{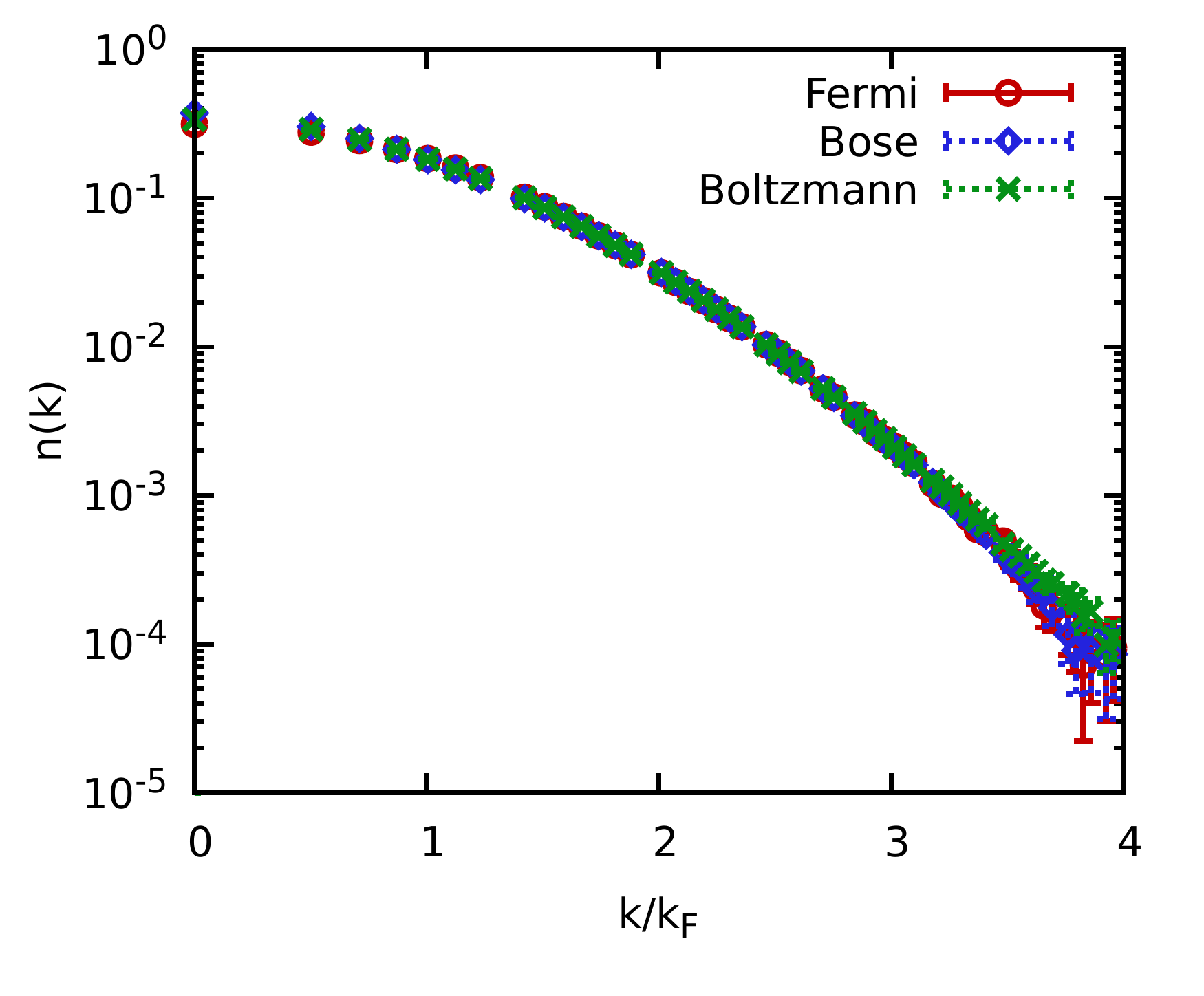}
\caption{\label{fig:quantum_statistics}
Effects of quantum statistics at the Fermi temperature: Shown are PIMC results for the momentum distribution function with Fermi (red circles), Bose (blue diamonds), and Boltzmann statistics (green crosses). The top, center, and bottom rows correspond to $r_s=2$, $r_s=10$, and $r_s=50$, and the left and right columns to a linear and logarithmic scale.
}
\end{figure*} 

The final physical phenomenon to be investigated in this work is the impact of quantum statistics. In fact, results for the momentum distribution of a Bose-system at the same conditions as the UEG can straightforwardly be obtained within a PIMC simulation of the latter, see Eq.~(\ref{eq:modified}) above. In addition, we have carried out independent PIMC simulations of boltzmannons, i.e., distinguishable particle, by disabling the sampling of exchange--cycles within our scheme. The results are shown in Fig.~\ref{fig:quantum_statistics} for $\theta=1$ at $r_s=2$ (top row), $r_s=10$ (center row), and $r_s=50$ (bottom row). Let us start with the highest density, where the impact of quantum statistics is expected to be most pronounced. Indeed, there appear striking differences in $n(\mathbf{k})$ between fermions (red circles), bosons (blue diamonds), and boltzmannons (green crosses) in particular for small $k$, with a difference of over $100\%$ between Bose- and Fermi-statistics for $n(0)$. This can be understood in the following way: for fermions, each orbital can at most be occupied once by an electron of the same spin due to the Pauli exclusion principle; boltzmannons do not feel this effect at all and simply follow a thermal occupation depending on the energy of each orbital; bosons, on the other hand, positively tend to cluster in the lowest lying momentum state, which is intimately connected to the onset of Bose-Einstein-condensation, e.g. in ultracold $^4$He~\cite{cep}.
On the logarithmic scale shown in the right panel, we further see that the order of occupation is reversed between the three types of particle statistics for large momenta, which is of course a direct consequence of the normalization, cf.~Eq.~(\ref{eq:normalization}) above.

\begin{figure}\centering
\includegraphics[width=0.475\textwidth]{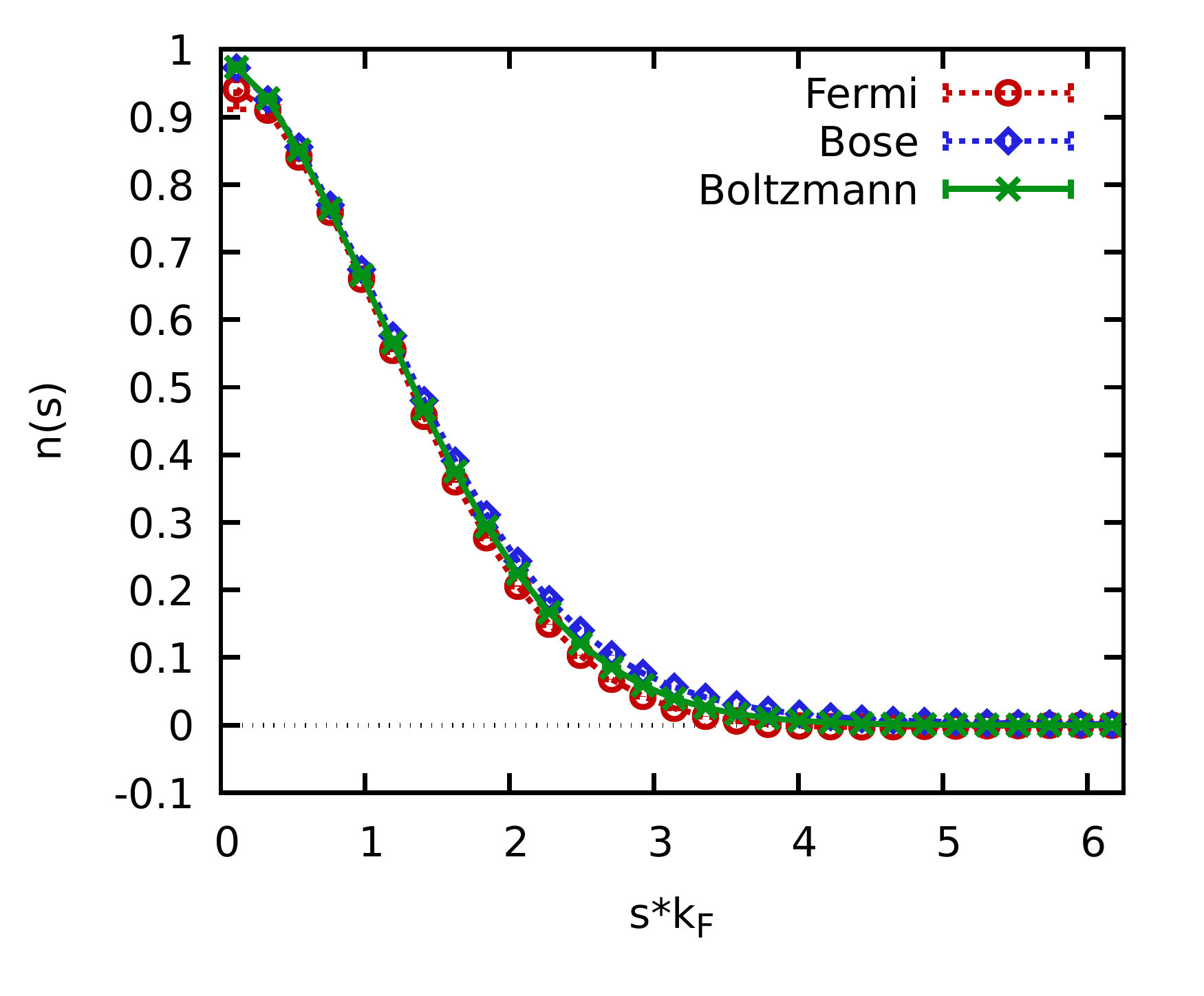}\\\vspace*{-1.3cm}
\hspace*{0.0133\textwidth}\includegraphics[width=0.46\textwidth]{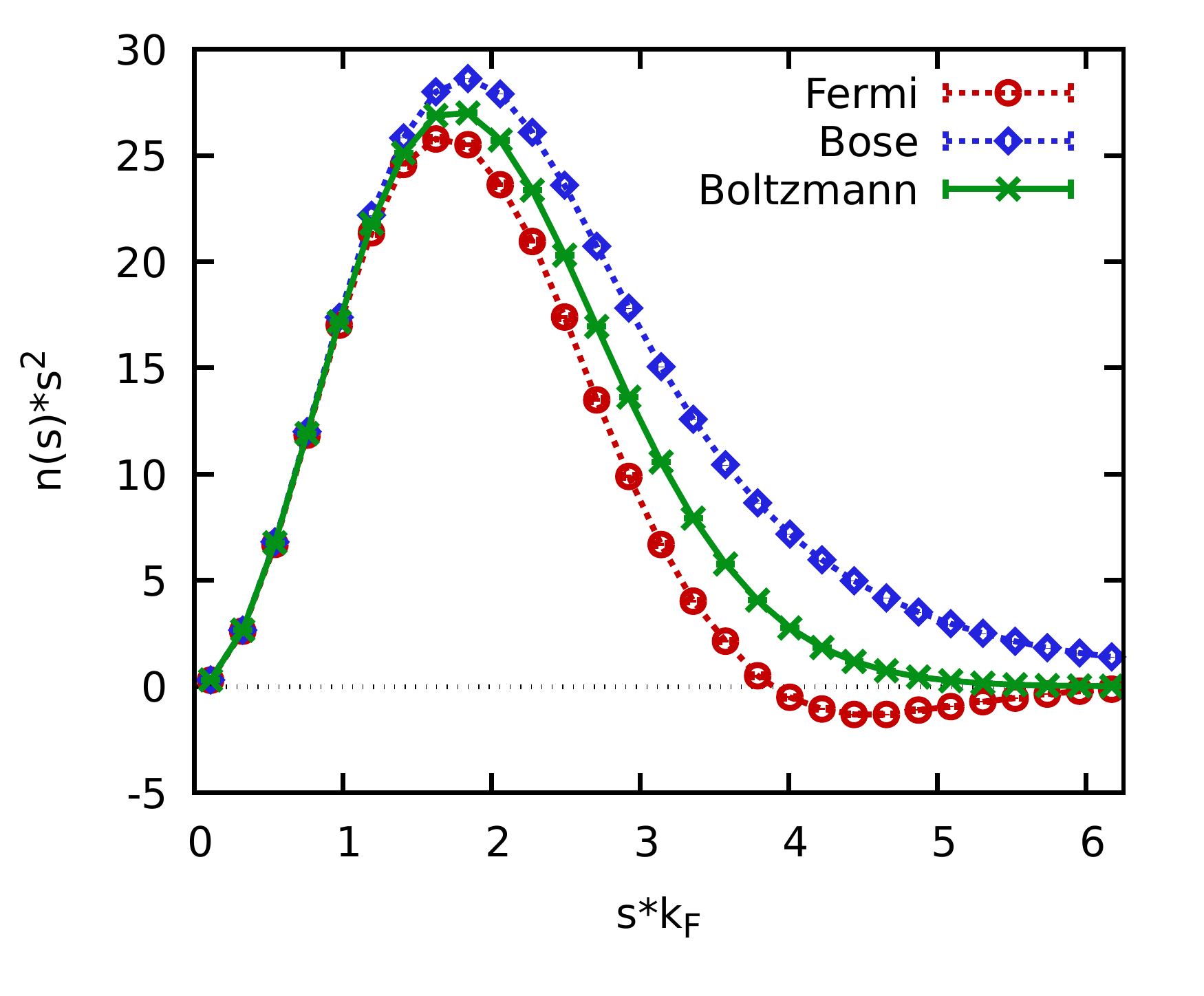}
\caption{\label{fig:offdiag_rs10_theta1} Off-diagonal density matrix [see Eq.~(\ref{eq:n_s})] as a function of the modulus distance $s=|\mathbf{r}-\mathbf{r'}|$. The red circles, blue diamonds, and green crosses have been obtained for Fermi, Bose, and Boltzmann statistics, respectively.
}
\end{figure}

Proceeding to $r_s=10$ shown in the center row of Fig.~\ref{fig:quantum_statistics}, we see a fairly similar behaviour compared to the top row, albeit with less pronounced deviations between the different particle types. This can directly be interpreted within the path-integral picture: the only impact of quantum statistics on the PIMC simulation is the formation of exchange--cycles, i.e., paths that wind multiple times around the imaginary time and, thus, have more than a single particle in it. For bosons, these macroscopic trajectories are connected to the onset of superfluidity, which is expressed as an off-diagonal long-range order~\cite{Shi_PRB_2005} in the density matrix $n(\mathbf{s})$ [see the discussion of Fig.~\ref{fig:offdiag_rs10_theta1} below]. For fermions, this results in the cancellation of positive and negative contributions to the partition function $Z$, which, in turn, leads to physical effects like Pauli blocking and the derivative degeneracy pressure. For exchange--cycles to be formed, the particles need to get sufficiently close to each other. Yet, the Coulomb repulsion between electrons counteracts this formation, and thus decreases the impact of quantum statistics on the system. Consequently, the average sign increases with $r_s$ and eventually approaches one when the system starts to crystallize and quantum exchange effects disappear.

Lastly, the bottom row of Fig.~\ref{fig:quantum_statistics} shows the momentum distribution function for $r_s=50$, which is a strongly coupled system in the center of the electron liquid regime~\cite{dornheim_electron_liquid}. Remarkably, even in this case the impact of quantum statistics has not yet vanished, and there appear significant deviations for small $k$.


Let us conclude our investigation by considering the off-diagonal density matrix in coordinate space, which is defined as~\cite{cep}
\begin{eqnarray}\label{eq:n_s}
n(\mathbf{s}) = \frac{1}{Z} \int \textnormal{d}\mathbf{R}\ \rho(\mathbf{r_1},\dots,\mathbf{r_N}, \mathbf{r_1}+\mathbf{s}, \mathbf{r_2},\dots,\mathbf{r_N})\ ,
\end{eqnarray}
where $\mathbf{s}=\mathbf{r}-\mathbf{r}'$ is the difference between the two open ends of a trajectory.
We note that Eq.~(\ref{eq:n_s}) is directly connected to the momentum distribution by a Fourier transform
\begin{eqnarray}
n(\mathbf{k}) = \int \textnormal{d}\mathbf{s}\ n(\mathbf{s}) e^{-i\mathbf{s}\cdot\mathbf{k}}\ .
\end{eqnarray}
Naturally, the PIMC estimation of Eq.~(\ref{eq:n_s}) has to be performed in the off-diagonal ensemble defined by $Z_{\mathbf{r},\mathbf{r'},\sigma}$ as well, which means that $n(\mathbf{s})$ requires the estimation of the same proportionality constant as $n(\mathbf{k})$. We recall that this is done automatically within our scheme, see Eq.~(\ref{eq:proportionality}) above.

The results for the off-diagonal density matrix for $r_s=10$ and $\theta=1$ are shown in Fig.~\ref{fig:offdiag_rs10_theta1} as a function of the absolute value of the distance vector $s=|\mathbf{s}|$. We note that this in principle only holds in the TDL, whereas the orientation of $\mathbf{s}$ towards the simulation cell matters for finite $N$~\cite{Militzer_momentum_HEDP_2019}. To avoid any associated inconsistencies, we only show $n(s)$ up to half the box length, $s\leq L/2$.

Let us first examine the top panel, where we show $n(s)$ itself.
The red circles have been obtained for the UEG with $N=66$ unpolarized electrons. The increasing error bars towards $s\to0$ are a direct consequence of the employed histogram within spherical shells around $\mathbf{r}$, the volume of which scales quadratically with $s$. This vividly illustrates the value of a direct PIMC estimation of the pre-factor of both $n(\mathbf{s})$ and $n(\mathbf{k})$, as a determination from the relation $n(\mathbf{s}=0)=1$ is potentially biased due the high noise level at small $s$. Moreover, the computation of $n(s)$ as histograms over spherical shells of finite volume introduces a binning error, which can potentially be removed by the sophisticated virtual-trajectory-estimator introduced in Ref.~\cite{boninsegni1} for the sampling of the Matsubara Green function. Yet, as the PIMC estimation of $n(\mathbf{k})$ is not subject to this error, we find the more simple estimator sufficient here.
The blue diamonds and green crosses in Fig.~\ref{fig:offdiag_rs10_theta1} have been computed for Bose- and Boltzmann-statistics, but can hardly be distinguished from the other curves on this scale.

To obtain a more detailed view, we show $n(s)s^2$ in the bottom panel of the same figure, which is a measure for the actual probability to find the two open ends in a distance $s$ from each other. In particular, the small value at small $s$ nicely illustrates the sampling problem of the histogram estimator. In addition, this plot allows us to clearly resolve the impact of quantum statistics on the off-diagonal density matrix, which is particularly large around $4/k_\textnormal{F}\leq s \leq 5/k_\textnormal{F}$ where $n(s)$ is actually negative in the case of fermions. Furthermore, we find that $n(s)$ only slowly decays in the case of Bose-statistics, which is a direct consequence of the presence of exchange--cycles within the PIMC simulation. For boltzmannons, there are always $P-1$ beads between $\mathbf{r}$ and $\mathbf{r'}$, which limits the $s$-range where $n(s)$ is nonzero approximately to the thermal wave-length, i.e., the quantum extension of a single particle. In contrast, an exchange cycle can potentially contain many particles, which gives potential contributions to $n(s)$ even for $s\gg\lambda_\beta$.


\begin{figure*}\centering
\includegraphics[width=0.475\textwidth]{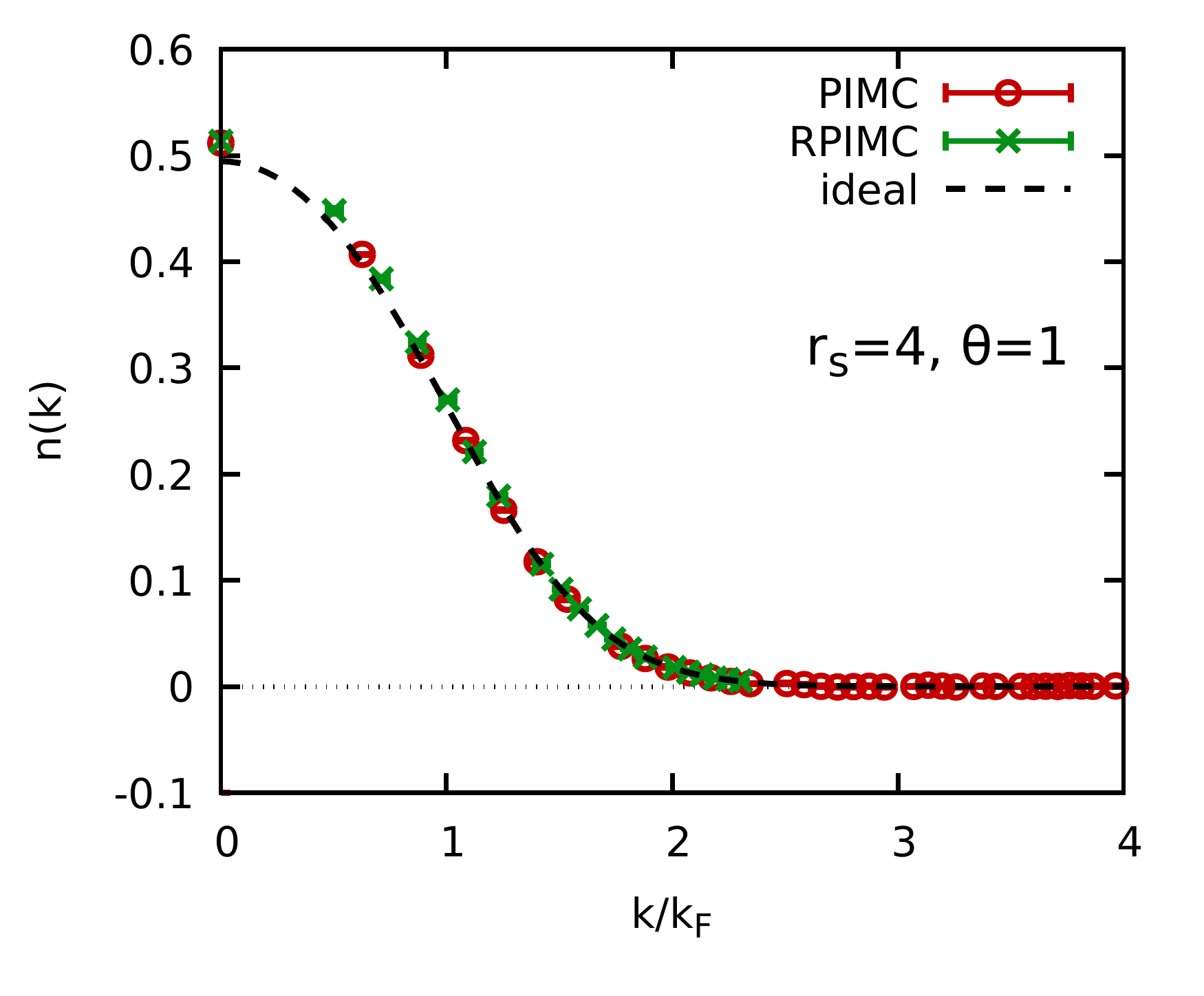}\includegraphics[width=0.475\textwidth]{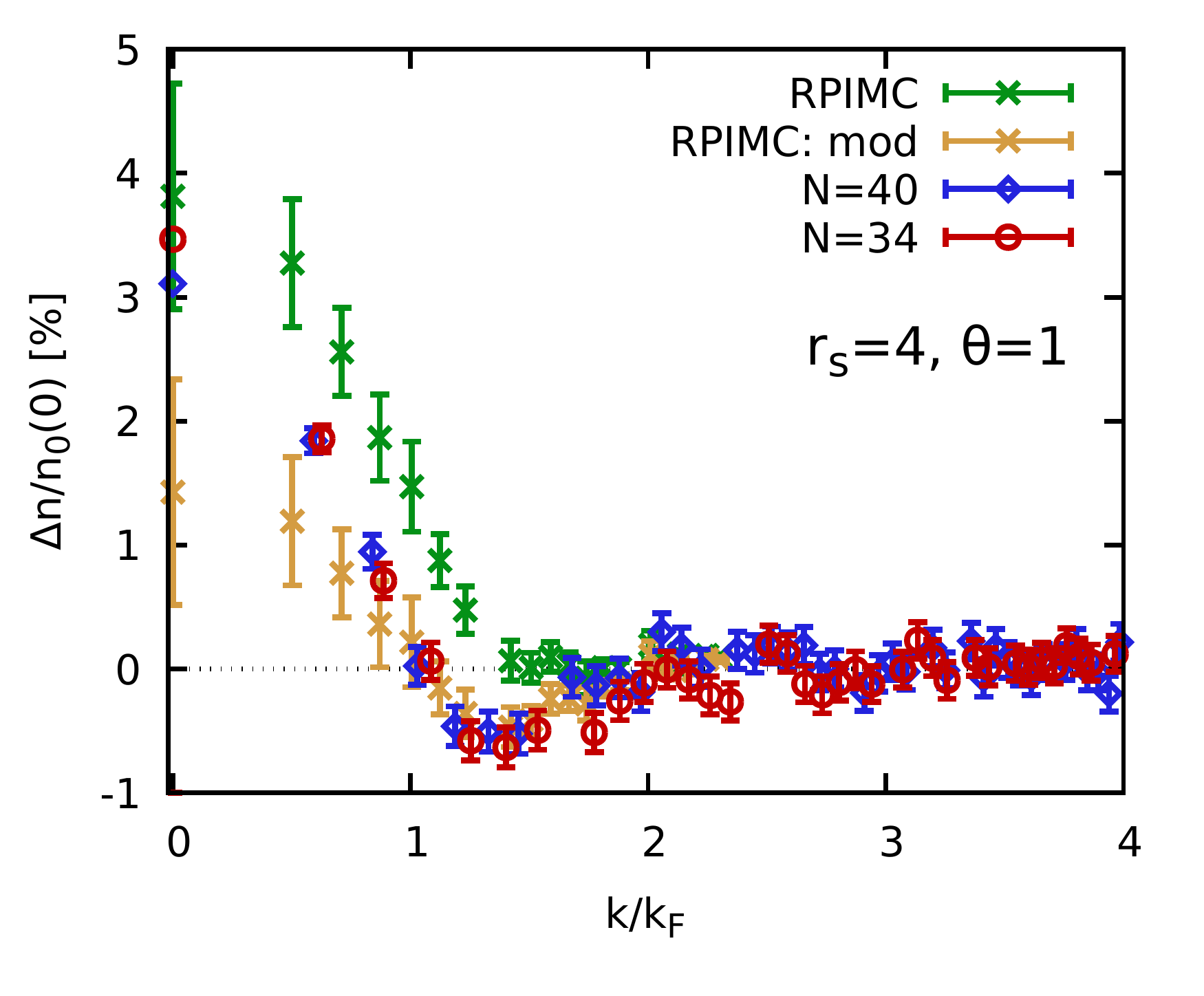}\\\vspace*{-1.35cm}
\includegraphics[width=0.475\textwidth]{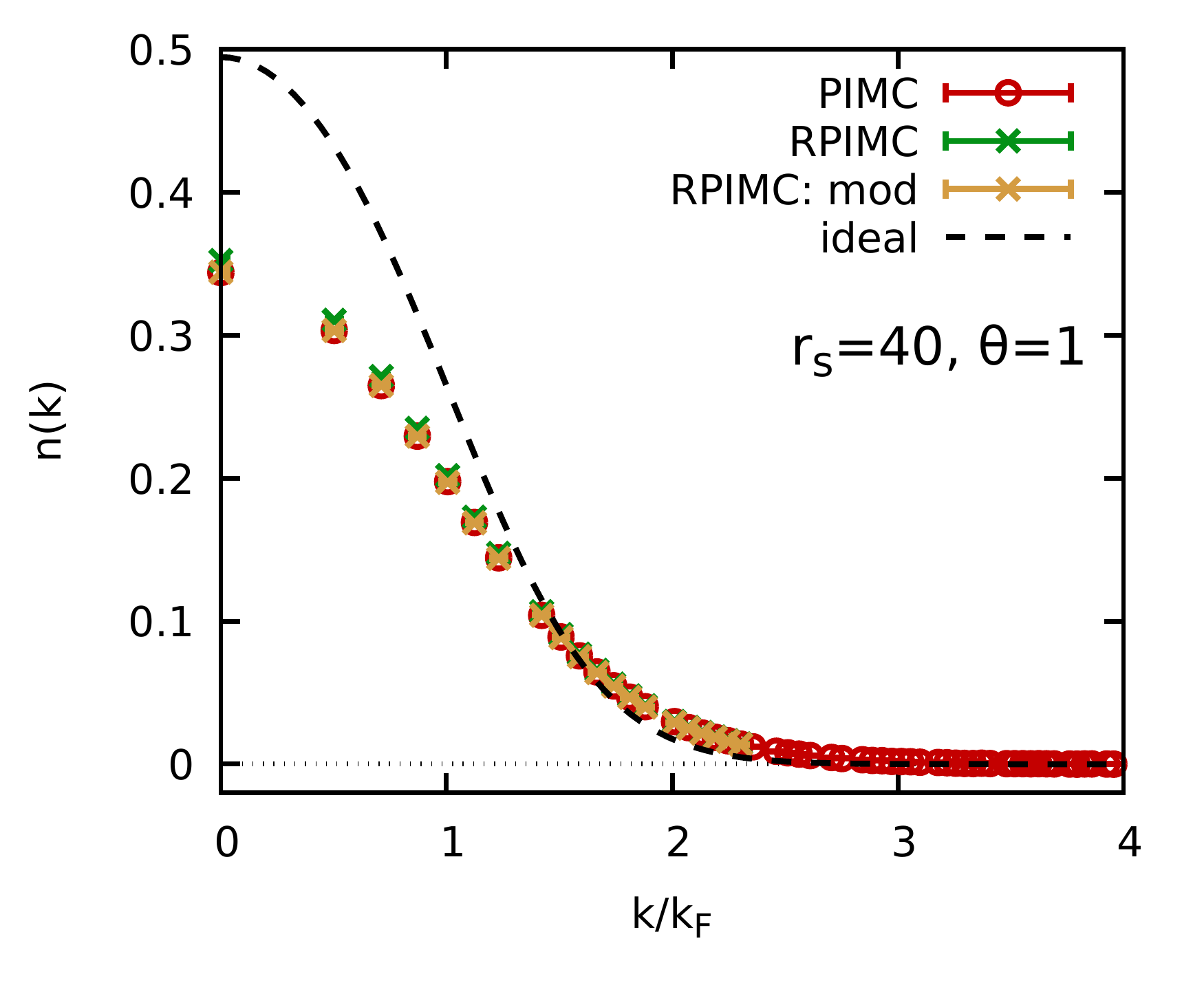}\hspace*{-0.012\textwidth}\includegraphics[width=0.482\textwidth]{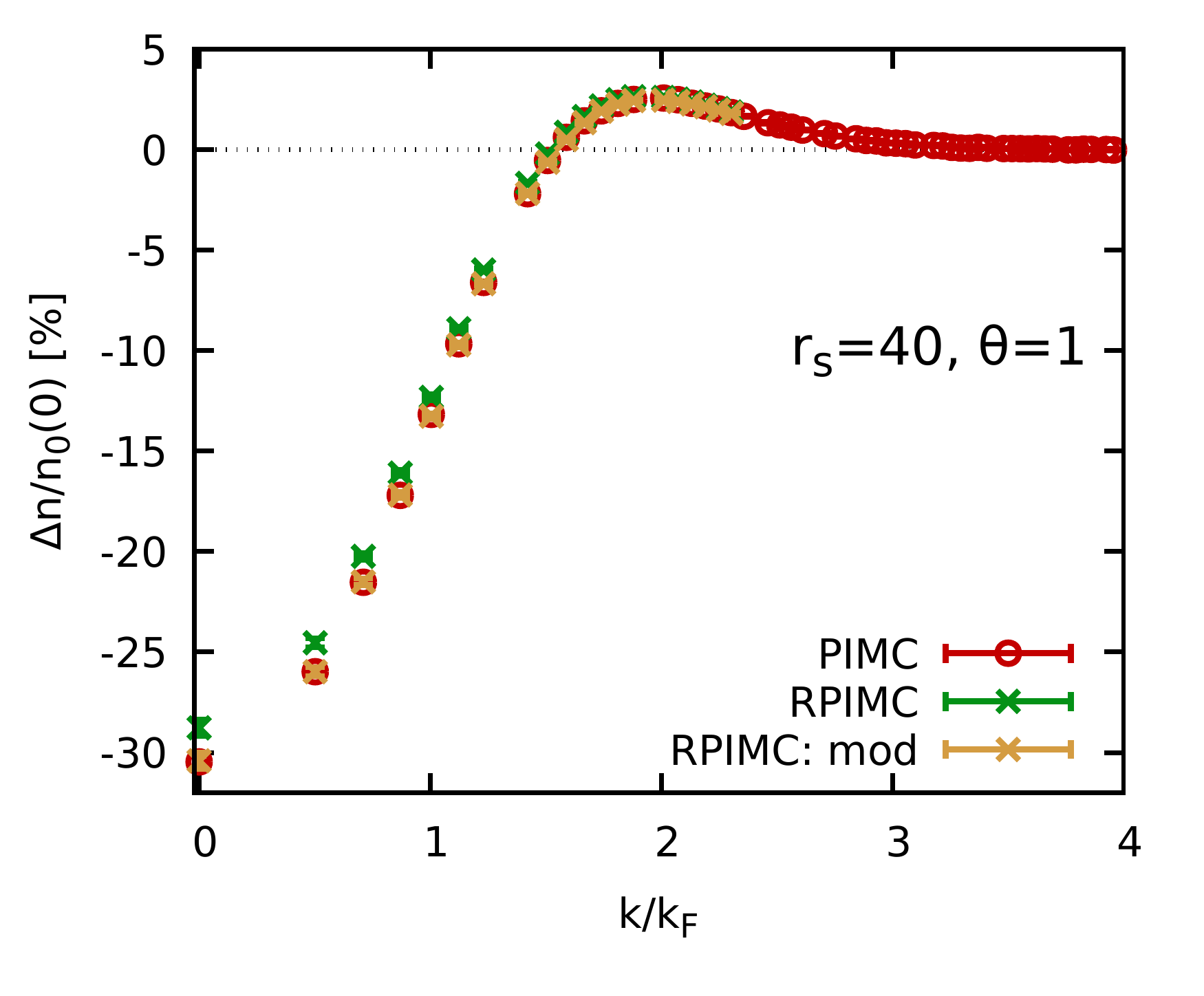}
\caption{\label{fig:RPIMC} Momentum distribution function of the unpolarized electron gas at $\theta=1$ and $r_s=4$ (top row) and $r_s=40$ (bottom row). Left column: $n(\mathbf{k})$; red circles: direct PIMC data from this work; green crosses: restricted PIMC data from Ref.~\cite{Militzer_momentum_HEDP_2019}; yellow crosses: RPIMC data with modified normalization constant, see the main text. Right column: relative deviation between different PIMC data and the Fermi function Eq.~(\ref{eq:Fermi}) in percent of $n_0(0)$.
}
\end{figure*}

\subsection{Comparison to restricted PIMC\label{sec:RPIMC}}

The final analysis presented in this work is the comparison of our new PIMC data to the restricted PIMC results from Ref.~\cite{Militzer_momentum_HEDP_2019}. Since the sign problem prevents direct PIMC simulations at high degeneracy~\cite{dornheim_sign_problem}, here we restrict ourselves to $\theta=1$. The results are shown in Fig.~\ref{fig:RPIMC}, with the top row corresponding to a metallic density, $r_s=4$. The left panel shows results for $n(\mathbf{k})$ and the red circles and green crosses are direct PIMC data from this work and RPIMC data from Ref.~\cite{Militzer_momentum_HEDP_2019}. First and foremost, we note that the RPIMC data have been obtained for $N=66$, whereas the direct PIMC method is restricted to $N\lesssim40$ at these conditions, see Sec.~\ref{sec:FSC} above. Consequently, the red circles have been obtained for $N=34$ and the results for $n(\mathbf{k})$ are thus available on a different grid of $\mathbf{k}$-points in this case. Still, we find good agreement between the two data sets over the entire depicted $k$-range.
To better resolve potential differences, we show the deviation of the direct PIMC and RPIMC data to the Fermi distribution [in percent of $n_0(0)$] in the right panel. As a reminder for the possible effect of the system size $N$, we have also included direct PIMC data for $N=40$, which are depicted by the blue diamonds. Evidently, no finite-size effects are present except for $\mathbf{k}=\mathbf{0}$; see the discussion of Fig.~\ref{fig:FSC} for a more complete analysis of this point.
At the same time, we find deviations between RPIMC and direct PIMC for $k\lesssim2k_\textnormal{F}$. In particular, the RPIMC data do not exhibit the minimum in $\Delta n$ around $k=1.2k_\textnormal{F}$. A likely explanation for this devation is the determination of the normalization in the RPIMC simulation, which required estimating the kinetic energy with an independent simulation~\cite{Militzer_momentum_HEDP_2019}, while only one simulation is required within our present scheme. We thus multiplied the RPIMC data by an empirical factor of $0.977$. The resulting yellow crosses are in very good agreement with our data except for very small wave numbers, where the statistical uncertainty in the RPIMC data is large.

Let us next consider a strongly coupled case, $r_s=40$, shown in the bottom row of Fig.~\ref{fig:RPIMC}. In this case, a direct PIMC simulation with $N=66$ is no problem, and we find an average sign of $S\approx0.77$ (simulations are feasible for $S\gtrsim10^{-2}$).
Considering the depiction of $n(\mathbf{k})$ itself (left panel), significant deviations between the red circles and green crosses are visible with the bare eye, as both data sets are available for the same wave numbers $k$. This is further confirmed by the right panel, showing again the relative deviation to $n_0(\mathbf{k})$. Finally, the yellow crosses have been obtained by multiplying the RPIMC data by the same empirical factor of $0.977$, which then leads to perfect agreement to the direct PIMC data within the given Monte Carlo error bars. This strongly indicates that the observed differences to the RPIMC data are not an inherent property of the fixed node approximation itself, but rather a consequence of the determination of the normalization constant from $n(\mathbf{r},\mathbf{r'})$.

\section{Summary and Discussion\label{sec:summary}}

In this work, we have presented extensive new \textit{ab initio} PIMC results for the momentum distribution of the UEG at finite temperature in the range of $2\leq r_s \leq 50$ and $0.75\leq\theta\leq4$. This was achieved using an extended PIMC configuration space consisting of both canonical (closed) and off-diagonal (open) configurations, which allows us to directly compute $n(\mathbf{k})$ without the subsequent need to determine a proportionality constant from a normalization condition. 
Since we have not imposed any nodal constraints, our simulations are subject to the fermion sign problem, leading to an exponential increase in computation time with decreasing temperature or increasing system-size. In particular, we have found that the sign problem is more severe in the off-diagonal configuration space, as the presence of an open trajectory makes the formation of exchange-cycles more likely.

From a physical perspective, we have investigated the nontrivial increase in the occupation of low-momentum states at the Fermi temperature due to exchange-correlation effects, and its connection to negative values of the XC-contribution to the kinetic energy $K_\textnormal{xc}$. More specifically, $K_\textnormal{xc}$ is negative at $\theta=1$ for $r_s\lesssim3$, and changes its sign for lower densities. The increased occupation of $n(\mathbf{0})$ compared to the ideal Fermi gas, on the other hand, persists for $r_s\lesssim7$.  
Therefore, we conclude that an increased occupation of low-momentum states is a necessary but not sufficient criterion for the XC-induced lowering of the kinetic energy reported in Refs.~\cite{Militzer_PRL_2002,Kraeft_PRE_2002}.
In addition, we have studied the dependence of $n(\mathbf{k})$ on the temperature and have found that the occupation of small-$\mathbf{k}$ states relative to the ideal Fermi distribution $n_0(\mathbf{k})$ is depleted when the temperature is decreased, as the system becomes more strongly correlated.
Furthermore, we have investigated the impact of quantum statistics on both the momentum distribution and the off-diagonal density matrix in coordinate space and have found that it cannot be neglected even in the strongly coupled electron liquid regime ($r_s=50$).

Finally, we have compared our new simulation results to previous data for $n(\mathbf{k})$ based on the fixed-node approximation. Here, we have found a constant factor between the two data sets, that can most likely be attributed to the inaccurate determination of the normalization of $n(\mathbf{k})$ in Ref.~\cite{Militzer_momentum_HEDP_2019}.

Considering future extensions of our work, we note that accurate results for different properties of the UEG and reliable parametrizations thereof are of paramount importance for many applications and, thus, constitute an important achievement in themselves. All PIMC data are freely available online~\cite{repo}, and can be used as input for other methods or as a benchmark for the development of new methods and to assess the accuracy of existing approximations.
A further interesting topic for future research is the investigation of spin effects on $n(\mathbf{k})$, which will be covered in a future publication. Moreover, the simulation scheme with the extended configuration space can straightforwardly be adapted to other methods like the permutation blocking PIMC (PB-PIMC) approach by Dornheim and co-workers~\cite{Dornheim_NJP_2015,dornheim_jcp,Dornheim_CPP_2019}. The PB-PIMC method employs approximations but it extends the direct PIMC method towards lower temperatures, and can thus further help to complete our current picture of the UEG as a fundamental model system.

\section*{Acknowledgments}
We gratefully acknowledge Kai Hunger for sharing his CPIMC data for $n(\mathbf{k})$ from Ref.~\cite{Hunger_PRE_2021}.

This work was partly funded by the Center of Advanced Systems Understanding (CASUS) which is financed by Germany's Federal Ministry of Education and Research (BMBF) and by the Saxon Ministry for Science, Culture and Tourism (SMWK) with tax funds on the basis of the budget approved by the Saxon State Parliament.
The PIMC calculations were carried out at the Norddeutscher Verbund f\"ur Hoch- und H\"ochstleistungsrechnen (HLRN) under grant shp00026, and on a Bull Cluster at the Center for Information Services and High Performace Computing (ZIH) at Technische Universit\"at Dresden.

\bibliography{bibliography}

\begin{thebibliography}{85}%
\makeatletter
\providecommand \@ifxundefined [1]{%
 \@ifx{#1\undefined}
}%
\providecommand \@ifnum [1]{%
 \ifnum #1\expandafter \@firstoftwo
 \else \expandafter \@secondoftwo
 \fi
}%
\providecommand \@ifx [1]{%
 \ifx #1\expandafter \@firstoftwo
 \else \expandafter \@secondoftwo
 \fi
}%
\providecommand \natexlab [1]{#1}%
\providecommand \enquote  [1]{``#1''}%
\providecommand \bibnamefont  [1]{#1}%
\providecommand \bibfnamefont [1]{#1}%
\providecommand \citenamefont [1]{#1}%
\providecommand \href@noop [0]{\@secondoftwo}%
\providecommand \href [0]{\begingroup \@sanitize@url \@href}%
\providecommand \@href[1]{\@@startlink{#1}\@@href}%
\providecommand \@@href[1]{\endgroup#1\@@endlink}%
\providecommand \@sanitize@url [0]{\catcode `\\12\catcode `\$12\catcode
  `\&12\catcode `\#12\catcode `\^12\catcode `\_12\catcode `\%12\relax}%
\providecommand \@@startlink[1]{}%
\providecommand \@@endlink[0]{}%
\providecommand \url  [0]{\begingroup\@sanitize@url \@url }%
\providecommand \@url [1]{\endgroup\@href {#1}{\urlprefix }}%
\providecommand \urlprefix  [0]{URL }%
\providecommand \Eprint [0]{\href }%
\providecommand \doibase [0]{http://dx.doi.org/}%
\providecommand \selectlanguage [0]{\@gobble}%
\providecommand \bibinfo  [0]{\@secondoftwo}%
\providecommand \bibfield  [0]{\@secondoftwo}%
\providecommand \translation [1]{[#1]}%
\providecommand \BibitemOpen [0]{}%
\providecommand \bibitemStop [0]{}%
\providecommand \bibitemNoStop [0]{.\EOS\space}%
\providecommand \EOS [0]{\spacefactor3000\relax}%
\providecommand \BibitemShut  [1]{\csname bibitem#1\endcsname}%
\let\auto@bib@innerbib\@empty
\bibitem [{\citenamefont {Fortov}(2009)}]{fortov_review}%
  \BibitemOpen
  \bibfield  {author} {\bibinfo {author} {\bibfnamefont {V.~E.}\ \bibnamefont
  {Fortov}},\ }\bibfield  {title} {\enquote {\bibinfo {title} {Extreme states
  of matter on earth and in space},}\ }\href
  {https://www.turpion.org/php/paper.phtml?journal_id=pu&paper_id=6821}
  {\bibfield  {journal} {\bibinfo  {journal} {Phys.-Usp}\ }\textbf {\bibinfo
  {volume} {52}},\ \bibinfo {pages} {615--647} (\bibinfo {year}
  {2009})}\BibitemShut {NoStop}%
\bibitem [{\citenamefont {Graziani}\ \emph {et~al.}(2014)\citenamefont
  {Graziani}, \citenamefont {Desjarlais}, \citenamefont {Redmer},\ and\
  \citenamefont {Trickey}}]{wdm_book}%
  \BibitemOpen
  \bibinfo {editor} {\bibfnamefont {F.}~\bibnamefont {Graziani}}, \bibinfo
  {editor} {\bibfnamefont {M.~P.}\ \bibnamefont {Desjarlais}}, \bibinfo
  {editor} {\bibfnamefont {R.}~\bibnamefont {Redmer}}, \ and\ \bibinfo {editor}
  {\bibfnamefont {S.~B.}\ \bibnamefont {Trickey}},\ eds.,\ \href@noop {} {\emph
  {\bibinfo {title} {Frontiers and Challenges in Warm Dense Matter}}}\
  (\bibinfo  {publisher} {Springer},\ \bibinfo {address} {International
  Publishing},\ \bibinfo {year} {2014})\BibitemShut {NoStop}%
\bibitem [{\citenamefont {Bonitz}\ \emph {et~al.}(2020)\citenamefont {Bonitz},
  \citenamefont {Dornheim}, \citenamefont {Moldabekov}, \citenamefont {Zhang},
  \citenamefont {Hamann}, \citenamefont {Kählert}, \citenamefont {Filinov},
  \citenamefont {Ramakrishna},\ and\ \citenamefont {Vorberger}}]{new_POP}%
  \BibitemOpen
  \bibfield  {author} {\bibinfo {author} {\bibfnamefont {M.}~\bibnamefont
  {Bonitz}}, \bibinfo {author} {\bibfnamefont {T.}~\bibnamefont {Dornheim}},
  \bibinfo {author} {\bibfnamefont {Zh.~A.}\ \bibnamefont {Moldabekov}},
  \bibinfo {author} {\bibfnamefont {S.}~\bibnamefont {Zhang}}, \bibinfo
  {author} {\bibfnamefont {P.}~\bibnamefont {Hamann}}, \bibinfo {author}
  {\bibfnamefont {H.}~\bibnamefont {Kählert}}, \bibinfo {author}
  {\bibfnamefont {A.}~\bibnamefont {Filinov}}, \bibinfo {author} {\bibfnamefont
  {K.}~\bibnamefont {Ramakrishna}}, \ and\ \bibinfo {author} {\bibfnamefont
  {J.}~\bibnamefont {Vorberger}},\ }\bibfield  {title} {\enquote {\bibinfo
  {title} {Ab initio simulation of warm dense matter},}\ }\href {\doibase
  10.1063/1.5143225} {\bibfield  {journal} {\bibinfo  {journal} {Physics of
  Plasmas}\ }\textbf {\bibinfo {volume} {27}},\ \bibinfo {pages} {042710}
  (\bibinfo {year} {2020})},\ \Eprint
  {http://arxiv.org/abs/https://doi.org/10.1063/1.5143225}
  {https://doi.org/10.1063/1.5143225} \BibitemShut {NoStop}%
\bibitem [{\citenamefont {Nettelmann}\ \emph {et~al.}(2008)\citenamefont
  {Nettelmann}, \citenamefont {Redmer},\ and\ \citenamefont
  {Blaschke}}]{Nettelmann2008}%
  \BibitemOpen
  \bibfield  {author} {\bibinfo {author} {\bibfnamefont {N.}~\bibnamefont
  {Nettelmann}}, \bibinfo {author} {\bibfnamefont {R.}~\bibnamefont {Redmer}},
  \ and\ \bibinfo {author} {\bibfnamefont {D.}~\bibnamefont {Blaschke}},\
  }\bibfield  {title} {\enquote {\bibinfo {title} {Warm dense matter in giant
  planets and exoplanets},}\ }\href {\doibase 10.1134/S1063779608070277}
  {\bibfield  {journal} {\bibinfo  {journal} {Physics of Particles and Nuclei}\
  }\textbf {\bibinfo {volume} {39}},\ \bibinfo {pages} {1122--1127} (\bibinfo
  {year} {2008})}\BibitemShut {NoStop}%
\bibitem [{\citenamefont {Militzer}\ \emph {et~al.}(2008)\citenamefont
  {Militzer}, \citenamefont {Hubbard}, \citenamefont {Vorberger}, \citenamefont
  {Tamblyn},\ and\ \citenamefont {Bonev}}]{Militzer_2008}%
  \BibitemOpen
  \bibfield  {author} {\bibinfo {author} {\bibfnamefont {B.}~\bibnamefont
  {Militzer}}, \bibinfo {author} {\bibfnamefont {W.~B.}\ \bibnamefont
  {Hubbard}}, \bibinfo {author} {\bibfnamefont {J.}~\bibnamefont {Vorberger}},
  \bibinfo {author} {\bibfnamefont {I.}~\bibnamefont {Tamblyn}}, \ and\
  \bibinfo {author} {\bibfnamefont {S.~A.}\ \bibnamefont {Bonev}},\ }\bibfield
  {title} {\enquote {\bibinfo {title} {A massive core in jupiter predicted from
  first-principles simulations},}\ }\href {\doibase 10.1086/594364} {\bibfield
  {journal} {\bibinfo  {journal} {The Astrophysical Journal}\ }\textbf
  {\bibinfo {volume} {688}},\ \bibinfo {pages} {L45--L48} (\bibinfo {year}
  {2008})}\BibitemShut {NoStop}%
\bibitem [{\citenamefont {Vorberger}\ \emph {et~al.}(2007)\citenamefont
  {Vorberger}, \citenamefont {Tamblyn}, \citenamefont {Militzer},\ and\
  \citenamefont {Bonev}}]{militzer1}%
  \BibitemOpen
  \bibfield  {author} {\bibinfo {author} {\bibfnamefont {J.}~\bibnamefont
  {Vorberger}}, \bibinfo {author} {\bibfnamefont {I.}~\bibnamefont {Tamblyn}},
  \bibinfo {author} {\bibfnamefont {B.}~\bibnamefont {Militzer}}, \ and\
  \bibinfo {author} {\bibfnamefont {S.~A.}\ \bibnamefont {Bonev}},\ }\bibfield
  {title} {\enquote {\bibinfo {title} {Hydrogen-helium mixtures in the
  interiors of giant planets},}\ }\href
  {https://journals.aps.org/prb/abstract/10.1103/PhysRevB.75.024206} {\bibfield
   {journal} {\bibinfo  {journal} {Phys. Rev. B}\ }\textbf {\bibinfo {volume}
  {75}},\ \bibinfo {pages} {024206} (\bibinfo {year} {2007})}\BibitemShut
  {NoStop}%
\bibitem [{\citenamefont {Benuzzi-Mounaix}\ \emph {et~al.}(2014)\citenamefont
  {Benuzzi-Mounaix}, \citenamefont {Mazevet}, \citenamefont {Ravasio},
  \citenamefont {Vinci}, \citenamefont {Denoeud}, \citenamefont {Koenig},
  \citenamefont {Amadou}, \citenamefont {Brambrink}, \citenamefont {Festa},
  \citenamefont {Levy}, \citenamefont {Harmand}, \citenamefont {Brygoo},
  \citenamefont {Huser}, \citenamefont {Recoules}, \citenamefont {Bouchet},
  \citenamefont {Morard}, \citenamefont {Guyot}, \citenamefont {de~Resseguier},
  \citenamefont {Myanishi}, \citenamefont {Ozaki}, \citenamefont {Dorchies},
  \citenamefont {Gaudin}, \citenamefont {Leguay}, \citenamefont {Peyrusse},
  \citenamefont {Henry}, \citenamefont {Raffestin}, \citenamefont {Pape},
  \citenamefont {Smith},\ and\ \citenamefont {Musella}}]{Benuzzi_Mounaix_2014}%
  \BibitemOpen
  \bibfield  {author} {\bibinfo {author} {\bibfnamefont {Alessandra}\
  \bibnamefont {Benuzzi-Mounaix}}, \bibinfo {author} {\bibfnamefont
  {St{\'{e}}phane}\ \bibnamefont {Mazevet}}, \bibinfo {author} {\bibfnamefont
  {Alessandra}\ \bibnamefont {Ravasio}}, \bibinfo {author} {\bibfnamefont
  {Tommaso}\ \bibnamefont {Vinci}}, \bibinfo {author} {\bibfnamefont {Adrien}\
  \bibnamefont {Denoeud}}, \bibinfo {author} {\bibfnamefont {Michel}\
  \bibnamefont {Koenig}}, \bibinfo {author} {\bibfnamefont {Nourou}\
  \bibnamefont {Amadou}}, \bibinfo {author} {\bibfnamefont {Erik}\ \bibnamefont
  {Brambrink}}, \bibinfo {author} {\bibfnamefont {Floriane}\ \bibnamefont
  {Festa}}, \bibinfo {author} {\bibfnamefont {Anna}\ \bibnamefont {Levy}},
  \bibinfo {author} {\bibfnamefont {Marion}\ \bibnamefont {Harmand}}, \bibinfo
  {author} {\bibfnamefont {St{\'{e}}phanie}\ \bibnamefont {Brygoo}}, \bibinfo
  {author} {\bibfnamefont {Gael}\ \bibnamefont {Huser}}, \bibinfo {author}
  {\bibfnamefont {Vanina}\ \bibnamefont {Recoules}}, \bibinfo {author}
  {\bibfnamefont {Johan}\ \bibnamefont {Bouchet}}, \bibinfo {author}
  {\bibfnamefont {Guillaume}\ \bibnamefont {Morard}}, \bibinfo {author}
  {\bibfnamefont {Fran{\c{c}}ois}\ \bibnamefont {Guyot}}, \bibinfo {author}
  {\bibfnamefont {Thibaut}\ \bibnamefont {de~Resseguier}}, \bibinfo {author}
  {\bibfnamefont {Kohei}\ \bibnamefont {Myanishi}}, \bibinfo {author}
  {\bibfnamefont {Norimasa}\ \bibnamefont {Ozaki}}, \bibinfo {author}
  {\bibfnamefont {Fabien}\ \bibnamefont {Dorchies}}, \bibinfo {author}
  {\bibfnamefont {Jer{\^{o}}me}\ \bibnamefont {Gaudin}}, \bibinfo {author}
  {\bibfnamefont {Pierre~Marie}\ \bibnamefont {Leguay}}, \bibinfo {author}
  {\bibfnamefont {Olivier}\ \bibnamefont {Peyrusse}}, \bibinfo {author}
  {\bibfnamefont {Olivier}\ \bibnamefont {Henry}}, \bibinfo {author}
  {\bibfnamefont {Didier}\ \bibnamefont {Raffestin}}, \bibinfo {author}
  {\bibfnamefont {Sebastien~Le}\ \bibnamefont {Pape}}, \bibinfo {author}
  {\bibfnamefont {Ray}\ \bibnamefont {Smith}}, \ and\ \bibinfo {author}
  {\bibfnamefont {Riccardo}\ \bibnamefont {Musella}},\ }\bibfield  {title}
  {\enquote {\bibinfo {title} {Progress in warm dense matter study with
  applications to planetology},}\ }\href {\doibase
  10.1088/0031-8949/2014/t161/014060} {\bibfield  {journal} {\bibinfo
  {journal} {Physica Scripta}\ }\textbf {\bibinfo {volume} {T161}},\ \bibinfo
  {pages} {014060} (\bibinfo {year} {2014})}\BibitemShut {NoStop}%
\bibitem [{\citenamefont {Saumon}\ \emph {et~al.}(1992)\citenamefont {Saumon},
  \citenamefont {Hubbard}, \citenamefont {Chabrier},\ and\ \citenamefont {van
  Horn}}]{saumon1}%
  \BibitemOpen
  \bibfield  {author} {\bibinfo {author} {\bibfnamefont {D.}~\bibnamefont
  {Saumon}}, \bibinfo {author} {\bibfnamefont {W.~B.}\ \bibnamefont {Hubbard}},
  \bibinfo {author} {\bibfnamefont {G.}~\bibnamefont {Chabrier}}, \ and\
  \bibinfo {author} {\bibfnamefont {H.~M.}\ \bibnamefont {van Horn}},\
  }\bibfield  {title} {\enquote {\bibinfo {title} {The role of the
  molecular-metallic transition of hydrogen in the evolution of jupiter,
  saturn, and brown dwarfs},}\ }\href
  {http://adsabs.harvard.edu/full/1992ApJ...391..827S} {\bibfield  {journal}
  {\bibinfo  {journal} {Astrophys. J}\ }\textbf {\bibinfo {volume} {391}},\
  \bibinfo {pages} {827--831} (\bibinfo {year} {1992})}\BibitemShut {NoStop}%
\bibitem [{\citenamefont {Becker}\ \emph {et~al.}(2014)\citenamefont {Becker},
  \citenamefont {Lorenzen}, \citenamefont {Fortney}, \citenamefont
  {Nettelmann}, \citenamefont {Sch\"ottler},\ and\ \citenamefont
  {Redmer}}]{becker}%
  \BibitemOpen
  \bibfield  {author} {\bibinfo {author} {\bibfnamefont {A.}~\bibnamefont
  {Becker}}, \bibinfo {author} {\bibfnamefont {W.}~\bibnamefont {Lorenzen}},
  \bibinfo {author} {\bibfnamefont {J.~J.}\ \bibnamefont {Fortney}}, \bibinfo
  {author} {\bibfnamefont {N.}~\bibnamefont {Nettelmann}}, \bibinfo {author}
  {\bibfnamefont {M.}~\bibnamefont {Sch\"ottler}}, \ and\ \bibinfo {author}
  {\bibfnamefont {R.}~\bibnamefont {Redmer}},\ }\bibfield  {title} {\enquote
  {\bibinfo {title} {Ab initio equations of state for hydrogen (h-reos.3) and
  helium (he-reos.3) and their implications for the interior of brown
  dwarfs},}\ }\href
  {https://iopscience.iop.org/article/10.1088/0067-0049/215/2/21/meta}
  {\bibfield  {journal} {\bibinfo  {journal} {Astrophys. J. Suppl. Ser}\
  }\textbf {\bibinfo {volume} {215}},\ \bibinfo {pages} {21} (\bibinfo {year}
  {2014})}\BibitemShut {NoStop}%
\bibitem [{\citenamefont {Chamel}\ and\ \citenamefont
  {Haensel}(2008)}]{Chamel2008}%
  \BibitemOpen
  \bibfield  {author} {\bibinfo {author} {\bibfnamefont {Nicolas}\ \bibnamefont
  {Chamel}}\ and\ \bibinfo {author} {\bibfnamefont {Pawel}\ \bibnamefont
  {Haensel}},\ }\bibfield  {title} {\enquote {\bibinfo {title} {Physics of
  neutron star crusts},}\ }\href {\doibase 10.12942/lrr-2008-10} {\bibfield
  {journal} {\bibinfo  {journal} {Living Reviews in Relativity}\ }\textbf
  {\bibinfo {volume} {11}},\ \bibinfo {pages} {10} (\bibinfo {year}
  {2008})}\BibitemShut {NoStop}%
\bibitem [{\citenamefont {Hu}\ \emph {et~al.}(2011)\citenamefont {Hu},
  \citenamefont {Militzer}, \citenamefont {Goncharov},\ and\ \citenamefont
  {Skupsky}}]{hu_ICF}%
  \BibitemOpen
  \bibfield  {author} {\bibinfo {author} {\bibfnamefont {S.~X.}\ \bibnamefont
  {Hu}}, \bibinfo {author} {\bibfnamefont {B.}~\bibnamefont {Militzer}},
  \bibinfo {author} {\bibfnamefont {V.~N.}\ \bibnamefont {Goncharov}}, \ and\
  \bibinfo {author} {\bibfnamefont {S.}~\bibnamefont {Skupsky}},\ }\bibfield
  {title} {\enquote {\bibinfo {title} {First-principles equation-of-state table
  of deuterium for inertial confinement fusion applications},}\ }\href
  {https://journals.aps.org/prb/abstract/10.1103/PhysRevB.84.224109} {\bibfield
   {journal} {\bibinfo  {journal} {Phys. Rev. B}\ }\textbf {\bibinfo {volume}
  {84}},\ \bibinfo {pages} {224109} (\bibinfo {year} {2011})}\BibitemShut
  {NoStop}%
\bibitem [{\citenamefont {Brongersma}\ \emph {et~al.}(2015)\citenamefont
  {Brongersma}, \citenamefont {Halas},\ and\ \citenamefont
  {Nordlander}}]{Brongersma2015}%
  \BibitemOpen
  \bibfield  {author} {\bibinfo {author} {\bibfnamefont {Mark~L.}\ \bibnamefont
  {Brongersma}}, \bibinfo {author} {\bibfnamefont {Naomi~J.}\ \bibnamefont
  {Halas}}, \ and\ \bibinfo {author} {\bibfnamefont {Peter}\ \bibnamefont
  {Nordlander}},\ }\bibfield  {title} {\enquote {\bibinfo {title}
  {Plasmon-induced hot carrier science and technology},}\ }\href {\doibase
  10.1038/nnano.2014.311} {\bibfield  {journal} {\bibinfo  {journal} {Nature
  Nanotechnology}\ }\textbf {\bibinfo {volume} {10}},\ \bibinfo {pages}
  {25--34} (\bibinfo {year} {2015})}\BibitemShut {NoStop}%
\bibitem [{\citenamefont {Kraus}\ \emph {et~al.}(2016)\citenamefont {Kraus},
  \citenamefont {Ravasio}, \citenamefont {Gauthier}, \citenamefont {Gericke},
  \citenamefont {Vorberger}, \citenamefont {Frydrych}, \citenamefont
  {Helfrich}, \citenamefont {Fletcher}, \citenamefont {Schaumann},
  \citenamefont {Nagler}, \citenamefont {Barbrel}, \citenamefont {Bachmann},
  \citenamefont {Gamboa}, \citenamefont {G{\"o}de}, \citenamefont {Granados},
  \citenamefont {Gregori}, \citenamefont {Lee}, \citenamefont {Neumayer},
  \citenamefont {Schumaker}, \citenamefont {D{\"o}ppner}, \citenamefont
  {Falcone}, \citenamefont {Glenzer},\ and\ \citenamefont {Roth}}]{Kraus2016}%
  \BibitemOpen
  \bibfield  {author} {\bibinfo {author} {\bibfnamefont {D.}~\bibnamefont
  {Kraus}}, \bibinfo {author} {\bibfnamefont {A.}~\bibnamefont {Ravasio}},
  \bibinfo {author} {\bibfnamefont {M.}~\bibnamefont {Gauthier}}, \bibinfo
  {author} {\bibfnamefont {D.~O.}\ \bibnamefont {Gericke}}, \bibinfo {author}
  {\bibfnamefont {J.}~\bibnamefont {Vorberger}}, \bibinfo {author}
  {\bibfnamefont {S.}~\bibnamefont {Frydrych}}, \bibinfo {author}
  {\bibfnamefont {J.}~\bibnamefont {Helfrich}}, \bibinfo {author}
  {\bibfnamefont {L.~B.}\ \bibnamefont {Fletcher}}, \bibinfo {author}
  {\bibfnamefont {G.}~\bibnamefont {Schaumann}}, \bibinfo {author}
  {\bibfnamefont {B.}~\bibnamefont {Nagler}}, \bibinfo {author} {\bibfnamefont
  {B.}~\bibnamefont {Barbrel}}, \bibinfo {author} {\bibfnamefont
  {B.}~\bibnamefont {Bachmann}}, \bibinfo {author} {\bibfnamefont {E.~J.}\
  \bibnamefont {Gamboa}}, \bibinfo {author} {\bibfnamefont {S.}~\bibnamefont
  {G{\"o}de}}, \bibinfo {author} {\bibfnamefont {E.}~\bibnamefont {Granados}},
  \bibinfo {author} {\bibfnamefont {G.}~\bibnamefont {Gregori}}, \bibinfo
  {author} {\bibfnamefont {H.~J.}\ \bibnamefont {Lee}}, \bibinfo {author}
  {\bibfnamefont {P.}~\bibnamefont {Neumayer}}, \bibinfo {author}
  {\bibfnamefont {W.}~\bibnamefont {Schumaker}}, \bibinfo {author}
  {\bibfnamefont {T.}~\bibnamefont {D{\"o}ppner}}, \bibinfo {author}
  {\bibfnamefont {R.~W.}\ \bibnamefont {Falcone}}, \bibinfo {author}
  {\bibfnamefont {S.~H.}\ \bibnamefont {Glenzer}}, \ and\ \bibinfo {author}
  {\bibfnamefont {M.}~\bibnamefont {Roth}},\ }\bibfield  {title} {\enquote
  {\bibinfo {title} {Nanosecond formation of diamond and lonsdaleite by shock
  compression of graphite},}\ }\href {\doibase 10.1038/ncomms10970} {\bibfield
  {journal} {\bibinfo  {journal} {Nature Communications}\ }\textbf {\bibinfo
  {volume} {7}},\ \bibinfo {pages} {10970} (\bibinfo {year}
  {2016})}\BibitemShut {NoStop}%
\bibitem [{\citenamefont {Kraus}\ \emph {et~al.}(2017)\citenamefont {Kraus},
  \citenamefont {Vorberger}, \citenamefont {Pak}, \citenamefont {Hartley},
  \citenamefont {Fletcher}, \citenamefont {Frydrych}, \citenamefont {Galtier},
  \citenamefont {Gamboa}, \citenamefont {Gericke}, \citenamefont {Glenzer},
  \citenamefont {Granados}, \citenamefont {MacDonald}, \citenamefont
  {MacKinnon}, \citenamefont {McBride}, \citenamefont {Nam}, \citenamefont
  {Neumayer}, \citenamefont {Roth}, \citenamefont {Saunders}, \citenamefont
  {Schuster}, \citenamefont {Sun}, \citenamefont {van Driel}, \citenamefont
  {D{\"o}ppner},\ and\ \citenamefont {Falcone}}]{Kraus2017}%
  \BibitemOpen
  \bibfield  {author} {\bibinfo {author} {\bibfnamefont {D.}~\bibnamefont
  {Kraus}}, \bibinfo {author} {\bibfnamefont {J.}~\bibnamefont {Vorberger}},
  \bibinfo {author} {\bibfnamefont {A.}~\bibnamefont {Pak}}, \bibinfo {author}
  {\bibfnamefont {N.~J.}\ \bibnamefont {Hartley}}, \bibinfo {author}
  {\bibfnamefont {L.~B.}\ \bibnamefont {Fletcher}}, \bibinfo {author}
  {\bibfnamefont {S.}~\bibnamefont {Frydrych}}, \bibinfo {author}
  {\bibfnamefont {E.}~\bibnamefont {Galtier}}, \bibinfo {author} {\bibfnamefont
  {E.~J.}\ \bibnamefont {Gamboa}}, \bibinfo {author} {\bibfnamefont {D.~O.}\
  \bibnamefont {Gericke}}, \bibinfo {author} {\bibfnamefont {S.~H.}\
  \bibnamefont {Glenzer}}, \bibinfo {author} {\bibfnamefont {E.}~\bibnamefont
  {Granados}}, \bibinfo {author} {\bibfnamefont {M.~J.}\ \bibnamefont
  {MacDonald}}, \bibinfo {author} {\bibfnamefont {A.~J.}\ \bibnamefont
  {MacKinnon}}, \bibinfo {author} {\bibfnamefont {E.~E.}\ \bibnamefont
  {McBride}}, \bibinfo {author} {\bibfnamefont {I.}~\bibnamefont {Nam}},
  \bibinfo {author} {\bibfnamefont {P.}~\bibnamefont {Neumayer}}, \bibinfo
  {author} {\bibfnamefont {M.}~\bibnamefont {Roth}}, \bibinfo {author}
  {\bibfnamefont {A.~M.}\ \bibnamefont {Saunders}}, \bibinfo {author}
  {\bibfnamefont {A.~K.}\ \bibnamefont {Schuster}}, \bibinfo {author}
  {\bibfnamefont {P.}~\bibnamefont {Sun}}, \bibinfo {author} {\bibfnamefont
  {T.}~\bibnamefont {van Driel}}, \bibinfo {author} {\bibfnamefont
  {T.}~\bibnamefont {D{\"o}ppner}}, \ and\ \bibinfo {author} {\bibfnamefont
  {R.~W.}\ \bibnamefont {Falcone}},\ }\bibfield  {title} {\enquote {\bibinfo
  {title} {Formation of diamonds in laser-compressed hydrocarbons at planetary
  interior conditions},}\ }\href {\doibase 10.1038/s41550-017-0219-9}
  {\bibfield  {journal} {\bibinfo  {journal} {Nature Astronomy}\ }\textbf
  {\bibinfo {volume} {1}},\ \bibinfo {pages} {606--611} (\bibinfo {year}
  {2017})}\BibitemShut {NoStop}%
\bibitem [{\citenamefont {Lazicki}\ \emph {et~al.}(2021)\citenamefont
  {Lazicki}, \citenamefont {McGonegle}, \citenamefont {Rygg}, \citenamefont
  {Braun}, \citenamefont {Swift}, \citenamefont {Gorman}, \citenamefont
  {Smith}, \citenamefont {Heighway}, \citenamefont {Higginbotham},
  \citenamefont {Suggit}, \citenamefont {Fratanduono}, \citenamefont {Coppari},
  \citenamefont {Wehrenberg}, \citenamefont {Kraus}, \citenamefont {Erskine},
  \citenamefont {Bernier}, \citenamefont {McNaney}, \citenamefont {Rudd},
  \citenamefont {Collins}, \citenamefont {Eggert},\ and\ \citenamefont
  {Wark}}]{Lazicki2021}%
  \BibitemOpen
  \bibfield  {author} {\bibinfo {author} {\bibfnamefont {A.}~\bibnamefont
  {Lazicki}}, \bibinfo {author} {\bibfnamefont {D.}~\bibnamefont {McGonegle}},
  \bibinfo {author} {\bibfnamefont {J.~R.}\ \bibnamefont {Rygg}}, \bibinfo
  {author} {\bibfnamefont {D.~G.}\ \bibnamefont {Braun}}, \bibinfo {author}
  {\bibfnamefont {D.~C.}\ \bibnamefont {Swift}}, \bibinfo {author}
  {\bibfnamefont {M.~G.}\ \bibnamefont {Gorman}}, \bibinfo {author}
  {\bibfnamefont {R.~F.}\ \bibnamefont {Smith}}, \bibinfo {author}
  {\bibfnamefont {P.~G.}\ \bibnamefont {Heighway}}, \bibinfo {author}
  {\bibfnamefont {A.}~\bibnamefont {Higginbotham}}, \bibinfo {author}
  {\bibfnamefont {M.~J.}\ \bibnamefont {Suggit}}, \bibinfo {author}
  {\bibfnamefont {D.~E.}\ \bibnamefont {Fratanduono}}, \bibinfo {author}
  {\bibfnamefont {F.}~\bibnamefont {Coppari}}, \bibinfo {author} {\bibfnamefont
  {C.~E.}\ \bibnamefont {Wehrenberg}}, \bibinfo {author} {\bibfnamefont
  {R.~G.}\ \bibnamefont {Kraus}}, \bibinfo {author} {\bibfnamefont
  {D.}~\bibnamefont {Erskine}}, \bibinfo {author} {\bibfnamefont {J.~V.}\
  \bibnamefont {Bernier}}, \bibinfo {author} {\bibfnamefont {J.~M.}\
  \bibnamefont {McNaney}}, \bibinfo {author} {\bibfnamefont {R.~E.}\
  \bibnamefont {Rudd}}, \bibinfo {author} {\bibfnamefont {G.~W.}\ \bibnamefont
  {Collins}}, \bibinfo {author} {\bibfnamefont {J.~H.}\ \bibnamefont {Eggert}},
  \ and\ \bibinfo {author} {\bibfnamefont {J.~S.}\ \bibnamefont {Wark}},\
  }\bibfield  {title} {\enquote {\bibinfo {title} {Metastability of diamond
  ramp-compressed to 2 terapascals},}\ }\href {\doibase
  10.1038/s41586-020-03140-4} {\bibfield  {journal} {\bibinfo  {journal}
  {Nature}\ }\textbf {\bibinfo {volume} {589}},\ \bibinfo {pages} {532--535}
  (\bibinfo {year} {2021})}\BibitemShut {NoStop}%
\bibitem [{\citenamefont {Falk}(2018)}]{falk_wdm}%
  \BibitemOpen
  \bibfield  {author} {\bibinfo {author} {\bibfnamefont {K.}~\bibnamefont
  {Falk}},\ }\bibfield  {title} {\enquote {\bibinfo {title} {Experimental
  methods for warm dense matter research},}\ }\href
  {https://www.cambridge.org/core/journals/high-power-laser-science-and-engineering/article/experimental-methods-for-warm-dense-matter-research/7205AE1029BEA0061044F84875F1CEDB}
  {\bibfield  {journal} {\bibinfo  {journal} {High Power Laser Sci. Eng}\
  }\textbf {\bibinfo {volume} {6}},\ \bibinfo {pages} {e59} (\bibinfo {year}
  {2018})}\BibitemShut {NoStop}%
\bibitem [{\citenamefont {Dornheim}\ \emph
  {et~al.}(2018{\natexlab{a}})\citenamefont {Dornheim}, \citenamefont {Groth},\
  and\ \citenamefont {Bonitz}}]{review}%
  \BibitemOpen
  \bibfield  {author} {\bibinfo {author} {\bibfnamefont {T.}~\bibnamefont
  {Dornheim}}, \bibinfo {author} {\bibfnamefont {S.}~\bibnamefont {Groth}}, \
  and\ \bibinfo {author} {\bibfnamefont {M.}~\bibnamefont {Bonitz}},\
  }\bibfield  {title} {\enquote {\bibinfo {title} {The uniform electron gas at
  warm dense matter conditions},}\ }\href
  {https://www.sciencedirect.com/science/article/abs/pii/S0370157318300516}
  {\bibfield  {journal} {\bibinfo  {journal} {Phys. Reports}\ }\textbf
  {\bibinfo {volume} {744}},\ \bibinfo {pages} {1--86} (\bibinfo {year}
  {2018}{\natexlab{a}})}\BibitemShut {NoStop}%
\bibitem [{\citenamefont {Giuliani}\ and\ \citenamefont
  {Vignale}(2008)}]{quantum_theory}%
  \BibitemOpen
  \bibfield  {author} {\bibinfo {author} {\bibfnamefont {G.}~\bibnamefont
  {Giuliani}}\ and\ \bibinfo {author} {\bibfnamefont {G.}~\bibnamefont
  {Vignale}},\ }\href@noop {} {\emph {\bibinfo {title} {Quantum Theory of the
  Electron Liquid}}}\ (\bibinfo  {publisher} {Cambridge University Press},\
  \bibinfo {address} {Cambridge},\ \bibinfo {year} {2008})\BibitemShut
  {NoStop}%
\bibitem [{\citenamefont {Ott}\ \emph {et~al.}(2018)\citenamefont {Ott},
  \citenamefont {Thomsen}, \citenamefont {Abraham}, \citenamefont {Dornheim},\
  and\ \citenamefont {Bonitz}}]{Ott2018}%
  \BibitemOpen
  \bibfield  {author} {\bibinfo {author} {\bibfnamefont {Torben}\ \bibnamefont
  {Ott}}, \bibinfo {author} {\bibfnamefont {Hauke}\ \bibnamefont {Thomsen}},
  \bibinfo {author} {\bibfnamefont {Jan~Willem}\ \bibnamefont {Abraham}},
  \bibinfo {author} {\bibfnamefont {Tobias}\ \bibnamefont {Dornheim}}, \ and\
  \bibinfo {author} {\bibfnamefont {Michael}\ \bibnamefont {Bonitz}},\
  }\bibfield  {title} {\enquote {\bibinfo {title} {Recent progress in the
  theory and simulation of strongly correlated plasmas: phase transitions,
  transport, quantum, and magnetic field effects},}\ }\href {\doibase
  10.1140/epjd/e2018-80385-7} {\bibfield  {journal} {\bibinfo  {journal} {The
  European Physical Journal D}\ }\textbf {\bibinfo {volume} {72}},\ \bibinfo
  {pages} {84} (\bibinfo {year} {2018})}\BibitemShut {NoStop}%
\bibitem [{\citenamefont {Ceperley}(1995)}]{cep}%
  \BibitemOpen
  \bibfield  {author} {\bibinfo {author} {\bibfnamefont {D.~M.}\ \bibnamefont
  {Ceperley}},\ }\bibfield  {title} {\enquote {\bibinfo {title} {Path integrals
  in the theory of condensed helium},}\ }\href
  {https://journals.aps.org/rmp/abstract/10.1103/RevModPhys.67.279} {\bibfield
  {journal} {\bibinfo  {journal} {Rev. Mod. Phys}\ }\textbf {\bibinfo {volume}
  {67}},\ \bibinfo {pages} {279} (\bibinfo {year} {1995})}\BibitemShut
  {NoStop}%
\bibitem [{\citenamefont {Ceperley}(1991)}]{Ce91}%
  \BibitemOpen
  \bibfield  {author} {\bibinfo {author} {\bibfnamefont {D.~M.}\ \bibnamefont
  {Ceperley}},\ }\bibfield  {title} {\enquote {\bibinfo {title} {{Fermion
  nodes}},}\ }\href@noop {} {\bibfield  {journal} {\bibinfo  {journal} {Journal
  of Statistical Physics}\ }\textbf {\bibinfo {volume} {63}},\ \bibinfo {pages}
  {1237--1267} (\bibinfo {year} {1991})}\BibitemShut {NoStop}%
\bibitem [{\citenamefont {Ceperley}(1996)}]{binder}%
  \BibitemOpen
  \bibfield  {author} {\bibinfo {author} {\bibfnamefont {D.~M.}\ \bibnamefont
  {Ceperley}},\ }\bibfield  {title} {\enquote {\bibinfo {title} {Path integral
  monte carlo methods for fermions},}\ }in\ \href@noop {} {\emph {\bibinfo
  {booktitle} {Monte Carlo and Molecular Dynamics of Condensed Matter}}},\
  \bibinfo {editor} {edited by\ \bibinfo {editor} {\bibfnamefont {G.~Ciccotti}\
  \bibnamefont {K.~Binder}}}\ (\bibinfo  {publisher} {Società Italiana di
  Fisica},\ \bibinfo {address} {Bologna, Italy},\ \bibinfo {year}
  {1996})\BibitemShut {NoStop}%
\bibitem [{\citenamefont {Dornheim}(2019)}]{dornheim_sign_problem}%
  \BibitemOpen
  \bibfield  {author} {\bibinfo {author} {\bibfnamefont {T.}~\bibnamefont
  {Dornheim}},\ }\bibfield  {title} {\enquote {\bibinfo {title} {Fermion sign
  problem in path integral {M}onte {C}arlo simulations: Quantum dots, ultracold
  atoms, and warm dense matter},}\ }\href
  {https://journals.aps.org/pre/abstract/10.1103/PhysRevE.100.023307}
  {\bibfield  {journal} {\bibinfo  {journal} {Phys. Rev. E}\ }\textbf {\bibinfo
  {volume} {100}},\ \bibinfo {pages} {023307} (\bibinfo {year}
  {2019})}\BibitemShut {NoStop}%
\bibitem [{\citenamefont {Troyer}\ and\ \citenamefont {Wiese}(2005)}]{troyer}%
  \BibitemOpen
  \bibfield  {author} {\bibinfo {author} {\bibfnamefont {M.}~\bibnamefont
  {Troyer}}\ and\ \bibinfo {author} {\bibfnamefont {U.~J.}\ \bibnamefont
  {Wiese}},\ }\bibfield  {title} {\enquote {\bibinfo {title} {Computational
  complexity and fundamental limitations to fermionic quantum {M}onte {C}arlo
  simulations},}\ }\href
  {http://link.aps.org/doi/10.1103/PhysRevLett.94.170201} {\bibfield  {journal}
  {\bibinfo  {journal} {Phys. Rev. Lett}\ }\textbf {\bibinfo {volume} {94}},\
  \bibinfo {pages} {170201} (\bibinfo {year} {2005})}\BibitemShut {NoStop}%
\bibitem [{\citenamefont {Pierleoni}\ \emph {et~al.}(1994)\citenamefont
  {Pierleoni}, \citenamefont {Ceperley}, \citenamefont {Bernu},\ and\
  \citenamefont {Magro}}]{PC94}%
  \BibitemOpen
  \bibfield  {author} {\bibinfo {author} {\bibfnamefont {C.}~\bibnamefont
  {Pierleoni}}, \bibinfo {author} {\bibfnamefont {D.~M.}\ \bibnamefont
  {Ceperley}}, \bibinfo {author} {\bibfnamefont {B.}~\bibnamefont {Bernu}}, \
  and\ \bibinfo {author} {\bibfnamefont {W.~R.}\ \bibnamefont {Magro}},\
  }\bibfield  {title} {\enquote {\bibinfo {title} {Equation of state of the
  hydrogen plasma by path integral monte carlo simulation},}\ }\href@noop {}
  {\bibfield  {journal} {\bibinfo  {journal} {Phys. Rev. Lett.}\ }\textbf
  {\bibinfo {volume} {73}},\ \bibinfo {pages} {2145--2149} (\bibinfo {year}
  {1994})}\BibitemShut {NoStop}%
\bibitem [{\citenamefont {Magro}\ \emph {et~al.}(1996)\citenamefont {Magro},
  \citenamefont {Ceperley}, \citenamefont {Pierleoni},\ and\ \citenamefont
  {Bernu}}]{Ma96}%
  \BibitemOpen
  \bibfield  {author} {\bibinfo {author} {\bibfnamefont {W.~R.}\ \bibnamefont
  {Magro}}, \bibinfo {author} {\bibfnamefont {D.~M.}\ \bibnamefont {Ceperley}},
  \bibinfo {author} {\bibfnamefont {C.}~\bibnamefont {Pierleoni}}, \ and\
  \bibinfo {author} {\bibfnamefont {B.}~\bibnamefont {Bernu}},\ }\bibfield
  {title} {\enquote {\bibinfo {title} {Molecular dissociation in hot, dense
  hydrogen},}\ }\href {\doibase 10.1103/PhysRevLett.76.1240} {\bibfield
  {journal} {\bibinfo  {journal} {Phys. Rev. Lett.}\ }\textbf {\bibinfo
  {volume} {76}},\ \bibinfo {pages} {1240--1243} (\bibinfo {year}
  {1996})}\BibitemShut {NoStop}%
\bibitem [{\citenamefont {Militzer}\ and\ \citenamefont
  {Ceperley}(2001)}]{MC01}%
  \BibitemOpen
  \bibfield  {author} {\bibinfo {author} {\bibfnamefont {B.}~\bibnamefont
  {Militzer}}\ and\ \bibinfo {author} {\bibfnamefont {D.~M.}\ \bibnamefont
  {Ceperley}},\ }\bibfield  {title} {\enquote {\bibinfo {title} {Path integral
  monte carlo simulation of the low-density hydrogen plasma},}\ }\href@noop {}
  {\bibfield  {journal} {\bibinfo  {journal} {Phys. Rev. E}\ }\textbf {\bibinfo
  {volume} {63}},\ \bibinfo {pages} {066404} (\bibinfo {year}
  {2001})}\BibitemShut {NoStop}%
\bibitem [{\citenamefont {Militzer}(2006)}]{Mi06}%
  \BibitemOpen
  \bibfield  {author} {\bibinfo {author} {\bibfnamefont {B.}~\bibnamefont
  {Militzer}},\ }\bibfield  {title} {\enquote {\bibinfo {title} {First
  principles calculations of shock compressed fluid helium},}\ }\href@noop {}
  {\bibfield  {journal} {\bibinfo  {journal} {Phys. Rev. Lett.}\ }\textbf
  {\bibinfo {volume} {97}},\ \bibinfo {pages} {175501} (\bibinfo {year}
  {2006})}\BibitemShut {NoStop}%
\bibitem [{\citenamefont {Driver}\ and\ \citenamefont
  {Militzer}(2012)}]{Driver2012}%
  \BibitemOpen
  \bibfield  {author} {\bibinfo {author} {\bibfnamefont {Kevin~P.}\
  \bibnamefont {Driver}}\ and\ \bibinfo {author} {\bibfnamefont {Burkhard}\
  \bibnamefont {Militzer}},\ }\bibfield  {title} {\enquote {\bibinfo {title}
  {{All-Electron Path Integral Monte Carlo Simulations of Warm Dense Matter:
  Application to Water and Carbon Plasmas}},}\ }\href@noop {} {\bibfield
  {journal} {\bibinfo  {journal} {Phys. Rev. Lett.}\ }\textbf {\bibinfo
  {volume} {108}},\ \bibinfo {pages} {115502} (\bibinfo {year}
  {2012})}\BibitemShut {NoStop}%
\bibitem [{\citenamefont {Driver}\ and\ \citenamefont
  {Militzer}(2015)}]{Driver2015}%
  \BibitemOpen
  \bibfield  {author} {\bibinfo {author} {\bibfnamefont {K.~P.}\ \bibnamefont
  {Driver}}\ and\ \bibinfo {author} {\bibfnamefont {B.}~\bibnamefont
  {Militzer}},\ }\bibfield  {title} {\enquote {\bibinfo {title}
  {{First-principles simulations and shock Hugoniot calculations of warm dense
  neon}},}\ }\href@noop {} {\bibfield  {journal} {\bibinfo  {journal} {Phys.
  Rev. B}\ }\textbf {\bibinfo {volume} {91}},\ \bibinfo {pages} {045103}
  (\bibinfo {year} {2015})}\BibitemShut {NoStop}%
\bibitem [{\citenamefont {Driver}\ \emph {et~al.}(2018)\citenamefont {Driver},
  \citenamefont {Soubiran},\ and\ \citenamefont {Militzer}}]{Driver2018}%
  \BibitemOpen
  \bibfield  {author} {\bibinfo {author} {\bibfnamefont {Kevin~P.}\
  \bibnamefont {Driver}}, \bibinfo {author} {\bibfnamefont {Fran{\c{c}}ois}\
  \bibnamefont {Soubiran}}, \ and\ \bibinfo {author} {\bibfnamefont {Burkhard}\
  \bibnamefont {Militzer}},\ }\bibfield  {title} {\enquote {\bibinfo {title}
  {{Path integral Monte Carlo simulations of warm dense aluminum}},}\
  }\href@noop {} {\bibfield  {journal} {\bibinfo  {journal} {Phys. Rev. E}\
  }\textbf {\bibinfo {volume} {97}},\ \bibinfo {pages} {063207} (\bibinfo
  {year} {2018})}\BibitemShut {NoStop}%
\bibitem [{\citenamefont {Militzer}\ and\ \citenamefont
  {Driver}(2015)}]{MilitzerDriver2015}%
  \BibitemOpen
  \bibfield  {author} {\bibinfo {author} {\bibfnamefont {Burkhard}\
  \bibnamefont {Militzer}}\ and\ \bibinfo {author} {\bibfnamefont {Kevin~P.}\
  \bibnamefont {Driver}},\ }\bibfield  {title} {\enquote {\bibinfo {title}
  {{Development of Path Integral Monte Carlo Simulations with Localized Nodal
  Surfaces for Second-Row Elements}},}\ }\href@noop {} {\bibfield  {journal}
  {\bibinfo  {journal} {Phys. Rev. Lett.}\ }\textbf {\bibinfo {volume} {115}},\
  \bibinfo {pages} {176403} (\bibinfo {year} {2015})}\BibitemShut {NoStop}%
\bibitem [{\citenamefont {B.~Militzer}\ \emph {et~al.}(2021)\citenamefont
  {B.~Militzer}, \citenamefont {Gonzalez-Cataldo}, \citenamefont {Zhang},
  \citenamefont {Driver},\ and\ \citenamefont {Soubiran}}]{FPEOS}%
  \BibitemOpen
  \bibfield  {author} {\bibinfo {author} {\bibfnamefont {B.}~\bibnamefont
  {B.~Militzer}}, \bibinfo {author} {\bibfnamefont {F.}~\bibnamefont
  {Gonzalez-Cataldo}}, \bibinfo {author} {\bibfnamefont {S.}~\bibnamefont
  {Zhang}}, \bibinfo {author} {\bibfnamefont {K~P.}\ \bibnamefont {Driver}}, \
  and\ \bibinfo {author} {\bibfnamefont {F.}~\bibnamefont {Soubiran}},\
  }\bibfield  {title} {\enquote {\bibinfo {title} {First-principles equation of
  state database for warm dense matter computation},}\ }\href@noop {}
  {\bibfield  {journal} {\bibinfo  {journal} {Phys. Rev. E}\ }\textbf {\bibinfo
  {volume} {103}},\ \bibinfo {pages} {013203} (\bibinfo {year}
  {2021})}\BibitemShut {NoStop}%
\bibitem [{\citenamefont {Kritcher}(2020)}]{Kritcher2020}%
  \BibitemOpen
  \bibfield  {author} {\bibinfo {author} {\bibfnamefont {A.~L. et~al}\
  \bibnamefont {Kritcher}},\ }\bibfield  {title} {\enquote {\bibinfo {title} {A
  measurement of the equation of state of carbon envelopes of white dwarfs},}\
  }\href {\doibase 10.1038/s41586-020-2535-y} {\bibfield  {journal} {\bibinfo
  {journal} {Nature}\ }\textbf {\bibinfo {volume} {584}},\ \bibinfo {pages}
  {51--54} (\bibinfo {year} {2020})}\BibitemShut {NoStop}%
\bibitem [{\citenamefont {Zhang}\ \emph {et~al.}(2017)\citenamefont {Zhang},
  \citenamefont {Driver}, \citenamefont {Soubiran},\ and\ \citenamefont
  {Militzer}}]{ZhangCH2017}%
  \BibitemOpen
  \bibfield  {author} {\bibinfo {author} {\bibfnamefont {Shuai}\ \bibnamefont
  {Zhang}}, \bibinfo {author} {\bibfnamefont {Kevin~P.}\ \bibnamefont
  {Driver}}, \bibinfo {author} {\bibfnamefont {Fran{\c{c}}ois}\ \bibnamefont
  {Soubiran}}, \ and\ \bibinfo {author} {\bibfnamefont {Burkhard}\ \bibnamefont
  {Militzer}},\ }\bibfield  {title} {\enquote {\bibinfo {title}
  {{First-principles equation of state and shock compression predictions of
  warm dense hydrocarbons}},}\ }\href@noop {} {\bibfield  {journal} {\bibinfo
  {journal} {Phys. Rev. E}\ }\textbf {\bibinfo {volume} {96}},\ \bibinfo
  {pages} {013204} (\bibinfo {year} {2017})}\BibitemShut {NoStop}%
\bibitem [{\citenamefont {Zhang}\ \emph {et~al.}(2018)\citenamefont {Zhang},
  \citenamefont {Militzer}, \citenamefont {Benedict}, \citenamefont {Soubiran},
  \citenamefont {Sterne},\ and\ \citenamefont {Driver}}]{ZhangCH2018}%
  \BibitemOpen
  \bibfield  {author} {\bibinfo {author} {\bibfnamefont {Shuai}\ \bibnamefont
  {Zhang}}, \bibinfo {author} {\bibfnamefont {Burkhard}\ \bibnamefont
  {Militzer}}, \bibinfo {author} {\bibfnamefont {Lorin~X.}\ \bibnamefont
  {Benedict}}, \bibinfo {author} {\bibfnamefont {Fran{\c{c}}ois}\ \bibnamefont
  {Soubiran}}, \bibinfo {author} {\bibfnamefont {Philip~A.}\ \bibnamefont
  {Sterne}}, \ and\ \bibinfo {author} {\bibfnamefont {Kevin~P.}\ \bibnamefont
  {Driver}},\ }\bibfield  {title} {\enquote {\bibinfo {title} {{Path integral
  Monte Carlo simulations of dense carbon-hydrogen plasmas}},}\ }\href@noop {}
  {\bibfield  {journal} {\bibinfo  {journal} {J. Chem. Phys.}\ }\textbf
  {\bibinfo {volume} {148}},\ \bibinfo {pages} {102318} (\bibinfo {year}
  {2018})}\BibitemShut {NoStop}%
\bibitem [{\citenamefont {Lee}\ \emph {et~al.}(2020)\citenamefont {Lee},
  \citenamefont {Morales},\ and\ \citenamefont {Malone}}]{lee2020phaseless}%
  \BibitemOpen
  \bibfield  {author} {\bibinfo {author} {\bibfnamefont {Joonho}\ \bibnamefont
  {Lee}}, \bibinfo {author} {\bibfnamefont {Miguel~A.}\ \bibnamefont
  {Morales}}, \ and\ \bibinfo {author} {\bibfnamefont {Fionn~D.}\ \bibnamefont
  {Malone}},\ }\href@noop {} {\enquote {\bibinfo {title} {A phaseless
  auxiliary-field quantum monte carlo perspective on the uniform electron gas
  at finite temperatures: Issues, observations, and benchmark study},}\ }
  (\bibinfo {year} {2020}),\ \Eprint {http://arxiv.org/abs/2012.12228}
  {arXiv:2012.12228 [physics.chem-ph]} \BibitemShut {NoStop}%
\bibitem [{\citenamefont {Dornheim}\ \emph
  {et~al.}(2017{\natexlab{a}})\citenamefont {Dornheim}, \citenamefont {Groth},
  \citenamefont {Malone}, \citenamefont {Schoof}, \citenamefont {Sjostrom},
  \citenamefont {Foulkes},\ and\ \citenamefont {Bonitz}}]{dornheim_POP}%
  \BibitemOpen
  \bibfield  {author} {\bibinfo {author} {\bibfnamefont {Tobias}\ \bibnamefont
  {Dornheim}}, \bibinfo {author} {\bibfnamefont {Simon}\ \bibnamefont {Groth}},
  \bibinfo {author} {\bibfnamefont {Fionn~D.}\ \bibnamefont {Malone}}, \bibinfo
  {author} {\bibfnamefont {Tim}\ \bibnamefont {Schoof}}, \bibinfo {author}
  {\bibfnamefont {Travis}\ \bibnamefont {Sjostrom}}, \bibinfo {author}
  {\bibfnamefont {W.~M.~C.}\ \bibnamefont {Foulkes}}, \ and\ \bibinfo {author}
  {\bibfnamefont {Michael}\ \bibnamefont {Bonitz}},\ }\bibfield  {title}
  {\enquote {\bibinfo {title} {Ab initio quantum monte carlo simulation of the
  warm dense electron gas},}\ }\href {\doibase 10.1063/1.4977920} {\bibfield
  {journal} {\bibinfo  {journal} {Physics of Plasmas}\ }\textbf {\bibinfo
  {volume} {24}},\ \bibinfo {pages} {056303} (\bibinfo {year}
  {2017}{\natexlab{a}})},\ \Eprint
  {http://arxiv.org/abs/https://doi.org/10.1063/1.4977920}
  {https://doi.org/10.1063/1.4977920} \BibitemShut {NoStop}%
\bibitem [{\citenamefont {Schoof}\ \emph {et~al.}(2015)\citenamefont {Schoof},
  \citenamefont {Groth}, \citenamefont {Vorberger},\ and\ \citenamefont
  {Bonitz}}]{Schoof_PRL_2015}%
  \BibitemOpen
  \bibfield  {author} {\bibinfo {author} {\bibfnamefont {T.}~\bibnamefont
  {Schoof}}, \bibinfo {author} {\bibfnamefont {S.}~\bibnamefont {Groth}},
  \bibinfo {author} {\bibfnamefont {J.}~\bibnamefont {Vorberger}}, \ and\
  \bibinfo {author} {\bibfnamefont {M.}~\bibnamefont {Bonitz}},\ }\bibfield
  {title} {\enquote {\bibinfo {title} {Ab initio thermodynamic results for the
  degenerate electron gas at finite temperature},}\ }\href {\doibase
  10.1103/PhysRevLett.115.130402} {\bibfield  {journal} {\bibinfo  {journal}
  {Phys. Rev. Lett.}\ }\textbf {\bibinfo {volume} {115}},\ \bibinfo {pages}
  {130402} (\bibinfo {year} {2015})}\BibitemShut {NoStop}%
\bibitem [{\citenamefont {Malone}\ \emph {et~al.}(2016)\citenamefont {Malone},
  \citenamefont {Blunt}, \citenamefont {Brown}, \citenamefont {Lee},
  \citenamefont {Spencer}, \citenamefont {Foulkes},\ and\ \citenamefont
  {Shepherd}}]{Malone_PRL_2016}%
  \BibitemOpen
  \bibfield  {author} {\bibinfo {author} {\bibfnamefont {Fionn~D.}\
  \bibnamefont {Malone}}, \bibinfo {author} {\bibfnamefont {N.~S.}\
  \bibnamefont {Blunt}}, \bibinfo {author} {\bibfnamefont {Ethan~W.}\
  \bibnamefont {Brown}}, \bibinfo {author} {\bibfnamefont {D.~K.~K.}\
  \bibnamefont {Lee}}, \bibinfo {author} {\bibfnamefont {J.~S.}\ \bibnamefont
  {Spencer}}, \bibinfo {author} {\bibfnamefont {W.~M.~C.}\ \bibnamefont
  {Foulkes}}, \ and\ \bibinfo {author} {\bibfnamefont {James~J.}\ \bibnamefont
  {Shepherd}},\ }\bibfield  {title} {\enquote {\bibinfo {title} {Accurate
  exchange-correlation energies for the warm dense electron gas},}\ }\href
  {\doibase 10.1103/PhysRevLett.117.115701} {\bibfield  {journal} {\bibinfo
  {journal} {Phys. Rev. Lett.}\ }\textbf {\bibinfo {volume} {117}},\ \bibinfo
  {pages} {115701} (\bibinfo {year} {2016})}\BibitemShut {NoStop}%
\bibitem [{\citenamefont {Fraser}\ \emph {et~al.}(1996)\citenamefont {Fraser},
  \citenamefont {Foulkes}, \citenamefont {Rajagopal}, \citenamefont {Needs},
  \citenamefont {Kenny},\ and\ \citenamefont
  {Williamson}}]{Fraser_Foulkes_PRB_1996}%
  \BibitemOpen
  \bibfield  {author} {\bibinfo {author} {\bibfnamefont {Louisa~M.}\
  \bibnamefont {Fraser}}, \bibinfo {author} {\bibfnamefont {W.~M.~C.}\
  \bibnamefont {Foulkes}}, \bibinfo {author} {\bibfnamefont {G.}~\bibnamefont
  {Rajagopal}}, \bibinfo {author} {\bibfnamefont {R.~J.}\ \bibnamefont
  {Needs}}, \bibinfo {author} {\bibfnamefont {S.~D.}\ \bibnamefont {Kenny}}, \
  and\ \bibinfo {author} {\bibfnamefont {A.~J.}\ \bibnamefont {Williamson}},\
  }\bibfield  {title} {\enquote {\bibinfo {title} {Finite-size effects and
  coulomb interactions in quantum monte carlo calculations for homogeneous
  systems with periodic boundary conditions},}\ }\href {\doibase
  10.1103/PhysRevB.53.1814} {\bibfield  {journal} {\bibinfo  {journal} {Phys.
  Rev. B}\ }\textbf {\bibinfo {volume} {53}},\ \bibinfo {pages} {1814--1832}
  (\bibinfo {year} {1996})}\BibitemShut {NoStop}%
\bibitem [{\citenamefont {Loos}\ and\ \citenamefont {Gill}(2016)}]{loos}%
  \BibitemOpen
  \bibfield  {author} {\bibinfo {author} {\bibfnamefont {P.-F.}\ \bibnamefont
  {Loos}}\ and\ \bibinfo {author} {\bibfnamefont {P.~M.~W.}\ \bibnamefont
  {Gill}},\ }\bibfield  {title} {\enquote {\bibinfo {title} {The uniform
  electron gas},}\ }\href
  {http://onlinelibrary.wiley.com/doi/10.1002/wcms.1257/abstract} {\bibfield
  {journal} {\bibinfo  {journal} {Comput. Mol. Sci}\ }\textbf {\bibinfo
  {volume} {6}},\ \bibinfo {pages} {410--429} (\bibinfo {year}
  {2016})}\BibitemShut {NoStop}%
\bibitem [{\citenamefont {Brown}\ \emph {et~al.}(2013)\citenamefont {Brown},
  \citenamefont {Clark}, \citenamefont {DuBois},\ and\ \citenamefont
  {Ceperley}}]{Brown_PRL_2013}%
  \BibitemOpen
  \bibfield  {author} {\bibinfo {author} {\bibfnamefont {Ethan~W.}\
  \bibnamefont {Brown}}, \bibinfo {author} {\bibfnamefont {Bryan~K.}\
  \bibnamefont {Clark}}, \bibinfo {author} {\bibfnamefont {Jonathan~L.}\
  \bibnamefont {DuBois}}, \ and\ \bibinfo {author} {\bibfnamefont {David~M.}\
  \bibnamefont {Ceperley}},\ }\bibfield  {title} {\enquote {\bibinfo {title}
  {Path-integral monte carlo simulation of the warm dense homogeneous electron
  gas},}\ }\href {\doibase 10.1103/PhysRevLett.110.146405} {\bibfield
  {journal} {\bibinfo  {journal} {Phys. Rev. Lett.}\ }\textbf {\bibinfo
  {volume} {110}},\ \bibinfo {pages} {146405} (\bibinfo {year}
  {2013})}\BibitemShut {NoStop}%
\bibitem [{\citenamefont {Malone}\ \emph {et~al.}(2015)\citenamefont {Malone},
  \citenamefont {Blunt}, \citenamefont {Shepherd}, \citenamefont {Lee},
  \citenamefont {Spencer},\ and\ \citenamefont {Foulkes}}]{Malone_JCP_2015}%
  \BibitemOpen
  \bibfield  {author} {\bibinfo {author} {\bibfnamefont {Fionn~D.}\
  \bibnamefont {Malone}}, \bibinfo {author} {\bibfnamefont {N.~S.}\
  \bibnamefont {Blunt}}, \bibinfo {author} {\bibfnamefont {James~J.}\
  \bibnamefont {Shepherd}}, \bibinfo {author} {\bibfnamefont {D.~K.~K.}\
  \bibnamefont {Lee}}, \bibinfo {author} {\bibfnamefont {J.~S.}\ \bibnamefont
  {Spencer}}, \ and\ \bibinfo {author} {\bibfnamefont {W.~M.~C.}\ \bibnamefont
  {Foulkes}},\ }\bibfield  {title} {\enquote {\bibinfo {title} {Interaction
  picture density matrix quantum monte carlo},}\ }\href {\doibase
  10.1063/1.4927434} {\bibfield  {journal} {\bibinfo  {journal} {The Journal of
  Chemical Physics}\ }\textbf {\bibinfo {volume} {143}},\ \bibinfo {pages}
  {044116} (\bibinfo {year} {2015})},\ \Eprint
  {http://arxiv.org/abs/https://doi.org/10.1063/1.4927434}
  {https://doi.org/10.1063/1.4927434} \BibitemShut {NoStop}%
\bibitem [{\citenamefont {Dornheim}\ \emph
  {et~al.}(2020{\natexlab{a}})\citenamefont {Dornheim}, \citenamefont
  {Vorberger},\ and\ \citenamefont {Bonitz}}]{Dornheim_PRL_2020}%
  \BibitemOpen
  \bibfield  {author} {\bibinfo {author} {\bibfnamefont {Tobias}\ \bibnamefont
  {Dornheim}}, \bibinfo {author} {\bibfnamefont {Jan}\ \bibnamefont
  {Vorberger}}, \ and\ \bibinfo {author} {\bibfnamefont {Michael}\ \bibnamefont
  {Bonitz}},\ }\bibfield  {title} {\enquote {\bibinfo {title} {Nonlinear
  electronic density response in warm dense matter},}\ }\href {\doibase
  10.1103/PhysRevLett.125.085001} {\bibfield  {journal} {\bibinfo  {journal}
  {Phys. Rev. Lett.}\ }\textbf {\bibinfo {volume} {125}},\ \bibinfo {pages}
  {085001} (\bibinfo {year} {2020}{\natexlab{a}})}\BibitemShut {NoStop}%
\bibitem [{\citenamefont {Dornheim}\ \emph
  {et~al.}(2016{\natexlab{a}})\citenamefont {Dornheim}, \citenamefont {Groth},
  \citenamefont {Schoof}, \citenamefont {Hann},\ and\ \citenamefont
  {Bonitz}}]{dornheim_prb_2016}%
  \BibitemOpen
  \bibfield  {author} {\bibinfo {author} {\bibfnamefont {T.}~\bibnamefont
  {Dornheim}}, \bibinfo {author} {\bibfnamefont {S.}~\bibnamefont {Groth}},
  \bibinfo {author} {\bibfnamefont {T.}~\bibnamefont {Schoof}}, \bibinfo
  {author} {\bibfnamefont {C.}~\bibnamefont {Hann}}, \ and\ \bibinfo {author}
  {\bibfnamefont {M.}~\bibnamefont {Bonitz}},\ }\bibfield  {title} {\enquote
  {\bibinfo {title} {Ab initio quantum monte carlo simulations of the uniform
  electron gas without fixed nodes: The unpolarized case},}\ }\href {\doibase
  10.1103/PhysRevB.93.205134} {\bibfield  {journal} {\bibinfo  {journal} {Phys.
  Rev. B}\ }\textbf {\bibinfo {volume} {93}},\ \bibinfo {pages} {205134}
  (\bibinfo {year} {2016}{\natexlab{a}})}\BibitemShut {NoStop}%
\bibitem [{\citenamefont {Dornheim}\ \emph
  {et~al.}(2016{\natexlab{b}})\citenamefont {Dornheim}, \citenamefont {Groth},
  \citenamefont {Sjostrom}, \citenamefont {Malone}, \citenamefont {Foulkes},\
  and\ \citenamefont {Bonitz}}]{dornheim_prl}%
  \BibitemOpen
  \bibfield  {author} {\bibinfo {author} {\bibfnamefont {T.}~\bibnamefont
  {Dornheim}}, \bibinfo {author} {\bibfnamefont {S.}~\bibnamefont {Groth}},
  \bibinfo {author} {\bibfnamefont {T.}~\bibnamefont {Sjostrom}}, \bibinfo
  {author} {\bibfnamefont {F.~D.}\ \bibnamefont {Malone}}, \bibinfo {author}
  {\bibfnamefont {W.~M.~C.}\ \bibnamefont {Foulkes}}, \ and\ \bibinfo {author}
  {\bibfnamefont {M.}~\bibnamefont {Bonitz}},\ }\bibfield  {title} {\enquote
  {\bibinfo {title} {Ab initio quantum {M}onte {C}arlo simulation of the warm
  dense electron gas in the thermodynamic limit},}\ }\href
  {http://link.aps.org/doi/10.1103/PhysRevLett.117.156403} {\bibfield
  {journal} {\bibinfo  {journal} {Phys. Rev. Lett.}\ }\textbf {\bibinfo
  {volume} {117}},\ \bibinfo {pages} {156403} (\bibinfo {year}
  {2016}{\natexlab{b}})}\BibitemShut {NoStop}%
\bibitem [{\citenamefont {Groth}\ \emph {et~al.}(2019)\citenamefont {Groth},
  \citenamefont {Dornheim},\ and\ \citenamefont
  {Vorberger}}]{dynamic_folgepaper}%
  \BibitemOpen
  \bibfield  {author} {\bibinfo {author} {\bibfnamefont {S.}~\bibnamefont
  {Groth}}, \bibinfo {author} {\bibfnamefont {T.}~\bibnamefont {Dornheim}}, \
  and\ \bibinfo {author} {\bibfnamefont {J.}~\bibnamefont {Vorberger}},\
  }\bibfield  {title} {\enquote {\bibinfo {title} {Ab initio path integral
  {M}onte {C}arlo approach to the static and dynamic density response of the
  uniform electron gas},}\ }\href
  {https://link.aps.org/doi/10.1103/PhysRevB.99.235122} {\bibfield  {journal}
  {\bibinfo  {journal} {Phys. Rev. B}\ }\textbf {\bibinfo {volume} {99}},\
  \bibinfo {pages} {235122} (\bibinfo {year} {2019})}\BibitemShut {NoStop}%
\bibitem [{\citenamefont {Dornheim}\ \emph
  {et~al.}(2018{\natexlab{b}})\citenamefont {Dornheim}, \citenamefont {Groth},
  \citenamefont {Vorberger},\ and\ \citenamefont {Bonitz}}]{dornheim_dynamic}%
  \BibitemOpen
  \bibfield  {author} {\bibinfo {author} {\bibfnamefont {T.}~\bibnamefont
  {Dornheim}}, \bibinfo {author} {\bibfnamefont {S.}~\bibnamefont {Groth}},
  \bibinfo {author} {\bibfnamefont {J.}~\bibnamefont {Vorberger}}, \ and\
  \bibinfo {author} {\bibfnamefont {M.}~\bibnamefont {Bonitz}},\ }\bibfield
  {title} {\enquote {\bibinfo {title} {Ab initio path integral {M}onte {C}arlo
  results for the dynamic structure factor of correlated electrons: From the
  electron liquid to warm dense matter},}\ }\href
  {https://journals.aps.org/prl/abstract/10.1103/PhysRevLett.121.255001}
  {\bibfield  {journal} {\bibinfo  {journal} {Phys. Rev. Lett.}\ }\textbf
  {\bibinfo {volume} {121}},\ \bibinfo {pages} {255001} (\bibinfo {year}
  {2018}{\natexlab{b}})}\BibitemShut {NoStop}%
\bibitem [{\citenamefont {Groth}\ \emph
  {et~al.}(2017{\natexlab{a}})\citenamefont {Groth}, \citenamefont {Dornheim},\
  and\ \citenamefont {Bonitz}}]{groth_jcp}%
  \BibitemOpen
  \bibfield  {author} {\bibinfo {author} {\bibfnamefont {S.}~\bibnamefont
  {Groth}}, \bibinfo {author} {\bibfnamefont {T.}~\bibnamefont {Dornheim}}, \
  and\ \bibinfo {author} {\bibfnamefont {M.}~\bibnamefont {Bonitz}},\
  }\bibfield  {title} {\enquote {\bibinfo {title} {Configuration path integral
  {M}onte {C}arlo approach to the static density response of the warm dense
  electron gas},}\ }\href {https://aip.scitation.org/doi/abs/10.1063/1.4999907}
  {\bibfield  {journal} {\bibinfo  {journal} {J. Chem. Phys}\ }\textbf
  {\bibinfo {volume} {147}},\ \bibinfo {pages} {164108} (\bibinfo {year}
  {2017}{\natexlab{a}})}\BibitemShut {NoStop}%
\bibitem [{\citenamefont {Dornheim}\ \emph
  {et~al.}(2017{\natexlab{b}})\citenamefont {Dornheim}, \citenamefont {Groth},
  \citenamefont {Vorberger},\ and\ \citenamefont {Bonitz}}]{dornheim_pre}%
  \BibitemOpen
  \bibfield  {author} {\bibinfo {author} {\bibfnamefont {T.}~\bibnamefont
  {Dornheim}}, \bibinfo {author} {\bibfnamefont {S.}~\bibnamefont {Groth}},
  \bibinfo {author} {\bibfnamefont {J.}~\bibnamefont {Vorberger}}, \ and\
  \bibinfo {author} {\bibfnamefont {M.}~\bibnamefont {Bonitz}},\ }\bibfield
  {title} {\enquote {\bibinfo {title} {Permutation blocking path integral
  {M}onte {C}arlo approach to the static density response of the warm dense
  electron gas},}\ }\href
  {https://journals.aps.org/pre/abstract/10.1103/PhysRevE.96.023203} {\bibfield
   {journal} {\bibinfo  {journal} {Phys. Rev. E}\ }\textbf {\bibinfo {volume}
  {96}},\ \bibinfo {pages} {023203} (\bibinfo {year}
  {2017}{\natexlab{b}})}\BibitemShut {NoStop}%
\bibitem [{\citenamefont {Groth}\ \emph
  {et~al.}(2017{\natexlab{b}})\citenamefont {Groth}, \citenamefont {Dornheim},
  \citenamefont {Sjostrom}, \citenamefont {Malone}, \citenamefont {Foulkes},\
  and\ \citenamefont {Bonitz}}]{groth_prl}%
  \BibitemOpen
  \bibfield  {author} {\bibinfo {author} {\bibfnamefont {S.}~\bibnamefont
  {Groth}}, \bibinfo {author} {\bibfnamefont {T.}~\bibnamefont {Dornheim}},
  \bibinfo {author} {\bibfnamefont {T.}~\bibnamefont {Sjostrom}}, \bibinfo
  {author} {\bibfnamefont {F.~D.}\ \bibnamefont {Malone}}, \bibinfo {author}
  {\bibfnamefont {W.~M.~C.}\ \bibnamefont {Foulkes}}, \ and\ \bibinfo {author}
  {\bibfnamefont {M.}~\bibnamefont {Bonitz}},\ }\bibfield  {title} {\enquote
  {\bibinfo {title} {Ab initio exchange--correlation free energy of the uniform
  electron gas at warm dense matter conditions},}\ }\href
  {https://journals.aps.org/prl/abstract/10.1103/PhysRevLett.119.135001}
  {\bibfield  {journal} {\bibinfo  {journal} {Phys. Rev. Lett.}\ }\textbf
  {\bibinfo {volume} {119}},\ \bibinfo {pages} {135001} (\bibinfo {year}
  {2017}{\natexlab{b}})}\BibitemShut {NoStop}%
\bibitem [{\citenamefont {Karasiev}\ \emph {et~al.}(2014)\citenamefont
  {Karasiev}, \citenamefont {Sjostrom}, \citenamefont {Dufty},\ and\
  \citenamefont {Trickey}}]{ksdt}%
  \BibitemOpen
  \bibfield  {author} {\bibinfo {author} {\bibfnamefont {Valentin~V.}\
  \bibnamefont {Karasiev}}, \bibinfo {author} {\bibfnamefont {Travis}\
  \bibnamefont {Sjostrom}}, \bibinfo {author} {\bibfnamefont {James}\
  \bibnamefont {Dufty}}, \ and\ \bibinfo {author} {\bibfnamefont {S.~B.}\
  \bibnamefont {Trickey}},\ }\bibfield  {title} {\enquote {\bibinfo {title}
  {Accurate homogeneous electron gas exchange-correlation free energy for local
  spin-density calculations},}\ }\href {\doibase
  10.1103/PhysRevLett.112.076403} {\bibfield  {journal} {\bibinfo  {journal}
  {Phys. Rev. Lett.}\ }\textbf {\bibinfo {volume} {112}},\ \bibinfo {pages}
  {076403} (\bibinfo {year} {2014})}\BibitemShut {NoStop}%
\bibitem [{\citenamefont {Mermin}(1965)}]{Mermin_DFT_1965}%
  \BibitemOpen
  \bibfield  {author} {\bibinfo {author} {\bibfnamefont {N.~David}\
  \bibnamefont {Mermin}},\ }\bibfield  {title} {\enquote {\bibinfo {title}
  {Thermal properties of the inhomogeneous electron gas},}\ }\href {\doibase
  10.1103/PhysRev.137.A1441} {\bibfield  {journal} {\bibinfo  {journal} {Phys.
  Rev.}\ }\textbf {\bibinfo {volume} {137}},\ \bibinfo {pages} {A1441--A1443}
  (\bibinfo {year} {1965})}\BibitemShut {NoStop}%
\bibitem [{\citenamefont {Sjostrom}\ and\ \citenamefont
  {Daligault}(2014)}]{Sjostrom_PRB_2014}%
  \BibitemOpen
  \bibfield  {author} {\bibinfo {author} {\bibfnamefont {Travis}\ \bibnamefont
  {Sjostrom}}\ and\ \bibinfo {author} {\bibfnamefont {J\'er\^ome}\ \bibnamefont
  {Daligault}},\ }\bibfield  {title} {\enquote {\bibinfo {title} {Gradient
  corrections to the exchange-correlation free energy},}\ }\href {\doibase
  10.1103/PhysRevB.90.155109} {\bibfield  {journal} {\bibinfo  {journal} {Phys.
  Rev. B}\ }\textbf {\bibinfo {volume} {90}},\ \bibinfo {pages} {155109}
  (\bibinfo {year} {2014})}\BibitemShut {NoStop}%
\bibitem [{\citenamefont {Ramakrishna}\ \emph {et~al.}(2020)\citenamefont
  {Ramakrishna}, \citenamefont {Dornheim},\ and\ \citenamefont
  {Vorberger}}]{kushal}%
  \BibitemOpen
  \bibfield  {author} {\bibinfo {author} {\bibfnamefont {Kushal}\ \bibnamefont
  {Ramakrishna}}, \bibinfo {author} {\bibfnamefont {Tobias}\ \bibnamefont
  {Dornheim}}, \ and\ \bibinfo {author} {\bibfnamefont {Jan}\ \bibnamefont
  {Vorberger}},\ }\bibfield  {title} {\enquote {\bibinfo {title} {Influence of
  finite temperature exchange-correlation effects in hydrogen},}\ }\href
  {\doibase 10.1103/PhysRevB.101.195129} {\bibfield  {journal} {\bibinfo
  {journal} {Phys. Rev. B}\ }\textbf {\bibinfo {volume} {101}},\ \bibinfo
  {pages} {195129} (\bibinfo {year} {2020})}\BibitemShut {NoStop}%
\bibitem [{\citenamefont {Karasiev}\ \emph {et~al.}(2016)\citenamefont
  {Karasiev}, \citenamefont {Calderin},\ and\ \citenamefont
  {Trickey}}]{karasiev_importance}%
  \BibitemOpen
  \bibfield  {author} {\bibinfo {author} {\bibfnamefont {V.~V.}\ \bibnamefont
  {Karasiev}}, \bibinfo {author} {\bibfnamefont {L.}~\bibnamefont {Calderin}},
  \ and\ \bibinfo {author} {\bibfnamefont {S.~B.}\ \bibnamefont {Trickey}},\
  }\bibfield  {title} {\enquote {\bibinfo {title} {Importance of
  finite-temperature exchange correlation for warm dense matter
  calculations},}\ }\href
  {https://journals.aps.org/pre/abstract/10.1103/PhysRevE.93.063207} {\bibfield
   {journal} {\bibinfo  {journal} {Phys. Rev. E}\ }\textbf {\bibinfo {volume}
  {93}},\ \bibinfo {pages} {063207} (\bibinfo {year} {2016})}\BibitemShut
  {NoStop}%
\bibitem [{\citenamefont {Dornheim}\ \emph
  {et~al.}(2019{\natexlab{a}})\citenamefont {Dornheim}, \citenamefont
  {Vorberger}, \citenamefont {Groth}, \citenamefont {Hoffmann}, \citenamefont
  {Moldabekov},\ and\ \citenamefont {Bonitz}}]{dornheim_ML}%
  \BibitemOpen
  \bibfield  {author} {\bibinfo {author} {\bibfnamefont {T.}~\bibnamefont
  {Dornheim}}, \bibinfo {author} {\bibfnamefont {J.}~\bibnamefont {Vorberger}},
  \bibinfo {author} {\bibfnamefont {S.}~\bibnamefont {Groth}}, \bibinfo
  {author} {\bibfnamefont {N.}~\bibnamefont {Hoffmann}}, \bibinfo {author}
  {\bibfnamefont {Zh.A.}\ \bibnamefont {Moldabekov}}, \ and\ \bibinfo {author}
  {\bibfnamefont {M.}~\bibnamefont {Bonitz}},\ }\bibfield  {title} {\enquote
  {\bibinfo {title} {The static local field correction of the warm dense
  electron gas: An ab initio path integral {M}onte {C}arlo study and machine
  learning representation},}\ }\href
  {https://aip.scitation.org/doi/full/10.1063/1.5123013} {\bibfield  {journal}
  {\bibinfo  {journal} {J. Chem. Phys}\ }\textbf {\bibinfo {volume} {151}},\
  \bibinfo {pages} {194104} (\bibinfo {year} {2019}{\natexlab{a}})}\BibitemShut
  {NoStop}%
\bibitem [{\citenamefont {Dornheim}\ \emph
  {et~al.}(2020{\natexlab{b}})\citenamefont {Dornheim}, \citenamefont
  {Sjostrom}, \citenamefont {Tanaka},\ and\ \citenamefont
  {Vorberger}}]{dornheim_electron_liquid}%
  \BibitemOpen
  \bibfield  {author} {\bibinfo {author} {\bibfnamefont {Tobias}\ \bibnamefont
  {Dornheim}}, \bibinfo {author} {\bibfnamefont {Travis}\ \bibnamefont
  {Sjostrom}}, \bibinfo {author} {\bibfnamefont {Shigenori}\ \bibnamefont
  {Tanaka}}, \ and\ \bibinfo {author} {\bibfnamefont {Jan}\ \bibnamefont
  {Vorberger}},\ }\bibfield  {title} {\enquote {\bibinfo {title} {Strongly
  coupled electron liquid: Ab initio path integral monte carlo simulations and
  dielectric theories},}\ }\href {\doibase 10.1103/PhysRevB.101.045129}
  {\bibfield  {journal} {\bibinfo  {journal} {Phys. Rev. B}\ }\textbf {\bibinfo
  {volume} {101}},\ \bibinfo {pages} {045129} (\bibinfo {year}
  {2020}{\natexlab{b}})}\BibitemShut {NoStop}%
\bibitem [{\citenamefont {Dornheim}\ \emph
  {et~al.}(2020{\natexlab{c}})\citenamefont {Dornheim}, \citenamefont
  {Moldabekov}, \citenamefont {Vorberger},\ and\ \citenamefont
  {Groth}}]{dornheim_HEDP}%
  \BibitemOpen
  \bibfield  {author} {\bibinfo {author} {\bibfnamefont {Tobias}\ \bibnamefont
  {Dornheim}}, \bibinfo {author} {\bibfnamefont {Zhandos~A}\ \bibnamefont
  {Moldabekov}}, \bibinfo {author} {\bibfnamefont {Jan}\ \bibnamefont
  {Vorberger}}, \ and\ \bibinfo {author} {\bibfnamefont {Simon}\ \bibnamefont
  {Groth}},\ }\bibfield  {title} {\enquote {\bibinfo {title} {Ab initio path
  integral monte carlo simulation of the uniform electron gas in the high
  energy density regime},}\ }\href {\doibase 10.1088/1361-6587/ab8bb4}
  {\bibfield  {journal} {\bibinfo  {journal} {Plasma Physics and Controlled
  Fusion}\ }\textbf {\bibinfo {volume} {62}},\ \bibinfo {pages} {075003}
  (\bibinfo {year} {2020}{\natexlab{c}})}\BibitemShut {NoStop}%
\bibitem [{\citenamefont {Dornheim}\ \emph
  {et~al.}(2020{\natexlab{d}})\citenamefont {Dornheim}, \citenamefont {Cangi},
  \citenamefont {Ramakrishna}, \citenamefont {B\"ohme}, \citenamefont
  {Tanaka},\ and\ \citenamefont {Vorberger}}]{Dornheim_PRL_2020_ESA}%
  \BibitemOpen
  \bibfield  {author} {\bibinfo {author} {\bibfnamefont {Tobias}\ \bibnamefont
  {Dornheim}}, \bibinfo {author} {\bibfnamefont {Attila}\ \bibnamefont
  {Cangi}}, \bibinfo {author} {\bibfnamefont {Kushal}\ \bibnamefont
  {Ramakrishna}}, \bibinfo {author} {\bibfnamefont {Maximilian}\ \bibnamefont
  {B\"ohme}}, \bibinfo {author} {\bibfnamefont {Shigenori}\ \bibnamefont
  {Tanaka}}, \ and\ \bibinfo {author} {\bibfnamefont {Jan}\ \bibnamefont
  {Vorberger}},\ }\bibfield  {title} {\enquote {\bibinfo {title} {Effective
  static approximation: A fast and reliable tool for warm-dense matter
  theory},}\ }\href {\doibase 10.1103/PhysRevLett.125.235001} {\bibfield
  {journal} {\bibinfo  {journal} {Phys. Rev. Lett.}\ }\textbf {\bibinfo
  {volume} {125}},\ \bibinfo {pages} {235001} (\bibinfo {year}
  {2020}{\natexlab{d}})}\BibitemShut {NoStop}%
\bibitem [{\citenamefont {Hamann}\ \emph
  {et~al.}(2020{\natexlab{a}})\citenamefont {Hamann}, \citenamefont
  {Vorberger}, \citenamefont {Dornheim}, \citenamefont {Moldabekov},\ and\
  \citenamefont {Bonitz}}]{Hamann_CPP_2020}%
  \BibitemOpen
  \bibfield  {author} {\bibinfo {author} {\bibfnamefont {Paul}\ \bibnamefont
  {Hamann}}, \bibinfo {author} {\bibfnamefont {Jan}\ \bibnamefont {Vorberger}},
  \bibinfo {author} {\bibfnamefont {Tobias}\ \bibnamefont {Dornheim}}, \bibinfo
  {author} {\bibfnamefont {Zhandos~A.}\ \bibnamefont {Moldabekov}}, \ and\
  \bibinfo {author} {\bibfnamefont {Michael}\ \bibnamefont {Bonitz}},\
  }\bibfield  {title} {\enquote {\bibinfo {title} {Ab initio results for the
  plasmon dispersion and damping of the warm dense electron gas},}\ }\href
  {\doibase https://doi.org/10.1002/ctpp.202000147} {\bibfield  {journal}
  {\bibinfo  {journal} {Contributions to Plasma Physics}\ }\textbf {\bibinfo
  {volume} {60}},\ \bibinfo {pages} {e202000147} (\bibinfo {year}
  {2020}{\natexlab{a}})}\BibitemShut {NoStop}%
\bibitem [{\citenamefont {Hamann}\ \emph
  {et~al.}(2020{\natexlab{b}})\citenamefont {Hamann}, \citenamefont {Dornheim},
  \citenamefont {Vorberger}, \citenamefont {Moldabekov},\ and\ \citenamefont
  {Bonitz}}]{Hamann_PRB_2020}%
  \BibitemOpen
  \bibfield  {author} {\bibinfo {author} {\bibfnamefont {Paul}\ \bibnamefont
  {Hamann}}, \bibinfo {author} {\bibfnamefont {Tobias}\ \bibnamefont
  {Dornheim}}, \bibinfo {author} {\bibfnamefont {Jan}\ \bibnamefont
  {Vorberger}}, \bibinfo {author} {\bibfnamefont {Zhandos~A.}\ \bibnamefont
  {Moldabekov}}, \ and\ \bibinfo {author} {\bibfnamefont {Michael}\
  \bibnamefont {Bonitz}},\ }\bibfield  {title} {\enquote {\bibinfo {title}
  {Dynamic properties of the warm dense electron gas based on $ab initio$ path
  integral monte carlo simulations},}\ }\href {\doibase
  10.1103/PhysRevB.102.125150} {\bibfield  {journal} {\bibinfo  {journal}
  {Phys. Rev. B}\ }\textbf {\bibinfo {volume} {102}},\ \bibinfo {pages}
  {125150} (\bibinfo {year} {2020}{\natexlab{b}})}\BibitemShut {NoStop}%
\bibitem [{\citenamefont {Dornheim}\ and\ \citenamefont
  {Vorberger}(2020)}]{Dornheim_PRE_2020}%
  \BibitemOpen
  \bibfield  {author} {\bibinfo {author} {\bibfnamefont {Tobias}\ \bibnamefont
  {Dornheim}}\ and\ \bibinfo {author} {\bibfnamefont {Jan}\ \bibnamefont
  {Vorberger}},\ }\bibfield  {title} {\enquote {\bibinfo {title} {Finite-size
  effects in the reconstruction of dynamic properties from ab initio path
  integral monte carlo simulations},}\ }\href {\doibase
  10.1103/PhysRevE.102.063301} {\bibfield  {journal} {\bibinfo  {journal}
  {Phys. Rev. E}\ }\textbf {\bibinfo {volume} {102}},\ \bibinfo {pages}
  {063301} (\bibinfo {year} {2020})}\BibitemShut {NoStop}%
\bibitem [{\citenamefont {Militzer}\ and\ \citenamefont
  {Pollock}(2002)}]{Militzer_PRL_2002}%
  \BibitemOpen
  \bibfield  {author} {\bibinfo {author} {\bibfnamefont {Burkhard}\
  \bibnamefont {Militzer}}\ and\ \bibinfo {author} {\bibfnamefont {E.~L.}\
  \bibnamefont {Pollock}},\ }\bibfield  {title} {\enquote {\bibinfo {title}
  {Lowering of the kinetic energy in interacting quantum systems},}\ }\href
  {\doibase 10.1103/PhysRevLett.89.280401} {\bibfield  {journal} {\bibinfo
  {journal} {Phys. Rev. Lett.}\ }\textbf {\bibinfo {volume} {89}},\ \bibinfo
  {pages} {280401} (\bibinfo {year} {2002})}\BibitemShut {NoStop}%
\bibitem [{\citenamefont {Militzer}\ \emph {et~al.}(2019)\citenamefont
  {Militzer}, \citenamefont {Pollock},\ and\ \citenamefont
  {Ceperley}}]{Militzer_momentum_HEDP_2019}%
  \BibitemOpen
  \bibfield  {author} {\bibinfo {author} {\bibfnamefont {B.}~\bibnamefont
  {Militzer}}, \bibinfo {author} {\bibfnamefont {E.L.}\ \bibnamefont
  {Pollock}}, \ and\ \bibinfo {author} {\bibfnamefont {D.M.}\ \bibnamefont
  {Ceperley}},\ }\bibfield  {title} {\enquote {\bibinfo {title} {Path integral
  monte carlo calculation of the momentum distribution of the homogeneous
  electron gas at finite temperature},}\ }\href {\doibase
  https://doi.org/10.1016/j.hedp.2018.12.004} {\bibfield  {journal} {\bibinfo
  {journal} {High Energy Density Physics}\ }\textbf {\bibinfo {volume} {30}},\
  \bibinfo {pages} {13--20} (\bibinfo {year} {2019})}\BibitemShut {NoStop}%
\bibitem [{\citenamefont {Hunger}\ \emph {et~al.}(2021)\citenamefont {Hunger},
  \citenamefont {Schoof}, \citenamefont {Dornheim}, \citenamefont {Bonitz},\
  and\ \citenamefont {Filinov}}]{Hunger_PRE_2021}%
  \BibitemOpen
  \bibfield  {author} {\bibinfo {author} {\bibfnamefont {Kai}\ \bibnamefont
  {Hunger}}, \bibinfo {author} {\bibfnamefont {Tim}\ \bibnamefont {Schoof}},
  \bibinfo {author} {\bibfnamefont {Tobias}\ \bibnamefont {Dornheim}}, \bibinfo
  {author} {\bibfnamefont {Michael}\ \bibnamefont {Bonitz}}, \ and\ \bibinfo
  {author} {\bibfnamefont {Alexey}\ \bibnamefont {Filinov}},\ }\href@noop {}
  {\enquote {\bibinfo {title} {Momentum distribution function and short-range
  correlations of the warm dense electron gas -- ab initio quantum monte carlo
  results},}\ } (\bibinfo {year} {2021}),\ \Eprint
  {http://arxiv.org/abs/2101.00842} {arXiv:2101.00842 [physics.plasm-ph]}
  \BibitemShut {NoStop}%
\bibitem [{\citenamefont {Boninsegni}\ \emph {et~al.}(2006)\citenamefont
  {Boninsegni}, \citenamefont {Prokofev},\ and\ \citenamefont
  {Svistunov}}]{boninsegni1}%
  \BibitemOpen
  \bibfield  {author} {\bibinfo {author} {\bibfnamefont {M.}~\bibnamefont
  {Boninsegni}}, \bibinfo {author} {\bibfnamefont {N.~V.}\ \bibnamefont
  {Prokofev}}, \ and\ \bibinfo {author} {\bibfnamefont {B.~V.}\ \bibnamefont
  {Svistunov}},\ }\bibfield  {title} {\enquote {\bibinfo {title} {Worm
  algorithm and diagrammatic {M}onte {C}arlo: A new approach to
  continuous-space path integral {M}onte {C}arlo simulations},}\ }\href
  {https://journals.aps.org/pre/abstract/10.1103/PhysRevE.74.036701} {\bibfield
   {journal} {\bibinfo  {journal} {Phys. Rev. E}\ }\textbf {\bibinfo {volume}
  {74}},\ \bibinfo {pages} {036701} (\bibinfo {year} {2006})}\BibitemShut
  {NoStop}%
\bibitem [{\citenamefont {Metropolis}\ \emph {et~al.}(1953)\citenamefont
  {Metropolis}, \citenamefont {Rosenbluth}, \citenamefont {Rosenbluth},
  \citenamefont {Teller},\ and\ \citenamefont {Teller}}]{metropolis}%
  \BibitemOpen
  \bibfield  {author} {\bibinfo {author} {\bibfnamefont {Nicholas}\
  \bibnamefont {Metropolis}}, \bibinfo {author} {\bibfnamefont {Arianna~W.}\
  \bibnamefont {Rosenbluth}}, \bibinfo {author} {\bibfnamefont {Marshall~N.}\
  \bibnamefont {Rosenbluth}}, \bibinfo {author} {\bibfnamefont {Augusta~H.}\
  \bibnamefont {Teller}}, \ and\ \bibinfo {author} {\bibfnamefont {Edward}\
  \bibnamefont {Teller}},\ }\bibfield  {title} {\enquote {\bibinfo {title}
  {Equation of state calculations by fast computing machines},}\ }\href
  {\doibase 10.1063/1.1699114} {\bibfield  {journal} {\bibinfo  {journal} {The
  Journal of Chemical Physics}\ }\textbf {\bibinfo {volume} {21}},\ \bibinfo
  {pages} {1087--1092} (\bibinfo {year} {1953})},\ \Eprint
  {http://arxiv.org/abs/https://doi.org/10.1063/1.1699114}
  {https://doi.org/10.1063/1.1699114} \BibitemShut {NoStop}%
\bibitem [{\citenamefont {Zastrau}\ \emph {et~al.}(2014)\citenamefont
  {Zastrau}, \citenamefont {Sperling}, \citenamefont {Harmand}, \citenamefont
  {Becker}, \citenamefont {Bornath}, \citenamefont {Bredow}, \citenamefont
  {Dziarzhytski}, \citenamefont {Fennel}, \citenamefont {Fletcher},
  \citenamefont {F{"o}rster}, \citenamefont {G{"o}de}, \citenamefont {Gregori},
  \citenamefont {Hilbert}, \citenamefont {Hochhaus}, \citenamefont {Holst},
  \citenamefont {Laarmann}, \citenamefont {Lee}, \citenamefont {Ma},
  \citenamefont {Mithen}, \citenamefont {Mitzner}, \citenamefont {Murphy},
  \citenamefont {Nakatsutsumi}, \citenamefont {Neumayer}, \citenamefont
  {Przystawik}, \citenamefont {Roling}, \citenamefont {Schulz}, \citenamefont
  {Siemer}, \citenamefont {Skruszewicz}, \citenamefont {Tiggesb{"a}umker},
  \citenamefont {Toleikis}, \citenamefont {Tschentscher}, \citenamefont
  {White}, \citenamefont {W{"o}stmann}, \citenamefont {Zacharias},
  \citenamefont {D{"o}ppner}, \citenamefont {Glenzer},\ and\ \citenamefont
  {Redmer}}]{Zastrau}%
  \BibitemOpen
  \bibfield  {author} {\bibinfo {author} {\bibfnamefont {U.}~\bibnamefont
  {Zastrau}}, \bibinfo {author} {\bibfnamefont {P.}~\bibnamefont {Sperling}},
  \bibinfo {author} {\bibfnamefont {M.}~\bibnamefont {Harmand}}, \bibinfo
  {author} {\bibfnamefont {A.}~\bibnamefont {Becker}}, \bibinfo {author}
  {\bibfnamefont {T.}~\bibnamefont {Bornath}}, \bibinfo {author} {\bibfnamefont
  {R.}~\bibnamefont {Bredow}}, \bibinfo {author} {\bibfnamefont
  {S.}~\bibnamefont {Dziarzhytski}}, \bibinfo {author} {\bibfnamefont
  {T.}~\bibnamefont {Fennel}}, \bibinfo {author} {\bibfnamefont {L.~B.}\
  \bibnamefont {Fletcher}}, \bibinfo {author} {\bibfnamefont {E.}~\bibnamefont
  {F{"o}rster}}, \bibinfo {author} {\bibfnamefont {S.}~\bibnamefont {G{"o}de}},
  \bibinfo {author} {\bibfnamefont {G.}~\bibnamefont {Gregori}}, \bibinfo
  {author} {\bibfnamefont {V.}~\bibnamefont {Hilbert}}, \bibinfo {author}
  {\bibfnamefont {D.}~\bibnamefont {Hochhaus}}, \bibinfo {author}
  {\bibfnamefont {B.}~\bibnamefont {Holst}}, \bibinfo {author} {\bibfnamefont
  {T.}~\bibnamefont {Laarmann}}, \bibinfo {author} {\bibfnamefont {H.~J.}\
  \bibnamefont {Lee}}, \bibinfo {author} {\bibfnamefont {T.}~\bibnamefont
  {Ma}}, \bibinfo {author} {\bibfnamefont {J.~P.}\ \bibnamefont {Mithen}},
  \bibinfo {author} {\bibfnamefont {R.}~\bibnamefont {Mitzner}}, \bibinfo
  {author} {\bibfnamefont {C.~D.}\ \bibnamefont {Murphy}}, \bibinfo {author}
  {\bibfnamefont {M.}~\bibnamefont {Nakatsutsumi}}, \bibinfo {author}
  {\bibfnamefont {P.}~\bibnamefont {Neumayer}}, \bibinfo {author}
  {\bibfnamefont {A.}~\bibnamefont {Przystawik}}, \bibinfo {author}
  {\bibfnamefont {S.}~\bibnamefont {Roling}}, \bibinfo {author} {\bibfnamefont
  {M.}~\bibnamefont {Schulz}}, \bibinfo {author} {\bibfnamefont
  {B.}~\bibnamefont {Siemer}}, \bibinfo {author} {\bibfnamefont
  {S.}~\bibnamefont {Skruszewicz}}, \bibinfo {author} {\bibfnamefont
  {J.}~\bibnamefont {Tiggesb{"a}umker}}, \bibinfo {author} {\bibfnamefont
  {S.}~\bibnamefont {Toleikis}}, \bibinfo {author} {\bibfnamefont
  {T.}~\bibnamefont {Tschentscher}}, \bibinfo {author} {\bibfnamefont
  {T.}~\bibnamefont {White}}, \bibinfo {author} {\bibfnamefont
  {M.}~\bibnamefont {W{"o}stmann}}, \bibinfo {author} {\bibfnamefont
  {H.}~\bibnamefont {Zacharias}}, \bibinfo {author} {\bibfnamefont
  {T.}~\bibnamefont {D{"o}ppner}}, \bibinfo {author} {\bibfnamefont {S.~H.}\
  \bibnamefont {Glenzer}}, \ and\ \bibinfo {author} {\bibfnamefont
  {R.}~\bibnamefont {Redmer}},\ }\bibfield  {title} {\enquote {\bibinfo {title}
  {Resolving ultrafast heating of dense cryogenic hydrogen},}\ }\href
  {https://journals.aps.org/prl/abstract/10.1103/PhysRevLett.112.105002}
  {\bibfield  {journal} {\bibinfo  {journal} {Phys. Rev. Lett}\ }\textbf
  {\bibinfo {volume} {112}},\ \bibinfo {pages} {105002} (\bibinfo {year}
  {2014})}\BibitemShut {NoStop}%
\bibitem [{\citenamefont {Takada}(2016)}]{Takada_PRB_2016}%
  \BibitemOpen
  \bibfield  {author} {\bibinfo {author} {\bibfnamefont {Yasutami}\
  \bibnamefont {Takada}},\ }\bibfield  {title} {\enquote {\bibinfo {title}
  {Emergence of an excitonic collective mode in the dilute electron gas},}\
  }\href {\doibase 10.1103/PhysRevB.94.245106} {\bibfield  {journal} {\bibinfo
  {journal} {Phys. Rev. B}\ }\textbf {\bibinfo {volume} {94}},\ \bibinfo
  {pages} {245106} (\bibinfo {year} {2016})}\BibitemShut {NoStop}%
\bibitem [{\citenamefont {Takada}\ and\ \citenamefont
  {Yasuhara}(2002)}]{Takada_PRL_2002}%
  \BibitemOpen
  \bibfield  {author} {\bibinfo {author} {\bibfnamefont {Yasutami}\
  \bibnamefont {Takada}}\ and\ \bibinfo {author} {\bibfnamefont {Hiroshi}\
  \bibnamefont {Yasuhara}},\ }\bibfield  {title} {\enquote {\bibinfo {title}
  {Dynamical structure factor of the homogeneous electron liquid: Its accurate
  shape and the interpretation of experiments on aluminum},}\ }\href {\doibase
  10.1103/PhysRevLett.89.216402} {\bibfield  {journal} {\bibinfo  {journal}
  {Phys. Rev. Lett.}\ }\textbf {\bibinfo {volume} {89}},\ \bibinfo {pages}
  {216402} (\bibinfo {year} {2002})}\BibitemShut {NoStop}%
\bibitem [{\citenamefont {Huotari}\ \emph {et~al.}(2010)\citenamefont
  {Huotari}, \citenamefont {Soininen}, \citenamefont {Pylkk\"anen},
  \citenamefont {H\"am\"al\"ainen}, \citenamefont {Issolah}, \citenamefont
  {Titov}, \citenamefont {McMinis}, \citenamefont {Kim}, \citenamefont {Esler},
  \citenamefont {Ceperley}, \citenamefont {Holzmann},\ and\ \citenamefont
  {Olevano}}]{Huotari_PRL_2010}%
  \BibitemOpen
  \bibfield  {author} {\bibinfo {author} {\bibfnamefont {Simo}\ \bibnamefont
  {Huotari}}, \bibinfo {author} {\bibfnamefont {J.~Aleksi}\ \bibnamefont
  {Soininen}}, \bibinfo {author} {\bibfnamefont {Tuomas}\ \bibnamefont
  {Pylkk\"anen}}, \bibinfo {author} {\bibfnamefont {Keijo}\ \bibnamefont
  {H\"am\"al\"ainen}}, \bibinfo {author} {\bibfnamefont {Arezki}\ \bibnamefont
  {Issolah}}, \bibinfo {author} {\bibfnamefont {Andrey}\ \bibnamefont {Titov}},
  \bibinfo {author} {\bibfnamefont {Jeremy}\ \bibnamefont {McMinis}}, \bibinfo
  {author} {\bibfnamefont {Jeongnim}\ \bibnamefont {Kim}}, \bibinfo {author}
  {\bibfnamefont {Ken}\ \bibnamefont {Esler}}, \bibinfo {author} {\bibfnamefont
  {David~M.}\ \bibnamefont {Ceperley}}, \bibinfo {author} {\bibfnamefont
  {Markus}\ \bibnamefont {Holzmann}}, \ and\ \bibinfo {author} {\bibfnamefont
  {Valerio}\ \bibnamefont {Olevano}},\ }\bibfield  {title} {\enquote {\bibinfo
  {title} {Momentum distribution and renormalization factor in sodium and the
  electron gas},}\ }\href {\doibase 10.1103/PhysRevLett.105.086403} {\bibfield
  {journal} {\bibinfo  {journal} {Phys. Rev. Lett.}\ }\textbf {\bibinfo
  {volume} {105}},\ \bibinfo {pages} {086403} (\bibinfo {year}
  {2010})}\BibitemShut {NoStop}%
\bibitem [{\citenamefont {Holzmann}\ \emph {et~al.}(2016)\citenamefont
  {Holzmann}, \citenamefont {Clay}, \citenamefont {Morales}, \citenamefont
  {Tubman}, \citenamefont {Ceperley},\ and\ \citenamefont
  {Pierleoni}}]{Holzmann_PRB_FSC_2016}%
  \BibitemOpen
  \bibfield  {author} {\bibinfo {author} {\bibfnamefont {Markus}\ \bibnamefont
  {Holzmann}}, \bibinfo {author} {\bibfnamefont {Raymond~C.}\ \bibnamefont
  {Clay}}, \bibinfo {author} {\bibfnamefont {Miguel~A.}\ \bibnamefont
  {Morales}}, \bibinfo {author} {\bibfnamefont {Norm~M.}\ \bibnamefont
  {Tubman}}, \bibinfo {author} {\bibfnamefont {David~M.}\ \bibnamefont
  {Ceperley}}, \ and\ \bibinfo {author} {\bibfnamefont {Carlo}\ \bibnamefont
  {Pierleoni}},\ }\bibfield  {title} {\enquote {\bibinfo {title} {Theory of
  finite size effects for electronic quantum monte carlo calculations of
  liquids and solids},}\ }\href {\doibase 10.1103/PhysRevB.94.035126}
  {\bibfield  {journal} {\bibinfo  {journal} {Phys. Rev. B}\ }\textbf {\bibinfo
  {volume} {94}},\ \bibinfo {pages} {035126} (\bibinfo {year}
  {2016})}\BibitemShut {NoStop}%
\bibitem [{\citenamefont {Chiesa}\ \emph {et~al.}(2006)\citenamefont {Chiesa},
  \citenamefont {Ceperley}, \citenamefont {Martin},\ and\ \citenamefont
  {Holzmann}}]{Chiesa_PRL_2006}%
  \BibitemOpen
  \bibfield  {author} {\bibinfo {author} {\bibfnamefont {Simone}\ \bibnamefont
  {Chiesa}}, \bibinfo {author} {\bibfnamefont {David~M.}\ \bibnamefont
  {Ceperley}}, \bibinfo {author} {\bibfnamefont {Richard~M.}\ \bibnamefont
  {Martin}}, \ and\ \bibinfo {author} {\bibfnamefont {Markus}\ \bibnamefont
  {Holzmann}},\ }\bibfield  {title} {\enquote {\bibinfo {title} {Finite-size
  error in many-body simulations with long-range interactions},}\ }\href
  {\doibase 10.1103/PhysRevLett.97.076404} {\bibfield  {journal} {\bibinfo
  {journal} {Phys. Rev. Lett.}\ }\textbf {\bibinfo {volume} {97}},\ \bibinfo
  {pages} {076404} (\bibinfo {year} {2006})}\BibitemShut {NoStop}%
\bibitem [{\citenamefont {Dornheim}\ and\ \citenamefont
  {Vorberger}(2021)}]{dornheim2021overcoming}%
  \BibitemOpen
  \bibfield  {author} {\bibinfo {author} {\bibfnamefont {Tobias}\ \bibnamefont
  {Dornheim}}\ and\ \bibinfo {author} {\bibfnamefont {Jan}\ \bibnamefont
  {Vorberger}},\ }\href@noop {} {\enquote {\bibinfo {title} {Overcoming
  finite-size effects in electronic structure simulations at extreme
  conditions},}\ } (\bibinfo {year} {2021}),\ \Eprint
  {http://arxiv.org/abs/2101.11364} {arXiv:2101.11364 [cond-mat.stat-mech]}
  \BibitemShut {NoStop}%
\bibitem [{\citenamefont {Hofmann}\ \emph {et~al.}()\citenamefont {Hofmann},
  \citenamefont {Barth},\ and\ \citenamefont
  {Zwerger}}]{hofmann_short-distance_2013}%
  \BibitemOpen
  \bibfield  {author} {\bibinfo {author} {\bibfnamefont {Johannes}\
  \bibnamefont {Hofmann}}, \bibinfo {author} {\bibfnamefont {Marcus}\
  \bibnamefont {Barth}}, \ and\ \bibinfo {author} {\bibfnamefont {Wilhelm}\
  \bibnamefont {Zwerger}},\ }\bibfield  {title} {\enquote {\bibinfo {title}
  {{Short-distance properties of Coulomb systems}},}\ }\href {\doibase
  10.1103/PhysRevB.87.235125} {\ \textbf {\bibinfo {volume} {87}},\ \bibinfo
  {pages} {235125}}\BibitemShut {NoStop}%
\bibitem [{\citenamefont {Yasuhara}\ and\ \citenamefont
  {Kawazoe}(1976)}]{yasuhara_note_1976}%
  \BibitemOpen
  \bibfield  {author} {\bibinfo {author} {\bibfnamefont {H.}~\bibnamefont
  {Yasuhara}}\ and\ \bibinfo {author} {\bibfnamefont {Y.}~\bibnamefont
  {Kawazoe}},\ }\bibfield  {title} {\enquote {\bibinfo {title} {A note on the
  momentum distribution function for an electron gas},}\ }\href {\doibase
  10.1016/0378-4371(76)90060-1} {\bibfield  {journal} {\bibinfo  {journal}
  {Physica A: Statistical Mechanics and its Applications}\ }\textbf {\bibinfo
  {volume} {85}},\ \bibinfo {pages} {416--424} (\bibinfo {year}
  {1976})}\BibitemShut {NoStop}%
\bibitem [{\citenamefont {Sperling}\ \emph {et~al.}(2015)\citenamefont
  {Sperling}, \citenamefont {Gamboa}, \citenamefont {Lee}, \citenamefont
  {Chung}, \citenamefont {Galtier}, \citenamefont {Omarbakiyeva}, \citenamefont
  {Reinholz}, \citenamefont {R\"opke}, \citenamefont {Zastrau}, \citenamefont
  {Hastings}, \citenamefont {Fletcher},\ and\ \citenamefont
  {Glenzer}}]{Sperling_PRL_2015}%
  \BibitemOpen
  \bibfield  {author} {\bibinfo {author} {\bibfnamefont {P.}~\bibnamefont
  {Sperling}}, \bibinfo {author} {\bibfnamefont {E.~J.}\ \bibnamefont
  {Gamboa}}, \bibinfo {author} {\bibfnamefont {H.~J.}\ \bibnamefont {Lee}},
  \bibinfo {author} {\bibfnamefont {H.~K.}\ \bibnamefont {Chung}}, \bibinfo
  {author} {\bibfnamefont {E.}~\bibnamefont {Galtier}}, \bibinfo {author}
  {\bibfnamefont {Y.}~\bibnamefont {Omarbakiyeva}}, \bibinfo {author}
  {\bibfnamefont {H.}~\bibnamefont {Reinholz}}, \bibinfo {author}
  {\bibfnamefont {G.}~\bibnamefont {R\"opke}}, \bibinfo {author} {\bibfnamefont
  {U.}~\bibnamefont {Zastrau}}, \bibinfo {author} {\bibfnamefont
  {J.}~\bibnamefont {Hastings}}, \bibinfo {author} {\bibfnamefont {L.~B.}\
  \bibnamefont {Fletcher}}, \ and\ \bibinfo {author} {\bibfnamefont {S.~H.}\
  \bibnamefont {Glenzer}},\ }\bibfield  {title} {\enquote {\bibinfo {title}
  {Free-electron x-ray laser measurements of collisional-damped plasmons in
  isochorically heated warm dense matter},}\ }\href {\doibase
  10.1103/PhysRevLett.115.115001} {\bibfield  {journal} {\bibinfo  {journal}
  {Phys. Rev. Lett.}\ }\textbf {\bibinfo {volume} {115}},\ \bibinfo {pages}
  {115001} (\bibinfo {year} {2015})}\BibitemShut {NoStop}%
\bibitem [{\citenamefont {Shi}(2005)}]{Shi_PRB_2005}%
  \BibitemOpen
  \bibfield  {author} {\bibinfo {author} {\bibfnamefont {Yu}~\bibnamefont
  {Shi}},\ }\bibfield  {title} {\enquote {\bibinfo {title} {Superfluidity or
  supersolidity as a consequence of off-diagonal long-range order},}\ }\href
  {\doibase 10.1103/PhysRevB.72.014533} {\bibfield  {journal} {\bibinfo
  {journal} {Phys. Rev. B}\ }\textbf {\bibinfo {volume} {72}},\ \bibinfo
  {pages} {014533} (\bibinfo {year} {2005})}\BibitemShut {NoStop}%
\bibitem [{\citenamefont {Kraeft}\ \emph {et~al.}(2002)\citenamefont {Kraeft},
  \citenamefont {Schlanges}, \citenamefont {Vorberger},\ and\ \citenamefont
  {DeWitt}}]{Kraeft_PRE_2002}%
  \BibitemOpen
  \bibfield  {author} {\bibinfo {author} {\bibfnamefont {W.~D.}\ \bibnamefont
  {Kraeft}}, \bibinfo {author} {\bibfnamefont {M.}~\bibnamefont {Schlanges}},
  \bibinfo {author} {\bibfnamefont {J.}~\bibnamefont {Vorberger}}, \ and\
  \bibinfo {author} {\bibfnamefont {H.~E.}\ \bibnamefont {DeWitt}},\ }\bibfield
   {title} {\enquote {\bibinfo {title} {Kinetic and correlation energies and
  distribution functions of dense plasmas},}\ }\href {\doibase
  10.1103/PhysRevE.66.046405} {\bibfield  {journal} {\bibinfo  {journal} {Phys.
  Rev. E}\ }\textbf {\bibinfo {volume} {66}},\ \bibinfo {pages} {046405}
  (\bibinfo {year} {2002})}\BibitemShut {NoStop}%
\bibitem [{rep()}]{repo}%
  \BibitemOpen
  \href@noop {} {}\bibinfo {note} {A link to a repository containing all PIMC
  raw data will be made available upon publication.}\BibitemShut {Stop}%
\bibitem [{\citenamefont {Dornheim}\ \emph
  {et~al.}(2015{\natexlab{a}})\citenamefont {Dornheim}, \citenamefont {Groth},
  \citenamefont {Filinov},\ and\ \citenamefont {Bonitz}}]{Dornheim_NJP_2015}%
  \BibitemOpen
  \bibfield  {author} {\bibinfo {author} {\bibfnamefont {Tobias}\ \bibnamefont
  {Dornheim}}, \bibinfo {author} {\bibfnamefont {Simon}\ \bibnamefont {Groth}},
  \bibinfo {author} {\bibfnamefont {Alexey}\ \bibnamefont {Filinov}}, \ and\
  \bibinfo {author} {\bibfnamefont {Michael}\ \bibnamefont {Bonitz}},\
  }\bibfield  {title} {\enquote {\bibinfo {title} {Permutation blocking path
  integral monte carlo: a highly efficient approach to the simulation of
  strongly degenerate non-ideal fermions},}\ }\href {\doibase
  10.1088/1367-2630/17/7/073017} {\bibfield  {journal} {\bibinfo  {journal}
  {New Journal of Physics}\ }\textbf {\bibinfo {volume} {17}},\ \bibinfo
  {pages} {073017} (\bibinfo {year} {2015}{\natexlab{a}})}\BibitemShut
  {NoStop}%
\bibitem [{\citenamefont {Dornheim}\ \emph
  {et~al.}(2015{\natexlab{b}})\citenamefont {Dornheim}, \citenamefont {Schoof},
  \citenamefont {Groth}, \citenamefont {Filinov},\ and\ \citenamefont
  {Bonitz}}]{dornheim_jcp}%
  \BibitemOpen
  \bibfield  {author} {\bibinfo {author} {\bibfnamefont {Tobias}\ \bibnamefont
  {Dornheim}}, \bibinfo {author} {\bibfnamefont {Tim}\ \bibnamefont {Schoof}},
  \bibinfo {author} {\bibfnamefont {Simon}\ \bibnamefont {Groth}}, \bibinfo
  {author} {\bibfnamefont {Alexey}\ \bibnamefont {Filinov}}, \ and\ \bibinfo
  {author} {\bibfnamefont {Michael}\ \bibnamefont {Bonitz}},\ }\bibfield
  {title} {\enquote {\bibinfo {title} {Permutation blocking path integral monte
  carlo approach to the uniform electron gas at finite temperature},}\ }\href
  {\doibase 10.1063/1.4936145} {\bibfield  {journal} {\bibinfo  {journal} {The
  Journal of Chemical Physics}\ }\textbf {\bibinfo {volume} {143}},\ \bibinfo
  {pages} {204101} (\bibinfo {year} {2015}{\natexlab{b}})},\ \Eprint
  {http://arxiv.org/abs/https://doi.org/10.1063/1.4936145}
  {https://doi.org/10.1063/1.4936145} \BibitemShut {NoStop}%
\bibitem [{\citenamefont {Dornheim}\ \emph
  {et~al.}(2019{\natexlab{b}})\citenamefont {Dornheim}, \citenamefont {Groth},\
  and\ \citenamefont {Bonitz}}]{Dornheim_CPP_2019}%
  \BibitemOpen
  \bibfield  {author} {\bibinfo {author} {\bibfnamefont {Tobias}\ \bibnamefont
  {Dornheim}}, \bibinfo {author} {\bibfnamefont {Simon}\ \bibnamefont {Groth}},
  \ and\ \bibinfo {author} {\bibfnamefont {Michael}\ \bibnamefont {Bonitz}},\
  }\bibfield  {title} {\enquote {\bibinfo {title} {Permutation blocking path
  integral monte carlo simulations of degenerate electrons at finite
  temperature},}\ }\href {\doibase https://doi.org/10.1002/ctpp.201800157}
  {\bibfield  {journal} {\bibinfo  {journal} {Contributions to Plasma Physics}\
  }\textbf {\bibinfo {volume} {59}},\ \bibinfo {pages} {e201800157} (\bibinfo
  {year} {2019}{\natexlab{b}})}\BibitemShut {NoStop}%
\end{thebibliography}%
\end{document}